\documentclass[longauth]{aa}  
\usepackage[utf8]{inputenc}
\usepackage{comment}
\usepackage{enumitem}
\usepackage{graphicx}   
\usepackage{caption}
\usepackage{subcaption}
\usepackage{xcolor}
\usepackage{gensymb, amsmath,amssymb}
\usepackage{booktabs}
\usepackage[hidelinks,colorlinks=true,linkcolor=blue,citecolor=blue,urlcolor=black]{hyperref}
\usepackage{makecell} 
\usepackage{lscape}
\usepackage{lastpage}

\begin{document}

    \title{CHEX-MATE: New detections and properties of the radio diffuse emission in massive clusters with MeerKAT}
    
    
    \titlerunning{CHEX-MATE: New detections and properties of the radio diffuse emission with MeerKA}
    \authorrunning{M. Balboni et al.}
    
    \author{M. Balboni
          \inst{\ref{unibo}, \ref{iasf}, \ref{insubria}},
          F. Gastaldello \inst{\ref{iasf}},
          A. Bonafede\inst{\ref{unibo},\ref{ira}},
          A. Botteon\inst{\ref{ira}},  
          I. Bartalucci\inst{\ref{iasf}},
          R. Cassano\inst{\ref{ira}},
          S. De Grandi\inst{\ref{brera}}, 
          S. Ettori\inst{\ref{oas},\ref{infn_bo}},
          M. Gaspari\inst{\ref{unimore}},
          S. Ghizzardi\inst{\ref{iasf}}, 
          M. Gitti\inst{\ref{unibo},\ref{ira}}, 
          M. Johnston-Hollitt\inst{\ref{curtin}},
          L. Lovisari\inst{\ref{iasf},\ref{cfa}}, 
          S. Molendi\inst{\ref{iasf}}, 
          E. Pointecouteau\inst{\ref{irap}}, 
          G.W. Pratt\inst{\ref{saclay}},
          G. Riva\inst{\ref{iasf}}, 
          M. Rossetti\inst{\ref{iasf}}, 
          J. Sayers\inst{\ref{caltech}},
          M. Sereno\inst{\ref{oas},\ref{infn_bo}},
          R.J. van Weeren\inst{\ref{leiden}}
          }

\institute{
        DIFA - Universit\`a di Bologna, via Gobetti 93/2, I-40129 Bologna, Italy\label{unibo}
        \and
        INAF - IASF Milano, via A. Corti 12, 20133 Milano, Italy \label{iasf}
        \and
        DiSAT, Universit\`a degli Studi dell’Insubria, via Valleggio 11, I-22100 Como, Italy \label{insubria}
        \and
        INAF, Osservatorio di Astrofisica e Scienza dello Spazio, via Piero Gobetti 93/3, 40129 Bologna, Italy\label{oas}
        \and
        INFN, Sezione di Bologna, viale Berti Pichat 6/2, 40127 Bologna, Italy\label{infn_bo}
        \and
        INAF - IRA, Via Gobetti 101, I-40129 Bologna, Italy\label{ira}
        \and
        Department of Physics, Informatics and Mathematics, University of Modena and Reggio Emilia, 41125 Modena, Italy\label{unimore}
        \and
        IRAP, Université de Toulouse, CNRS, CNES, UT3-UPS, (Toulouse), France\label{irap}
        \and
        Universit\'e Paris-Saclay, Universit\'e Paris Cit\'e, CEA, CNRS, AIM, 91191, Gif-sur-Yvette, France \label{saclay}
        \and
        INAF - Osservatorio Astronomico di Brera, via E. Bianchi 46, I-23807 Merate (LC), Italy\label{brera}
        \and
        Leiden Observatory, Leiden University, PO Box 9513, 2300 RA Leiden, The Netherlands\label{leiden}
        \and
        California Institute of Technology, 1200 East California Boulevard, Pasadena, CA 91125, USA\label{caltech}
        \and
        Center for Astrophysics $|$ Harvard $\&$ Smithsonian, 60 Garden Street, Cambridge, MA 02138, USA\label{cfa}
        \and
        Curtin Institute for Data Science, Curtin University, GPO Box U1987, Perth, WA 6845, Australia\label{curtin}
        }

   \abstract{Modern radio telescopes are revolutionising our understanding of non-thermal phenomena in galaxy clusters, collecting large samples of extended sources with unprecedented sensitivity and angular resolution.
   In this work, we present novel MeerKAT observations for a sample of 21 galaxy clusters that are part of the CHEX-MATE project.
   These systems were selected based on their high mass and displaying signs of dynamical activity.
   Thanks to the high-quality data at hand, we were able to detect extended radio emission in every target considered.
   We report two new halos, one new relic, and two new candidate relics. We also confirm a previous candidate halo and two candidate relics.
   After investigating the scaling relations with the cluster properties, we confirmed the presence of a radio halo power-mass correlation and relate it to a higher radio halo emissivity in more massive clusters.
   For radio relics, we highlight the MeerKAT capabilities to significantly extend the depth of radio observations to a new, unexplored field of low-radio power sources ($\lesssim 10^{23} ~ {\rm W~Hz^{-1}} $ at 1.28 GHz).
   Thanks to such high-sensitivity data, we have found that the radio relic power can be characterised by a wide range of values for a given cluster mass and relic size. 
   Ultimately, we discuss how current radio observations, in combination with large radio surveys, are increasingly capable of testing numerical simulation predictions and coming close to performing direct comparisons with their data, enabling new insights on the evolution of radio relics.}
   
\maketitle
\section{Introduction}\label{sec:intro}

Galaxy clusters allow us to study a wide range of processes, from structure formation and cosmology to particle re-acceleration processes and radiative mechanisms. 
They are the end product of hierarchical accretion in the Universe, composed mainly of dark matter forming deep potential wells ($10^{14-15} M_{\odot}$). 
Baryons account for only $\approx 20\%$ of the total cluster mass and they are mostly in the form of a diffuse plasma called the intra-cluster medium \citep[ICM; e.g.][]{Voit2005}.
This gas is accreted during the cluster formation process and it is virialised at $10^{7-8}$ K by strong shocks.
ICM properties can be studied through observations in the X-rays, thanks to its thermal bremsstrahlung emission, and/or in the microwave, exploiting the Sunyaev-Zel'dovich (SZ) effect between the hot plasma and CMB photons \citep[e.g.][]{Bohringer10, Mroczkowski2019}. 
Alongside thermal plasma studies, analyses in the radio band enable us to explore a complementary cluster medium, known as the non-thermal component. 
In fact, while X-ray data offer valuable information on the well-studied virialised cluster matter, radio observations trace cosmic ray electrons (CRes) and magnetic fields within the cluster extension, for which physical constraints are more challenging to derive.

At present, it has been well established that within the cluster environment, many sources of radio synchrotron emission are present \citep[see e.g.][for an observational review]{vanWeeren19}. Giant radio halos are among the most puzzling and extended cases of such emission.
They are megaparsec-scale sources characterised by a steep spectral index ($\alpha < -1$, where $S_{\nu} \propto \nu^{\alpha}$), permeating the cluster volume and mainly found in massive merging clusters. Historically, two main mechanisms for the production of radio halos have been proposed, namely, either a hadronic or a turbulent origin (see \citealt{Brunetti14} for a review).
 {In the former, CRes form as secondary particles during collisions between Cosmic ray protons and thermal protons, while in the latter, CRes are produced via the re-acceleration of pre-existing, relativistic electrons in the ICM operated by merger-induced turbulence.}
In recent years, several observational results have been found to stand in contrast to pure hadronic model predictions \citep[e.g.][]{Thierbach2003, Cassano2010, Brunetti2017, Osinga2024} and, nowadays, a significant effort is being made to constrain the role of turbulence for the radio halo production.

During the structure formation process, galaxy clusters experience the accretion of sub-clusters and groups through mergers, which release up to $\sim 10^{64}$ ergs in the ICM in a cluster crossing time \citep[$\sim$ Gyr, ][]{Tormen2004}. This energy is mainly dissipated as gas heating through shocks, enhancing the ICM thermal bremsstrahlung emission in the X-rays. A small fraction of this energy can be channelled into turbulence that re-accelerates relativistic electrons via a Fermi-II like mechanism, producing synchrotron emission \citep{Brunetti2001,Petrosian2001}. 
Since this process is driven by gravity, the quantity that sets the initial energy budget is the cluster mass. Therefore, we expect that more massive clusters to host the most powerful radio halos (emitting up to GHz frequencies), whereas less massive systems (i.e. which have experienced less energetic mergers) will give rise to less luminous halos \citep[e.g.][]{Cassano06}. 
Turbulent re-acceleration models can reproduce the general properties of radio halos, such as the number and redshift distribution, and their expected flux density \citep{Cassano2023}. It also provides a natural explanation for two of the most important observational properties of radio halos, namely, the observed radio power-mass relation and the halo merger connection \citep[e.g.][]{Cassano2010,Cuciti2023}. 
Indeed, these two findings, which have been possible thanks to X-ray and SZ studies of clusters, have been crucial to constrain the role of galaxy cluster mergers in the formation of radio halos \citep[e.g.][]{Cassano06, Venturi2008, Basu2012}.
However, the details of such a process are not yet fully understood (especially with respect to the precise turbulent re-acceleration mechanism occurring). More systematic studies on the halo properties, such as their radio spectra, are required to shed light on this topic.\par
Alongside the formation of radio halos, a small fraction of the merger shock energy is also spent to accelerate electrons at relativistic energies through diffusive shock acceleration (DSA), where charged particles gain energy by being scattered back and forth across the shock front \citep[e.g.][]{Brunetti14}. 
Due to the presence of magnetic fields within the ICM, the formed CRes emit via a synchrotron process in the radio band, illuminating the shock front as a diffuse, arc-like structure called radio relic.
{The DSA model predictions have indeed been confirmed by observations, finding these sources to display power-law radio spectra and observing a high ($\sim 20-60\%$) polarisation fraction caused by the alignment of magnetic field lines over the relic surface \citep[see e.g.][and references therein]{vanWeeren19}.
However, some discrepancies were found between the observed and expected properties of radio relics.
For instance, the DSA of thermal electrons in the ICM is severely challenged by the typical low ($\lesssim 3$) Mach numbers of cluster shocks. Indeed, the high luminosities observed in many radio relics generally imply unphysically high acceleration efficiencies \citep{Botteon2020RR}.
To overcome this efficiency issue, it has been proposed that relics are generated by the re-acceleration of relativistic electrons already present in the ICM \citep[e.g.][]{Markevitch2005, Macario2011, Kang2011, Pinzke2013}. 
Alternatively, modifications of the standard DSA mechanism could also be invoked to explain the observations \citep{Kang2018, Zimbardo2018}.}\par
Thanks to the modern radio facilities (e.g. LOFAR, \citealt{LOFAR2013}, MeerKAT, \citealt{Jonas2016,Camilo2018}, ASKAP, \citealt{Hotan2021}, uGMRT, \citealt{Gupta2017}, JVLA, \citealt{Perley2011}, MWA \citealt{Tingay2013}), a new era for the study of diffuse radio sources is taking place, in terms of sensitivity, frequency coverage and angular resolution.
These capabilities are proving a new picture of radio sources and improving our understanding of the non-trivial distribution of CRes and magnetic fields within the cluster environment.

In this paper, we present MeerKAT L-band ($\sim 1.28$ GHz) observations of a sample of massive galaxy clusters taken from the Cluster HEritage project with \textit{XMM-Newton} - Mass Assembly and Thermodynamics at the Endpoint of structure formation \citep[CHEX-MATE,][]{Arnaud21}. 
We focus on the radio halos and relics within the considered systems, characterising their radio emission properties and integrating radio data with deep and homogeneous X-ray observations from \textit{XMM-Newton}, which provide information on the cluster dynamical activity.
First, we present our new MeerKAT observations to demonstrate the quality of the data and report new radio detections. Next, we use our novel results to perform statistical analyses of radio halo clusters and to expand upon previous literature studies.

The paper is organised as follows. In Sect.~\ref{sec:obs-data}, we present the cluster sample and the data reduction process. In Sect.~\ref{sec:sample}, we present new MeerKAT observations of CHEX-MATE clusters and briefly describe each object. In Sects.~\ref{sec:radio_halos} and \ref{sec:radio_relics}, we report and discuss the results of halo and relic analyses and perform a comparison with recent literature works. In Sect.~\ref{sec:sum&concl}, we conclude and summarise our results.
Throughout the paper we assume a flat, $\Lambda$CDM Universe cosmology with $H_0 = 70 \rm ~ km/s/Mpc$ and $\Omega_{m,0}=0.3$.
\section{Sample observations and data reduction}\label{sec:obs-data}
\subsection{CHEX-MATE}
The CHEX-MATE project \citep{Arnaud21} is a three mega-second \textit{XMM-Newton} Multi-Year Heritage Programme to obtain X-ray observations of a minimally biased, signal-to-noise limited sample of 118 galaxy clusters detected by Planck through the Sunyaev-Zeldovich effect. The program aims to study the ultimate products of structure formation in time and mass, using a census of the most recent objects to have formed (Tier-1: $0.05 < \rm{z} < 0.2$ ; $M_{500}> 2 \times 10^{14}~M_{\odot}$), together with a sample of the highest mass objects in the Universe (Tier-2: z$<0.6$; $M_{500}>7.25 \times 10^{14}~M_{\odot}$\footnote{As explained in \cite{Arnaud21}, the cluster $M_{500}$ estimates are those derived by the MMF3 Planck cluster catalogue \citep{Melin2006,Planck2016}, and $\rm M_{500} \propto 500 \rho_c R_{500}^3$ with $\rho_c$ as the critical density and $R_{500}$ as the radius within which the average cluster density is 500 $\rho_c$.)}.
The project acquired uniform depth X-ray exposures that ensure a detailed mapping of the X-ray emission in the cluster volume, making it the best choice for a systematic and statistical analysis of the thermodynamic properties of the cluster population.
A major objective of the CHEX-MATE collaboration is to ensure a multiwavelength (from radio to optical) coverage of the cluster sample \citep[e.g.][]{Sereno2025, Pizzuti2025} and to carry out numerical simulations. 
In this respect, we present homogeneous MeerKAT radio observations of CHEX-MATE clusters, highlighting the importance of sample studies performed with the new radio facilities and the capabilities of forthcoming radio observations of the CHEX-MATE objects. 

In the next section, we describe the data reduction and calibration for the radio data as it is the main focus of this work. However, we will also exploit X-ray information of each cluster using \textit{XMM-Newton} images in the 0.7-1.2 keV to perform a comparison between the two bands. The X-ray data reduction is extensively described in \cite{Bartalucci23} and \cite{Rossetti2024}. We refer to the cited works for more information on how such X-ray images are obtained.

{We also exploited the systematic investigations made on the CHEX-MATE sample by \citet[hereafter C22]{Campitiello2022}, who focused on the X-ray morphological cluster classification.
In fact, the analyses on the cluster morphological state are of crucial importance in the determination of the cluster dynamical state and in the characterisation of the diffuse radio emission.}
\subsection{Sample description}

The considered objects are all part of the CHEX-MATE Tier-2 sub-sample and, by construction, they are all massive clusters spanning a relatively wide redshift range: $0.15 \leq {\rm z} \leq 0.43 $ with $M_{500}$ between $ 7.86 \times 10^{14}~M_{\odot} < M_{500} < 13.74 \times 10^{14} ~ M_{\odot}$. 
Specifically, we selected targets at $\delta<0^{\degree}$, allowing for longer ($\sim 7.5$ h) tracking in the sky and an optimal uv-coverage by the MeerKAT telescope. 
Here, we considered all the MeerKAT observations of the southern CHEX-MATE clusters available to date.
In total, 22 targets have L-band MeerKAT coverage, out of 36 CHEX-MATE objects at $\delta<0^{\degree}$.

{To obtain the final sample, we combined the publicly available MeerKAT Galaxy Cluster Legacy Survey data \citep[MGCLS;][]{Knowles2022} with single-target observations  obtained via dedicated proposals (see Sect.~\ref{sec:obs_radio}).
Specifically, there were ten CHEX-MATE clusters presented in the MGCLS, of which we include nine here (i.e. removing one, PSZ2G106.87-83.23, due to poor image quality). 
We note that two targets in our dataset, namely, PSZ2G172.98-53.55 and PSZ2G262.27-35.38, were also part of the MGCLS, but we elect to present our deeper data here, thanks to an improved calibration strategy (see Sect.~\ref{sec:radio_data}).
For these two targets, we report the results found by the MGCLS and discuss the findings of our observations.}\par
 {In the top panel of Fig.~\ref{fig:sample_info}, we display the distribution of the CHEX-MATE sample in the mass-redshift plane, highlighting the targets considered in this work.
By construction, our sub-sample comprises massive and dynamically disturbed clusters of the CHEX-MATE sample. In addition, the latter is naturally less biased towards relaxed objects as it is an SZ-selected sample.
Therefore, we expect to find a high fraction of these systems to display radio halos and relics, which originate as a consequence of cluster dynamical activities. 
In this respect, our sample it is not meant to be representative of the cluster population and cannot be used to derive the fraction of radio sources in clusters. Instead, our aim is to use it to study the connection between the radio diffuse emission and the cluster dynamical activity and to discover new radio sources.
In the bottom panel of Fig. \ref{fig:sample_info}, we show the $c-w$ plot of the whole CHEX-MATE sample (using the values from C22), where $c$ is the concentration parameter \citep{Santos2008} and $w$ is the centroid shift \citep{OHara2006,Poole2006}. 
We note how our targets are preferentially found in the lower right quadrant of the $c-w$ figure.
However, some of them are also found in the relaxed part of the plot. This suggests a minor disturbance of the systems, though it is possible, as in the case of PSZ2G313.33-17.11 (one of the two objects classified as the most relaxed in the CHEX-MATE sample by C22), that projection effects are affecting the X-ray morphological classification.
\begin{figure}
    \centering
    \includegraphics[width=\linewidth]{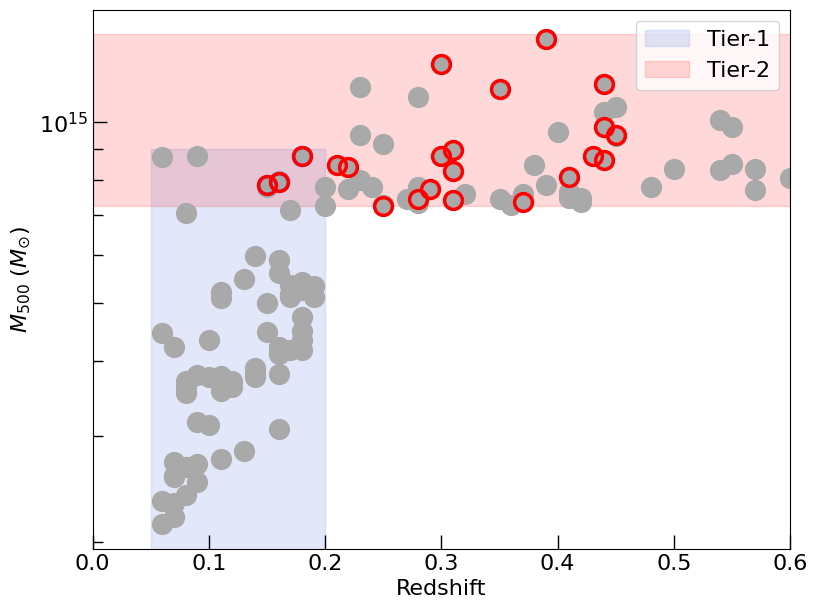}
    \includegraphics[width=\linewidth]{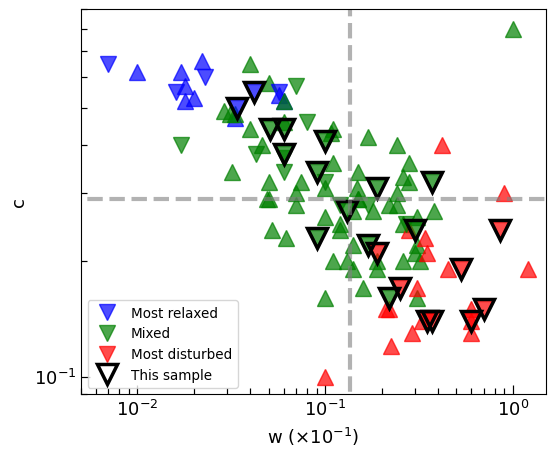}
    \caption{Visual comparison of the whole CHEX-MATE sample and the studied cluster sample. Top: Mass-redshift distribution of the CHEX-MATE clusters (grey) with highlighted in red the targets considered in this work, alongside Tier-1 ($0.05 < \rm{z} < 0.2$ ; $M_{500}> 2 \times 10^{14}~M_{\odot}$) and Tier-2 (z$<0.6$; $\rm{M_{500}}>7.25 \times 10^{14}~M_{\odot}$ sub-samples.
    Bottom: Same comparison as the top panel, but for the $c-w$ plane. The dynamical classification is taken from C22, with up (down) triangles indicating objects with $\delta>0{\degree} ~ (<0{\degree})$, while the two dashed lines are the median value of the $c$ and $w$ parameters of CHEX-MATE.}
    \label{fig:sample_info}
\end{figure}
General data on the targets analysed are presented in Table~\ref{tab:targets_info}.
\begin{table*}[ht]
    \centering
    \caption{Summary of the new radio observations used here.}
    \begin{tabular}{lccccc}
    \toprule
    Name & PID &\thead{ Taper \\ ($''$)} & \thead{Beam \\ ($'' ~ \times ~ ''$)} & \thead{$\sigma_{RMS}$ ~ \\ $(\mu {\rm Jy ~ beam^{-1}})$} & Diffuse sources \\
    \midrule
    PSZ2G008.31-64.74 & SCI-20230907-MB-02 & 21.9 & $22 \times 22$ & 12.6 & \thead{ RH* \\ SE-RR}\\
    PSZ2G056.93-55.08 & SCI-20220822-MB-01 & 17.4 & $18 \times 18$ & 11.6 & \thead{RH \\ RR*}\\
    PSZ2G172.98-53.55 & SCI-20230907-MB-02 & 38.9 & $40 \times 40$ & 40.8 & RH\\
    PSZ2G225.93-19.99 & SCI-20230907-MB-02 & 8.9 & $9 \times 9$ & 4.6 & RH$^\dagger$\\
    PSZ2G239.27-26.01 & SCI-20220822-MB-01 & 8.9 & $11 \times 10$ & 5.3 & RH \\
    PSZ2G243.15-73.84 & SCI-20230907-MB-02 & 9.2 & $10 \times 10$ & 3.7 & \thead{RH \\ RR-W/N \\ cRR-E$^\dagger$}\\
    PSZ2G262.27-35.38 & SCI-20230907-MB-02 & 22.7 & $23 \times 23$ & 11.0 & \thead{RH \\ RR-S* \\ RR-N + Tailed RG$^\dagger$}\\
    PSZ2G277.76-51.74 & SCI-20230907-MB-02 & 25.0 & $25 \times 25$ & 12.6 & \thead{RH$^\dagger$ \\ cRR$^\dagger$}\\
    PSZ2G278.58+39.16 & SCI-20220822-MB-01 & 11.1 & $12 \times 12$ & 8.5 & \thead{RH \\ RR \\ Elongation$^\dagger$}\\
    PSZ2G286.98+32.90 & SCI-20220822-MB-01 & 9.6 & $10 \times 10$ & 3.9 & \thead{RH \\  RR-NW+SE}\\
    PSZ2G313.33+61.13 & SCI-20220822-MB-01 & 16.2 & $17 \times 16$ & 9.9 & RH\\
    PSZ2G346.61+35.06 & SCI-20220822-MB-01 & 13.9 & $14 \times 14$ & 7.2 & \thead{RH \\ RR$^\dagger$}\\
    \bottomrule
    \end{tabular}
    \tablebib{
    (1) cluster name; (2) MeerKAT project ID; (3-5) properties of the source-subtracted, low-resolution images used to study the cluster diffuse emission, namely the Gaussian taper size applied, image resolution, and $\sigma_{\rm RMS}$; (6) classification of the detected diffuse sources as radio halo (RH), radio relic (RR and candidate cRR), or radio galaxy (RG), while  $\dagger$ indicates whether this source was first reported here and * if it was previously indicated as a candidate.
    }
    \label{tab:radio_info}
\end{table*}
\subsection{Observations}\label{sec:obs_radio}
The MeerKAT data used here comprise both the public products of the MGCLS and proprietary observations. 
We refer to \cite{Knowles2022} for a detailed description of MGCLS observations.
MeerKAT data for the remaining 12 clusters, instead, were collected through dedicated proposals in Cycles 3 and 4 (P.I.: M. Balboni, PIDs: SCI-20220822-MB-01 and SCI-20230907-MB-02). The first proposal focused on clusters known to host a radio halo and the second targeted the most dynamically perturbed objects among the ones left unobserved (with a morphological parameter of M>0, defined by C22).
The requested observations were carried out to ensure the same depth as the ones of the MCGLS, in the same band (950-1600 MHz) and with same time and frequency resolution (8 seconds and 4K channels), allowing for a consistent comparison of their non-thermal properties. 
We ensured at least 5.5~hr of on-source time for each target in order to reach a noise level of $\sim 10 ~{\rm \mu Jy ~ beam^{-1}}$ at $15-20^{''}$ (see Table~\ref{tab:radio_info}). 
The observing schedules were organised according to  standard MeerKAT practice, with a continuous target tracking of 30 min, followed by 2 min of secondary calibrator observations. The phase calibrator was set to be the closest to the target, with a typical angular separation below $13\degree$.
The primary calibrator was observed every three hours, with bandpass solutions and the absolute flux scale obtained using the model of either J0408-6545 or J1939-6342 provided by \cite{Hugo2021}. For these datasets, we do not present the polarisation results in this work, as it will be given in forthcoming works aimed to study the magnetised properties of galaxy clusters.
\subsection{MeerKAT data reduction}\label{sec:radio_data}
The following calibration and imaging process was  carried out only for the data collected by our dedicated observations. Instead, for those targets observed by the MGCLS, we used the low-resolution images at 15\arcsec\ that were made publicly available by \cite{Knowles2022}.

The data processing consists of two steps: we first exploited and applied the calibration solutions obtained by the SARAO Science Data Processor (SDP)\footnote{https://skaafrica.atlassian.net/wiki/spaces/ESDKB/
pages/338723406/SDP+pipelines+overview} and then we performed further self-calibration cycles via \texttt{facetselfcal} \citep{vanWeeren2021}, broadly following the steps detailed in \cite{Botteon2024-A754}.
In particular, we started by applying the Science Data Processor ``default calibration'' to our data using the \texttt{mvftoms.py} script of the \texttt{katdal}\footnote{https://archive.sarao.ac.za} package.
In this way, we obtained a calibrated measurement set for the target visibilities, corrected for delay, bandpass, and gain calibration \citep{Hugo2021}. 
We also compressed the data using \textsc{Dysco} \citep{Offringa2016} and averaged the visibilities by a factor of two both in time and frequency.
We performed the self-calibration procedure through the \texttt{facetselfcal} algorithm \citep{vanWeeren2021}. It uses \texttt{DP3} \citep{vanDiepen2018,Dijkema2023} and WSClean \citep{Offringa2014} for calibration and imaging, respectively. We typically performed four self-calibration cycles (two phase-only and two phase and amplitude) on the full MeerKAT field of view (FoV). Then, to speed up the subsequent steps, we adopted the extraction and self-calibration technique described in \cite{vanWeeren2021}.
First, we subtracted all the sources outside a box region (typically $0.4 \degree$, which was found to be a good compromise between having enough flux in the region and FoV reduction) containing the target. 
Then, we performed four other self-calibration cycles on the subtracted dataset, following the previous calibration strategy scheme.
For PSZ2G278.58+39.16 (Abell 1300) and PSZ2G286.98+32.90, after the extraction step, additional cycles of direction-dependent (DD) calibration were required due to residual artefacts of bright sources.
We carried out these DD self-calibrations  using \texttt{facetselfcal} every time, dividing the sky into four or five facets and performing two phase-only and two phase+amplitude cycles of calibrations.

The imaging was done with WSClean v3.4 \citep{Offringa2014} and with a specific focus on the extended emission. We first created an image with a uv cut corresponding to a physical size of 250 kpc at the cluster redshift to obtain a model for the compact sources. 
Then we subtracted such a model from the visibility, obtaining a source-subtracted dataset. 
Then, we produced low-resolution images using the \cite{Briggs1995} weighting scheme with \texttt{robust=-0.5} and applying a Gaussian uv taper in the visibility plane equivalent approximately to 50 kpc at the cluster redshift. For those objects showing a faint radio halo emission, we applied a taper equivalent to 100 kpc at the cluster redshift to increase the sensitivity to the diffuse emission.
More details on the final images obtained are reported in Table~\ref{tab:radio_info}.

In this work, the errors on the image flux densities take into account both the statistical and systematic uncertainty. To derive the latter, we exploited the data provided by the Rapid ASKAP Continuum Survey (RACS) at 1.37 GHz \citep{McConnell2020, Duchesne2024-RACSmid}. We compared the flux density values for the bright, compact sources present in our final images that were detected by the RACS observations. We rescaled our flux values, assuming a fiducial $\alpha=-0.7$ \citep{Knowles2022}, and found an average agreement within 6\%. Therefore, we considered a systematic flux uncertainty of 6\% for our data.
Additionally, we performed a visual inspection checking for contaminating sources, both compact and extended. 
We searched for residual source emission due to poor subtraction, such as the emission from tailed active galactic nuclei (AGNs), and we also searched for very low (high) surface brightness regions clearly not associated with the radio halo emission (e.g. diffuse emission detached from the central halo, revived fossil plasma or cluster's sub-components).
The detected contaminated regions were then masked out from the final image.
The results of the described data reduction are presented in Fig.~\ref{fig:RXimgs}, where we show both the radio and X-ray data for the new MeerKAT observations presented in this work.
\begin{figure*}[ht]
    \centering
    \includegraphics[width=0.305\linewidth]{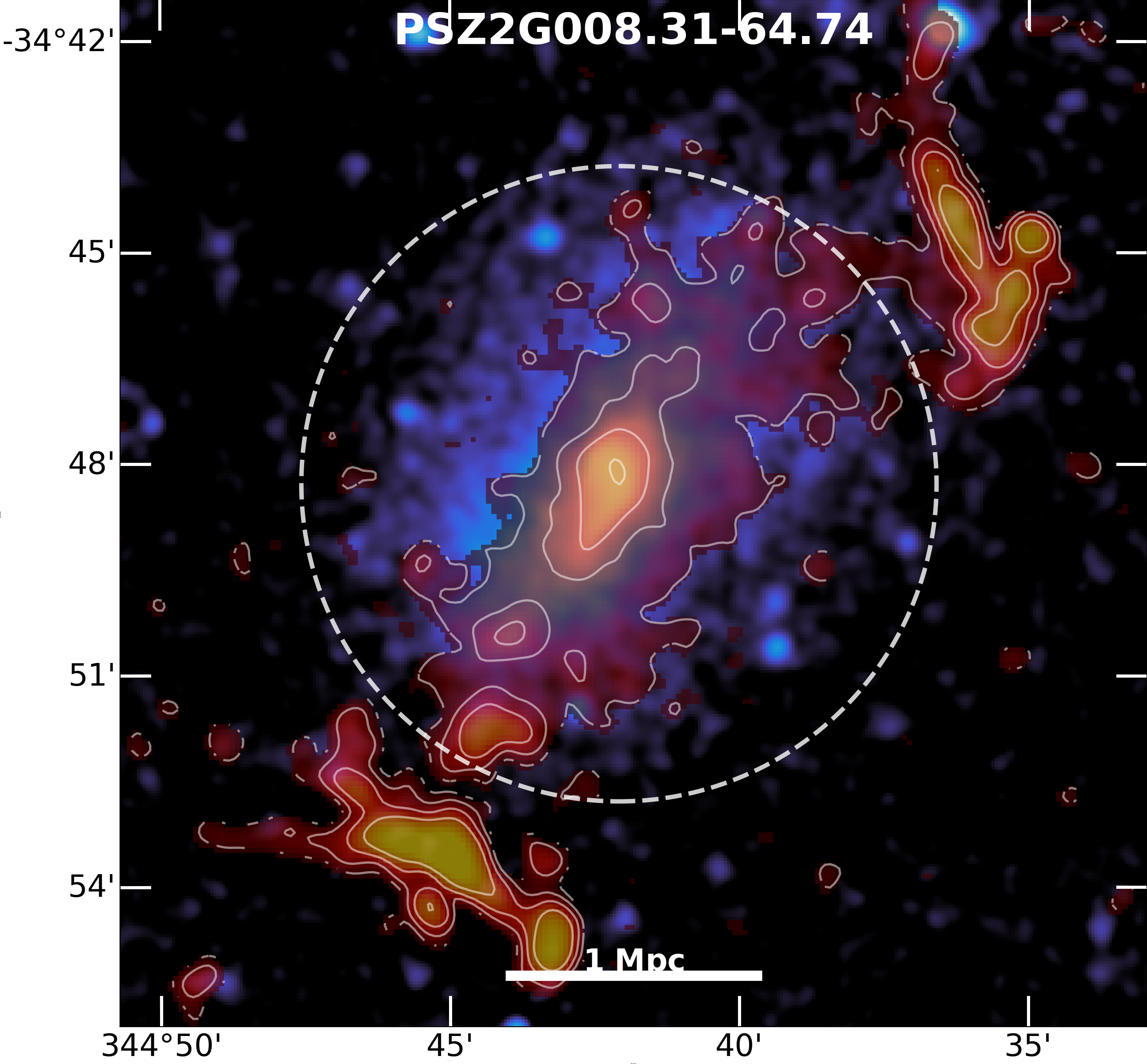}
    \includegraphics[width=0.305\linewidth]{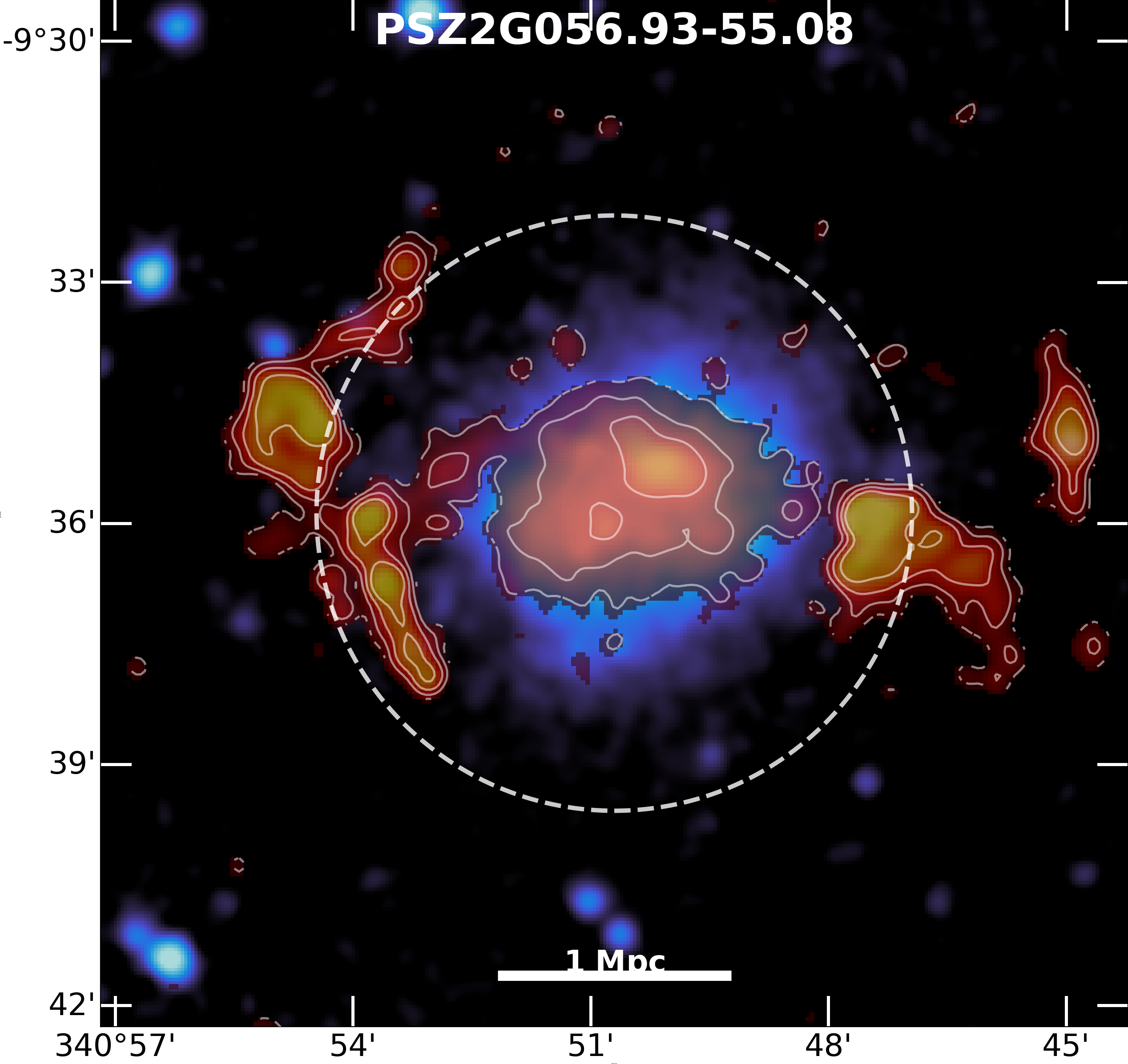}
    \includegraphics[width=0.305\linewidth]{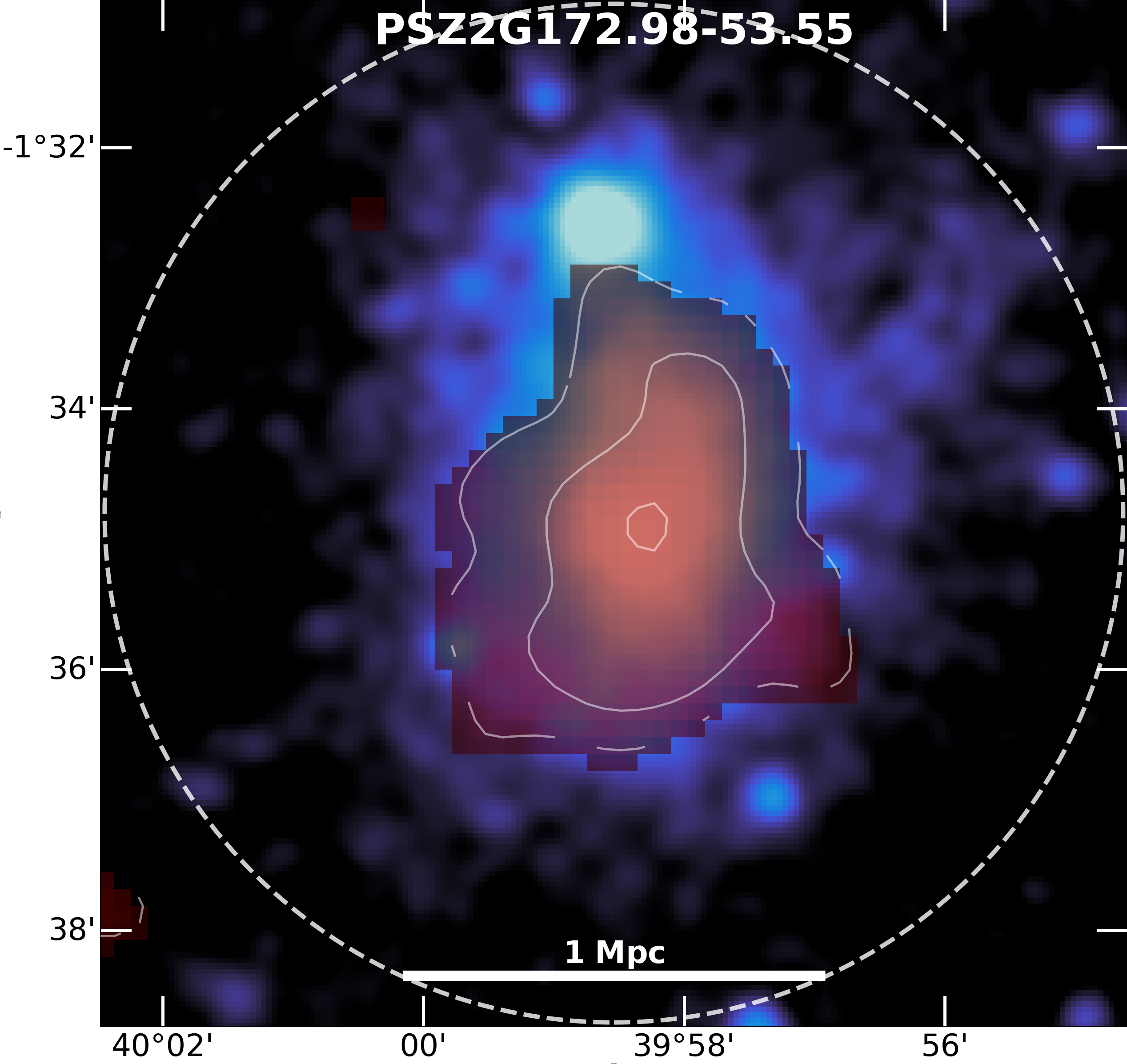}
    \includegraphics[width=0.305\linewidth]{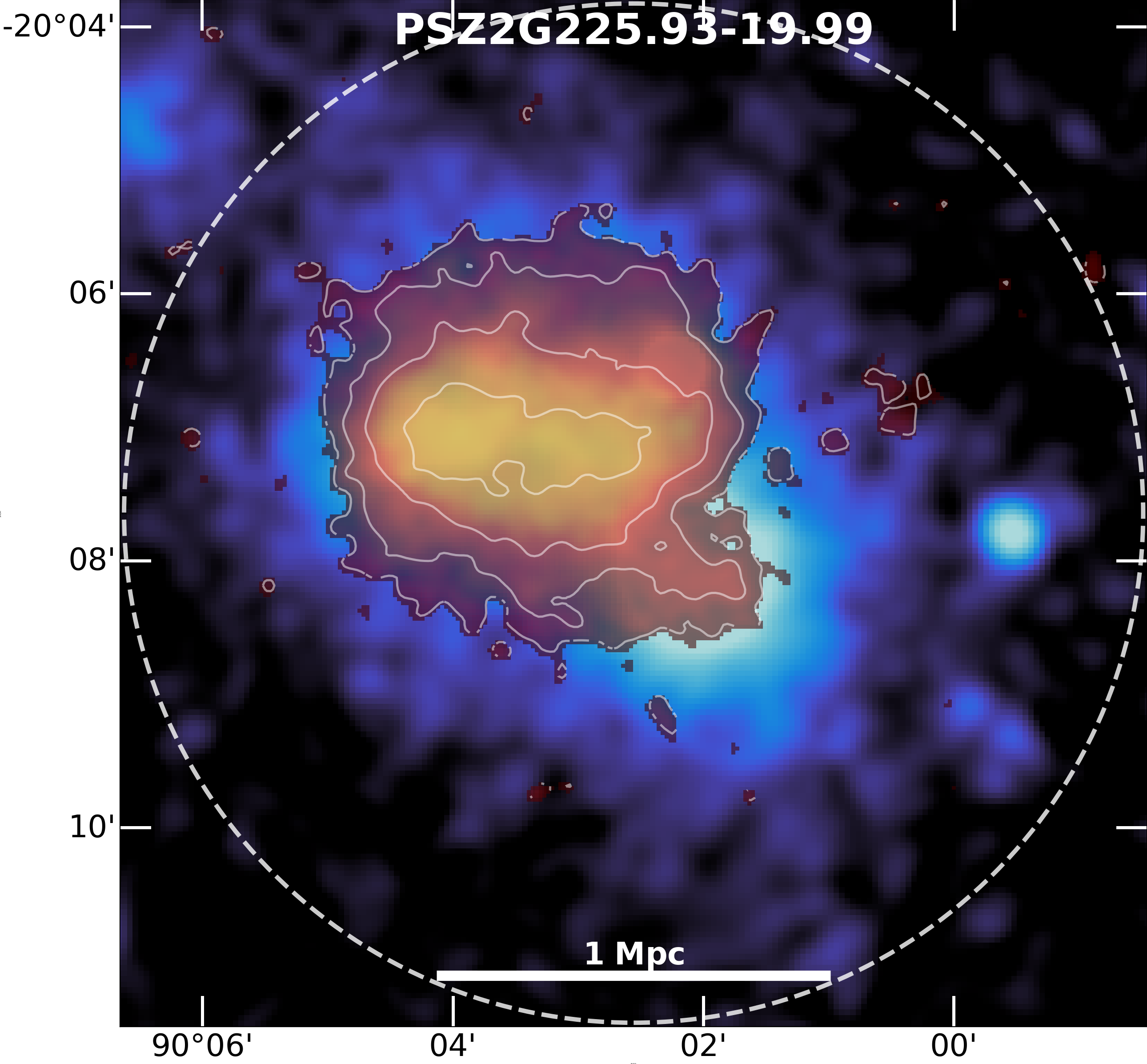}
    \includegraphics[width=0.305\linewidth]{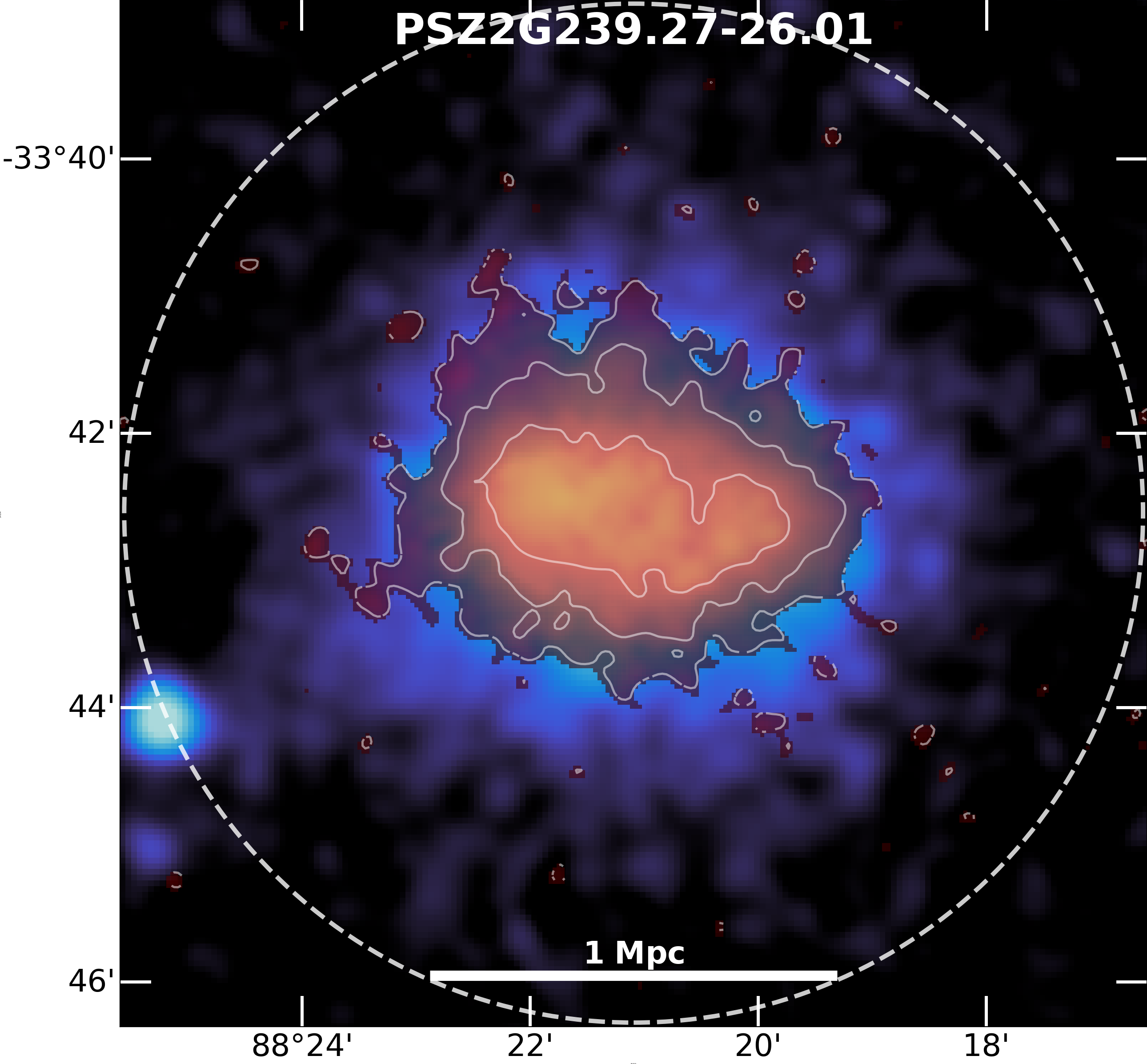}
    \includegraphics[width=0.305\linewidth]{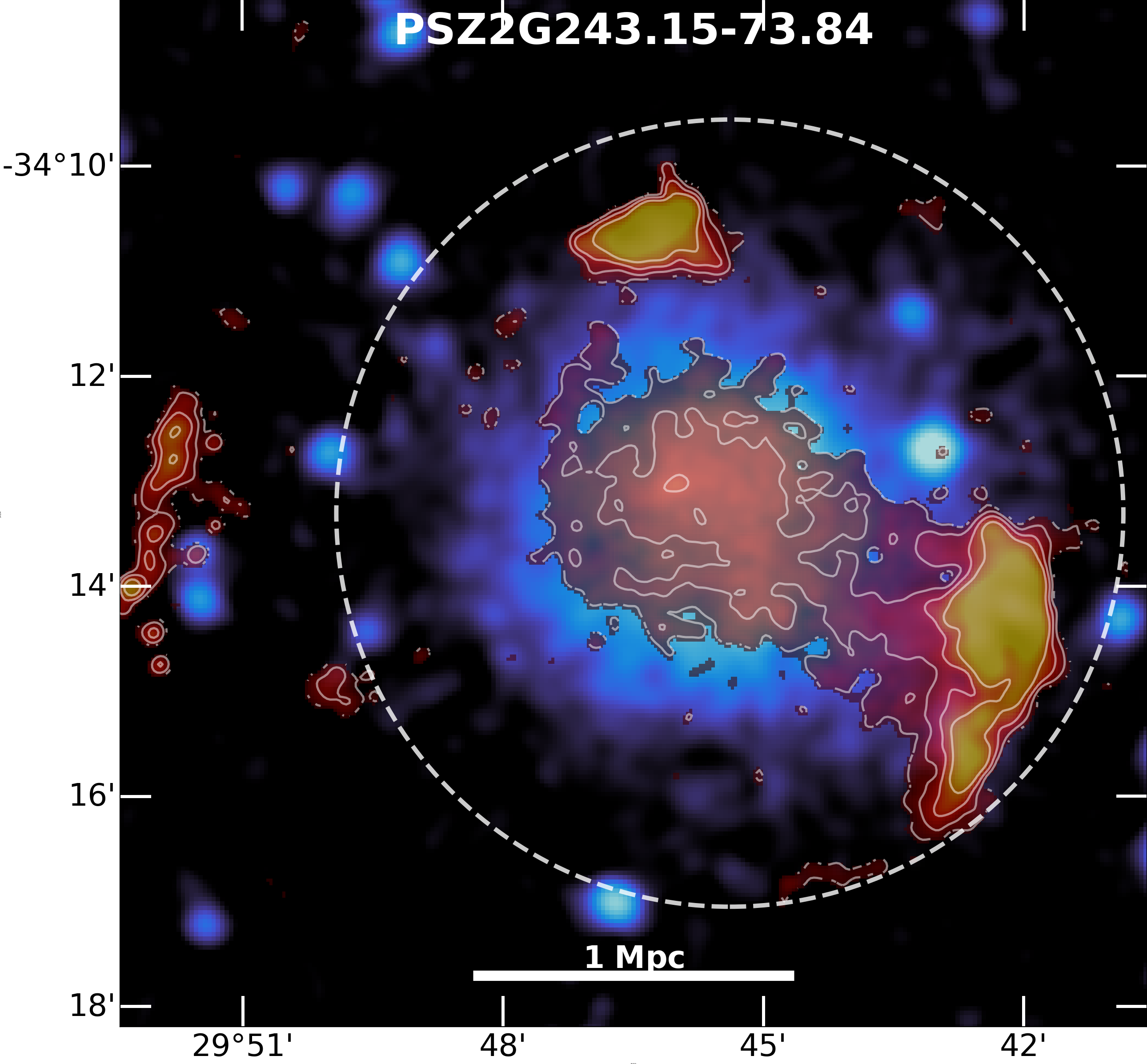}
    \includegraphics[width=0.305\linewidth]{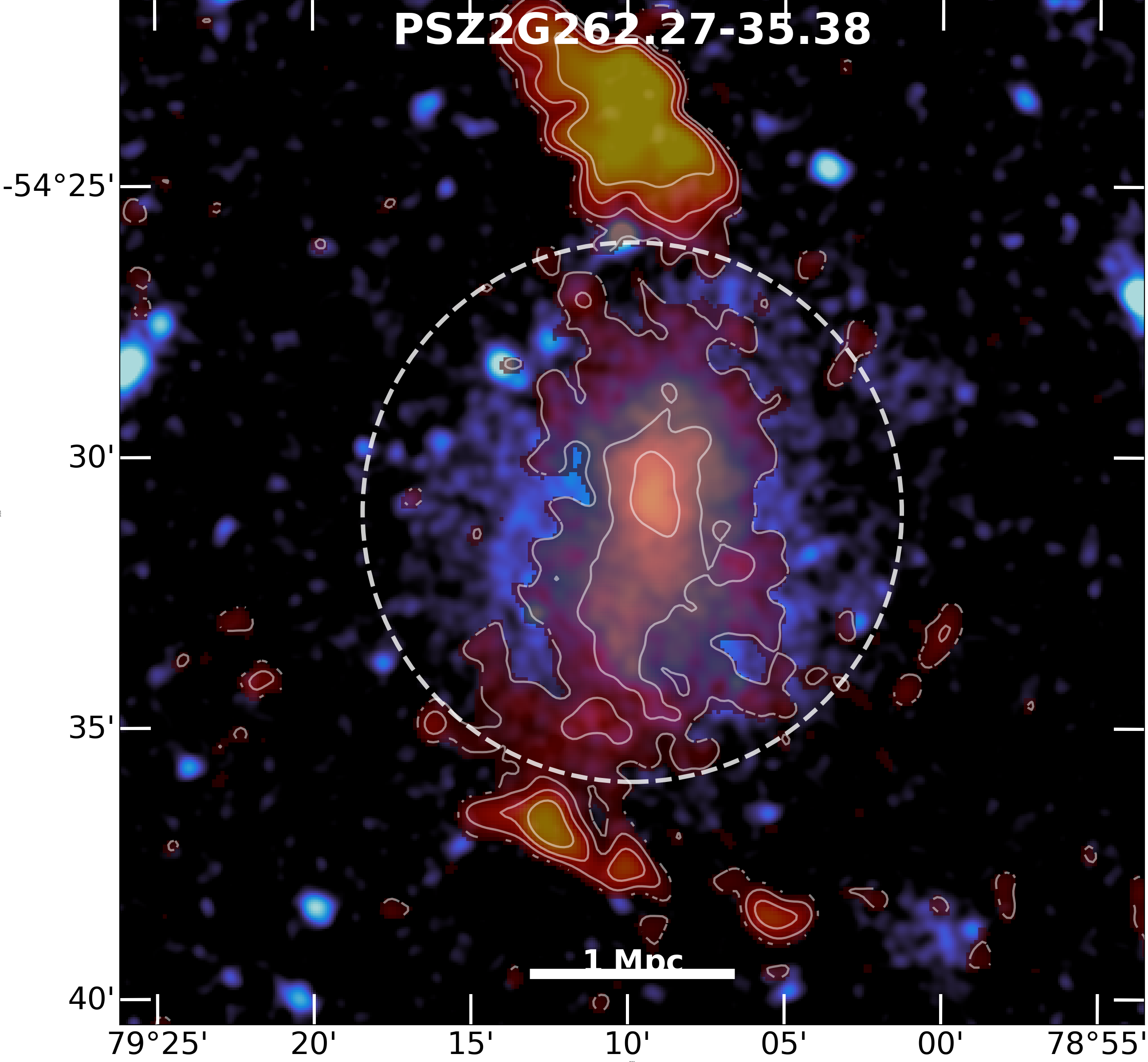}
    \includegraphics[width=0.305\linewidth]{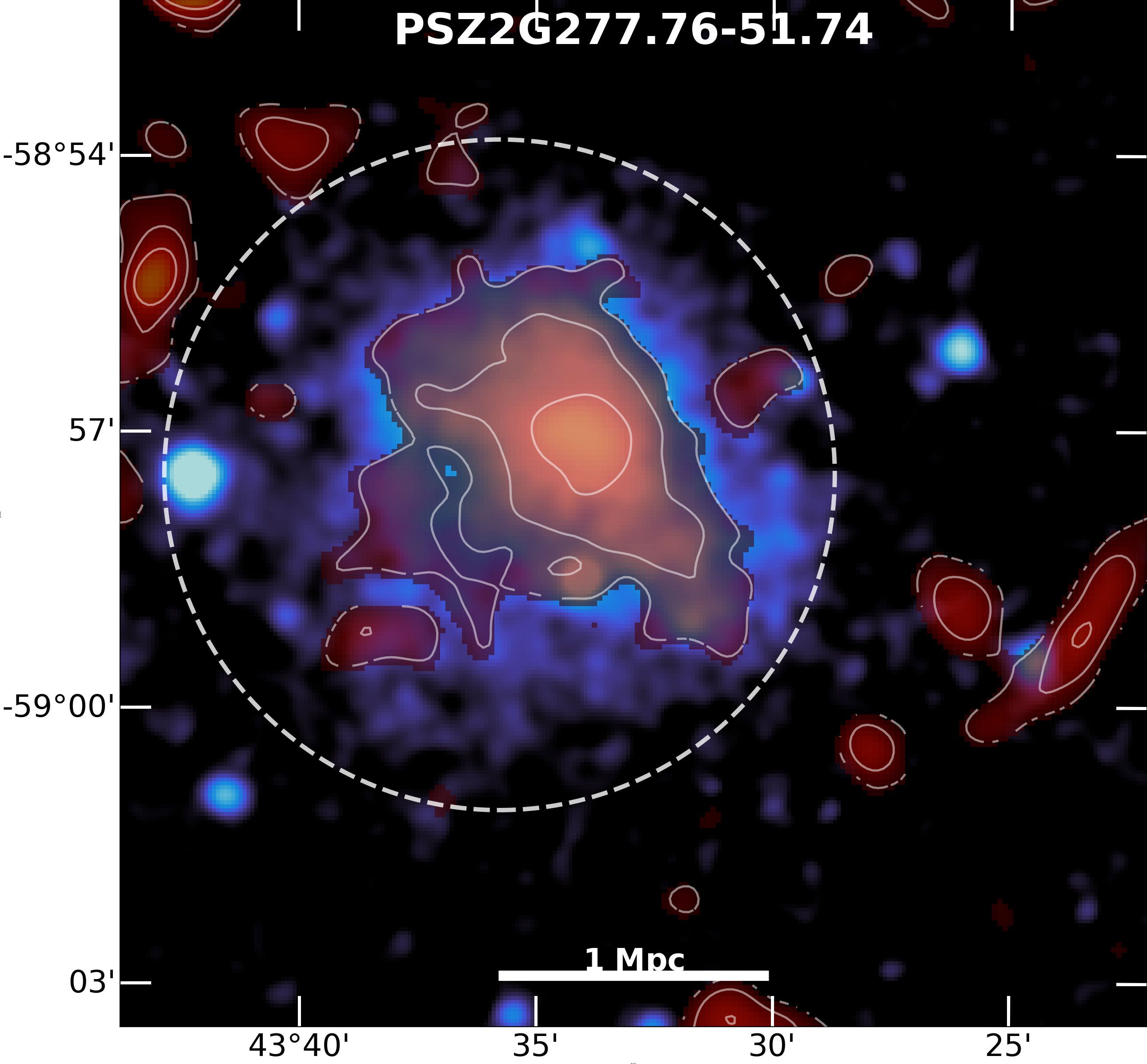}
    \includegraphics[width=0.305\linewidth]{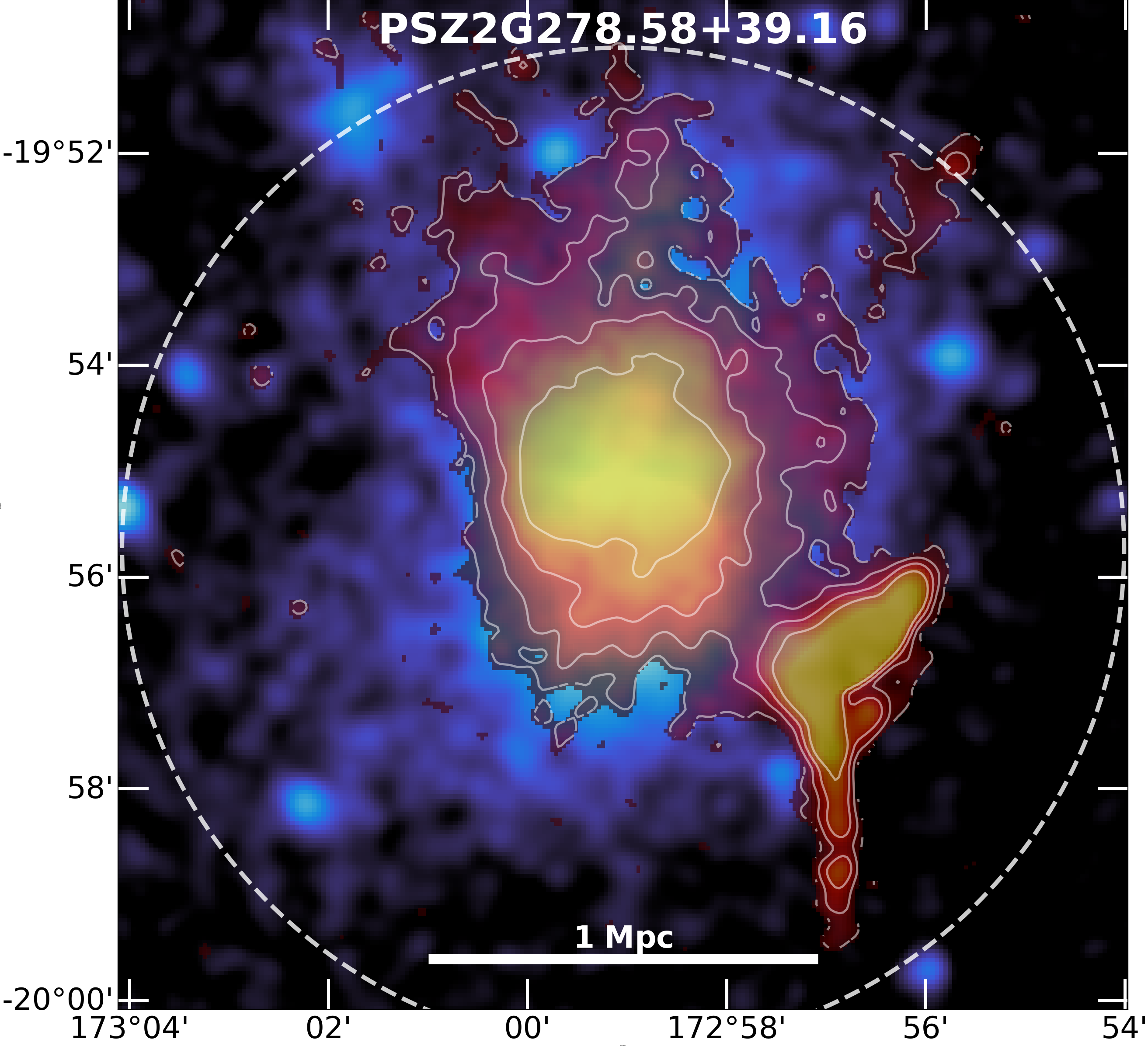}
    \includegraphics[width=0.305\linewidth]{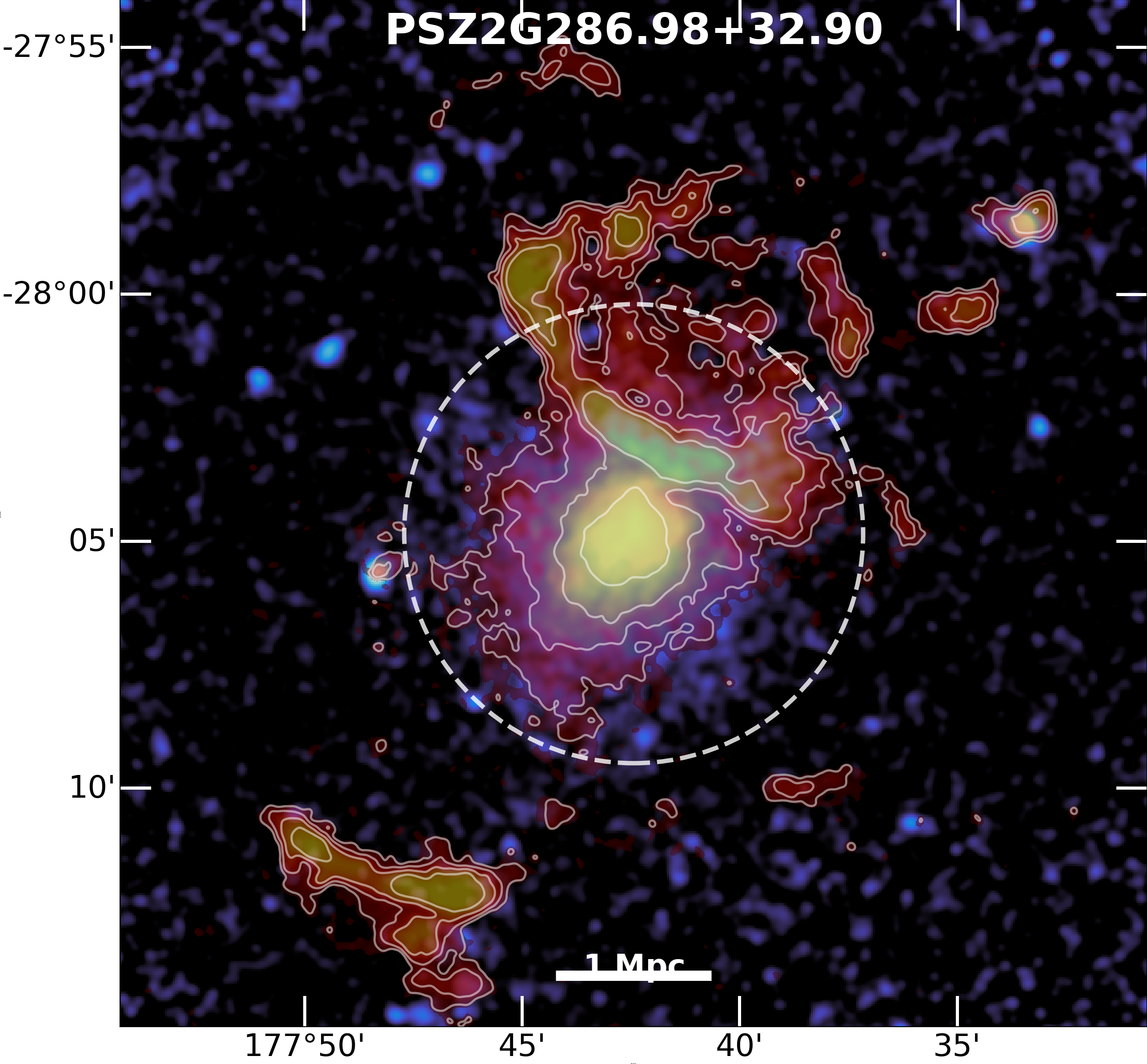}
    \includegraphics[width=0.305\linewidth]{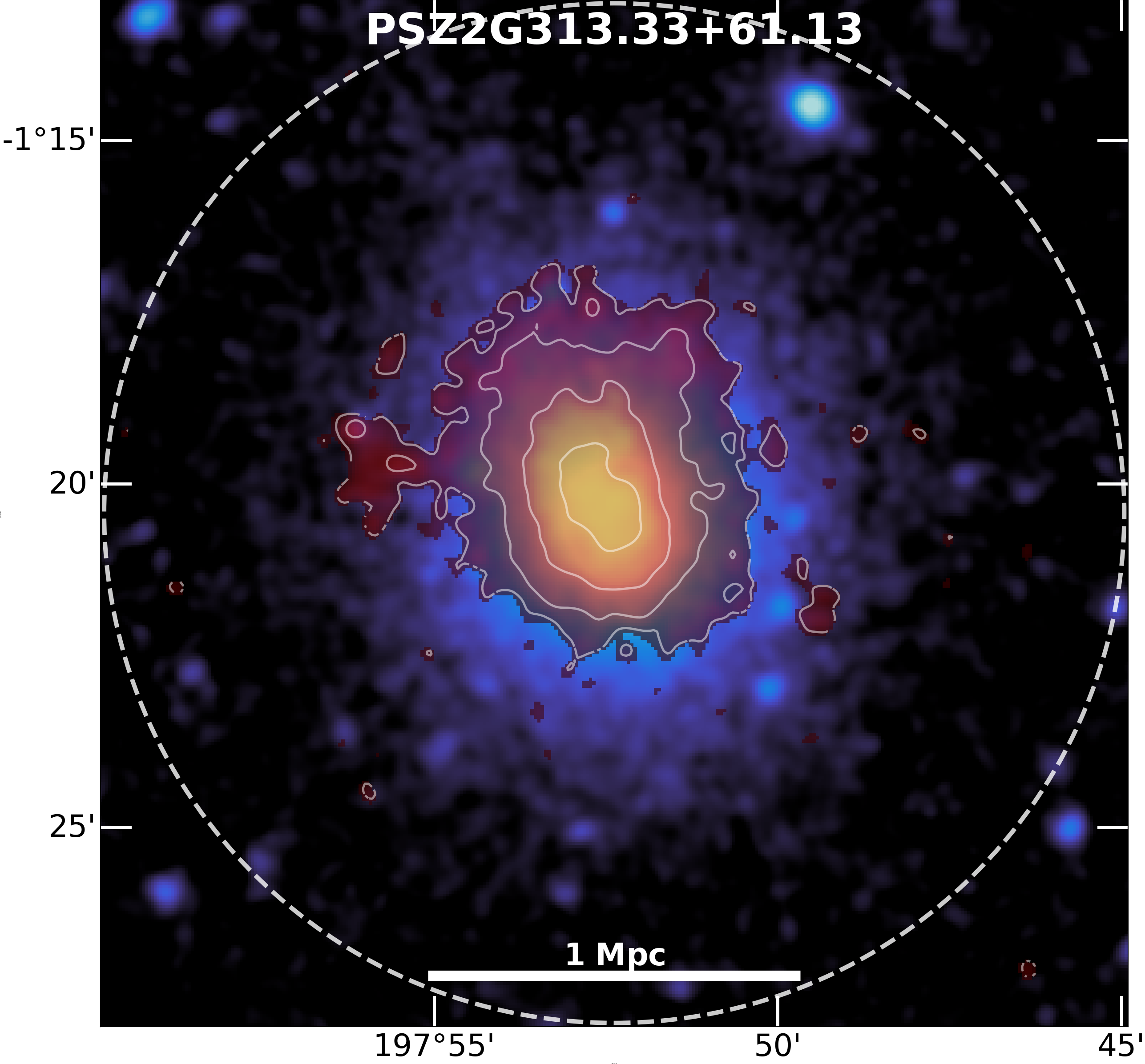}
    \includegraphics[width=0.305\linewidth]{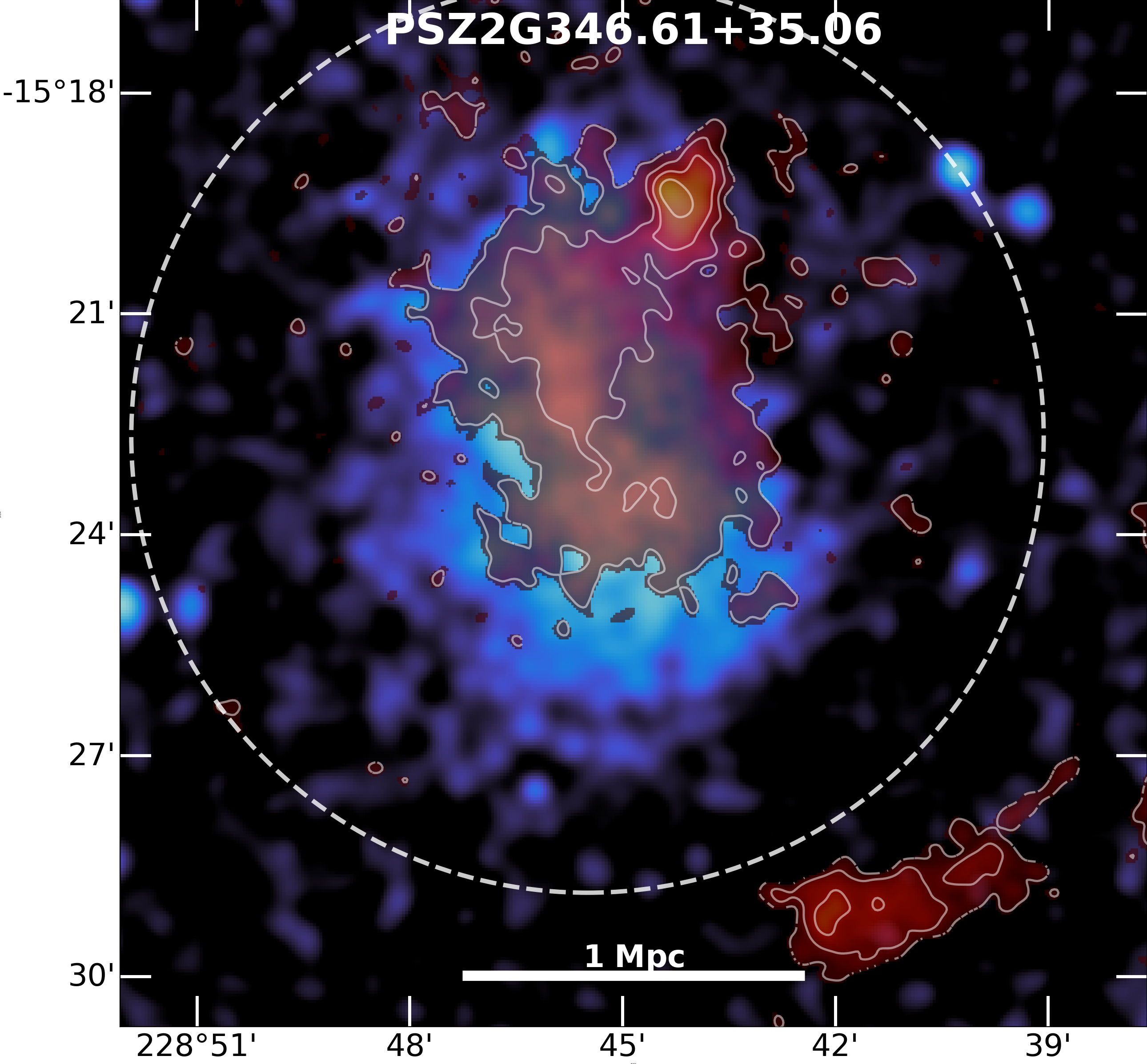}
    \caption{Radio (red with white contours) and X-ray (blue) overlays of the diffuse emission in the galaxy clusters covered by our new MeerKAT observations. We also report the cluster $R_{500}$ with a dashed white line.}
    \label{fig:RXimgs}
\end{figure*}
\section{Sample overview and results on individual clusters}\label{sec:sample}
In the following, we briefly present each target observed by our new dedicated observations and discuss the results obtained using the data reduction procedure described in Sect.~\ref{sec:obs-data}.
We highlight the newly discovered sources (see Table~\ref{tab:radio_info}), compare the radio information with X-ray data, and (when required to classify the detected radio emission) present the derived spectral maps, computed by producing four in-band images of the whole 950-1570 MHz bandwidth and applying a inner uv cut of 200 $\lambda$.
{We do not further exploit the in-band spectral index maps in this work, as a dedicated analysis on the radio and X-ray spectral properties for this cluster sample will be made in a forthcoming work (Balboni et al. in prep.).}

The radio images are obtained as described in Sect.~\ref{sec:radio_data} (unless otherwise specified).
We computed radio spectral index maps using lower resolution images, typically obtained applying a Gaussian uv taper equivalent to 100 or 150 kpc at the cluster redshift. 
Uncertainty maps for all spectral index maps shown in this paper are provided in Appendix~\ref{appendix:spidx_err}, whereas we report MGCLS images in Appendix~\ref{appendix:mgcls_imgs}.
\paragraph{PSZ2G008.31-64.74 (AC114n, Fig.~\ref{fig:1})}
This is a dynamically active cluster, as demonstrated in prior X-ray and optical studies (\citealt{DeFilippis2004,Sereno2010, Lovisari2024}). 
{The X-ray emission is elongated in the south-east to north-west (SE-NW) direction, with two surface brightness discontinuities detected by \cite{DeFilippis2004}.
\cite{Duchesne2024-EMU} reported central diffuse emission, which they classified as a candidate halo. Our source-subtracted images clearly reveal central diffuse radio emission that is well aligned with the thermal emission (Fig.~\ref{fig:1}), allowing for its classification as a radio halo.
We also detected the relic in the south-east (SE) region reported by \cite{Duchesne2024-EMU}, which shows a clear spectral steepening towards the cluster centre (Fig.~\ref{fig:1}). 
Furthermore, we recovered the elongated radio emission in the NW part of the cluster that \cite{Duchesne2024-EMU} classified as a radio relic. Thanks to the MeerKAT high-resolution images and the spectral index imaging, we argue that it is caused by a complex of radio galaxy tails that extend up to $\sim 1$ Mpc in size.
\begin{figure}[h]
    \center{\includegraphics[width=\linewidth]{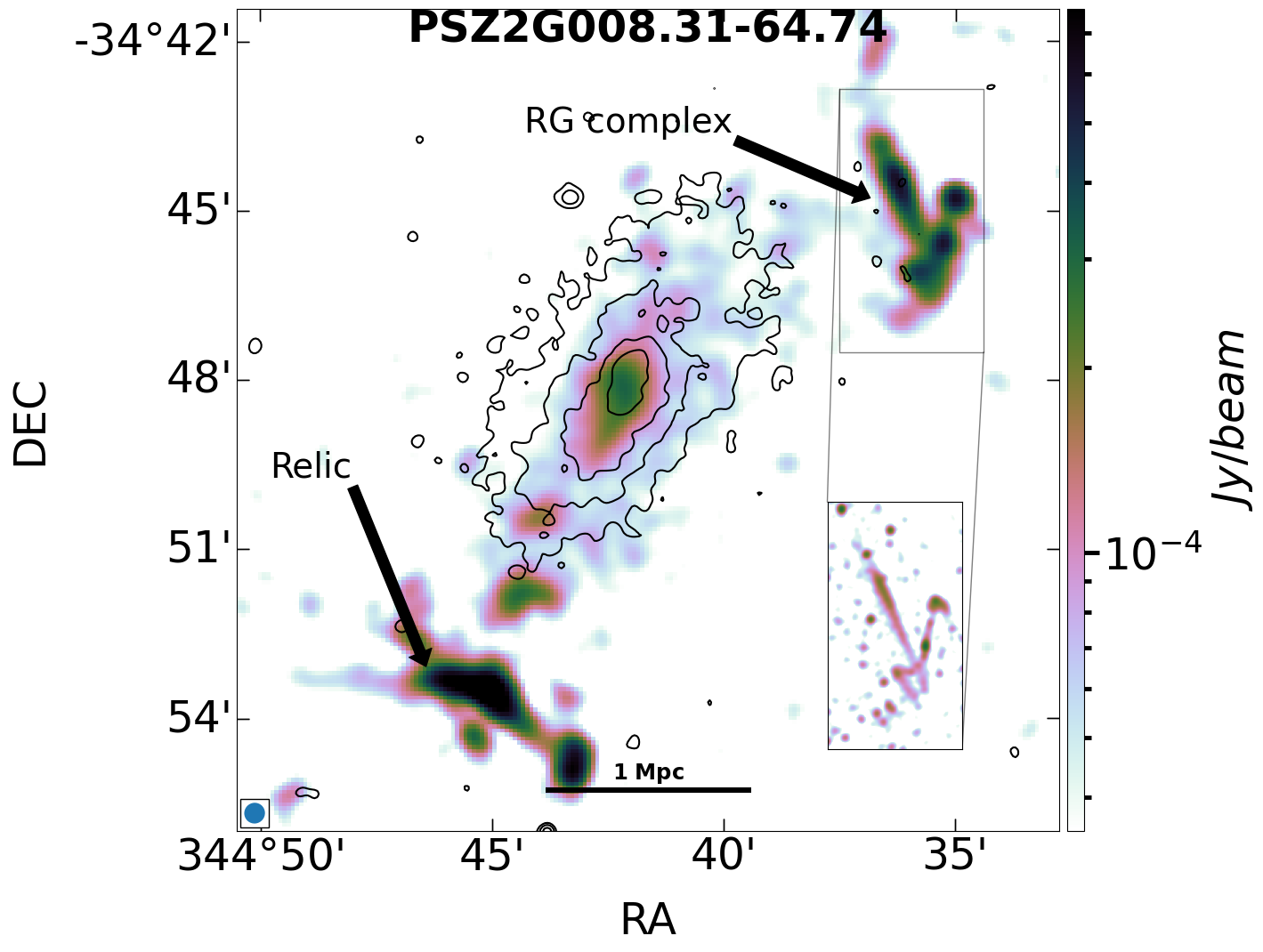}}
    \includegraphics[scale=0.12]{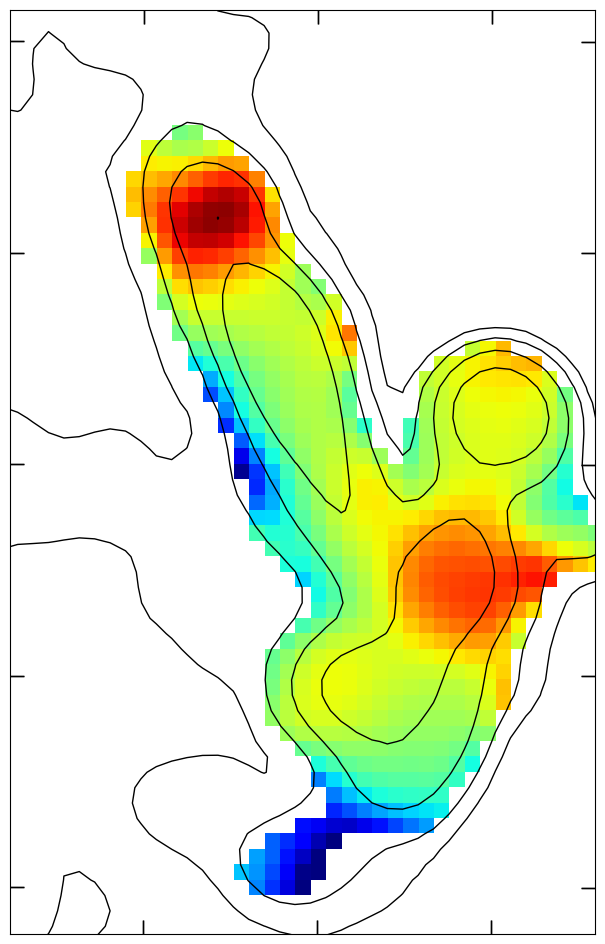}
    \includegraphics[scale=0.2]{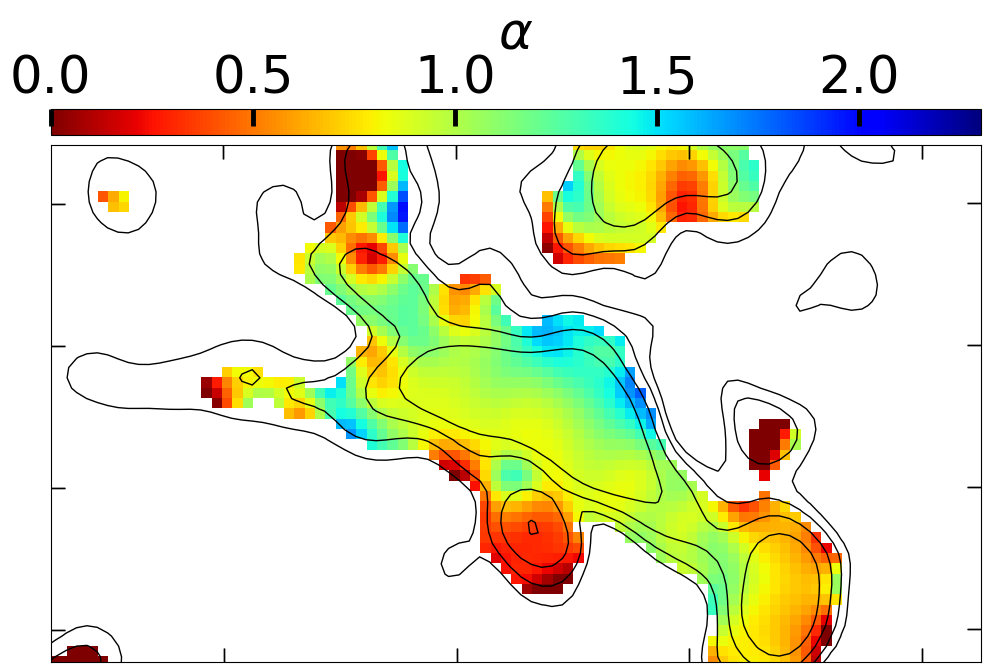}    \caption{MeerKAT image of PSZ2G008.31-04.74. Top: Radio emission above $3\sigma_{\rm RMS}$ after the source subtraction of discrete sources (at 22\arcsec resolution and $\sigma_{RMS} = 12.6 ~\mu \rm{Jy ~ beam^{-1}}$) and with an inset showing the high-resolution ($\sim$ 7\arcsec) map for the radio galaxy complex. X-ray contours are overlayed in black, starting at $2\times10^{-6}$ cts/s and spaced by a factor of $2$. Bottom: Spectral index map considering only the signal above $3\sigma_{\rm RMS}$ in the two bands used to derive $\alpha$, with overlayed the contours of the radio emission of the left image. Spectral index uncertainty maps are shown in Fig~\ref{fig:1err}.}
    \label{fig:1}
\end{figure}
\paragraph{PSZ2G056.93-55.08 (MACSJ2243.3-0935, Fig.~\ref{fig:2})}
This system is found at the centre of the supercluster SCL2243-0935 \citep{Schrimer2011} and its dynamical state has been probed at different wavelengths \citep{Ebeling2010,Mann2012,Wen2013}. 
It has been classified as a disturbed object with an elongated X-ray emission; however, this emission is misaligned with the merger axis suggested by the galaxy projected distribution. 
Its radio emission has been investigated by \cite{Cantwell2016} and \cite{Parekh2017}, who detected the central halo emission and the brightest part of the other diffuse sources, two of which were classified as complex of radio galaxies (labelled as B and C in Fig.~\ref{fig:2}, following \citealt{Cantwell2016}). 
\cite{Cantwell2016} also discussed the possibility of the most western source being a candidate radio relic originating from the infalling material.
We recovered all those sources at high significance, in addition to new diffuse emission. 
In particular, we find the halo emission to display an elongation in the (east-west) EW direction, well aligned with the X-ray emission, which shows an asymmetric morphology.
We also report a connection with the central part of source B, which, however, is likely to come from the superimposition of background sources, as pointed out by \cite{Cantwell2016}.
Our spectral index maps confirm the presence of radio galaxy emission of sources B and C, with a steepening of the spectrum moving away from the emission peak.
Thanks to the spectral information, we can definitely classify the elongated Western source as a radio relic. However, unlike what was  proposed by \cite{Cantwell2016}, the spectral steepening shows an inward direction, which is consistent with a merger shock.
\begin{figure}[ht!]
\includegraphics[width=\linewidth]{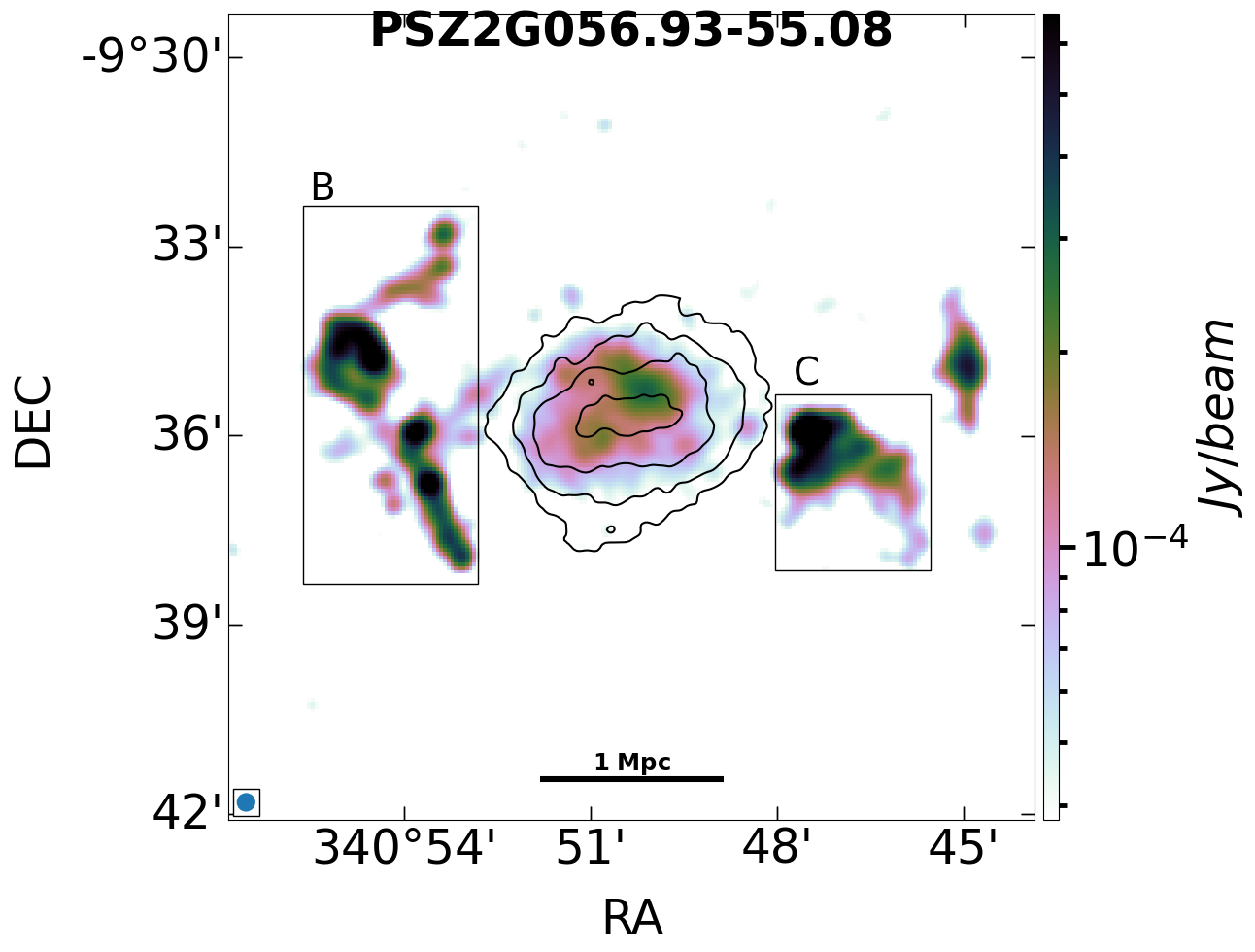}
\includegraphics[height=3.4cm]{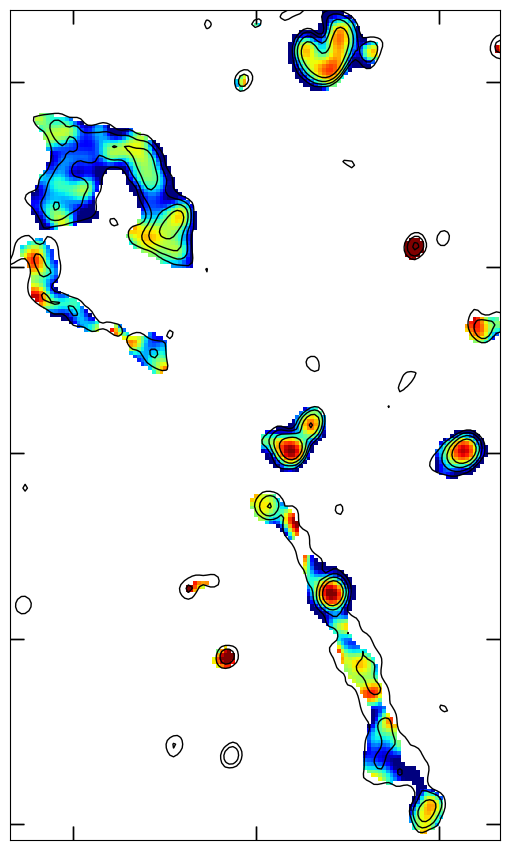}
\includegraphics[height=3.35cm]{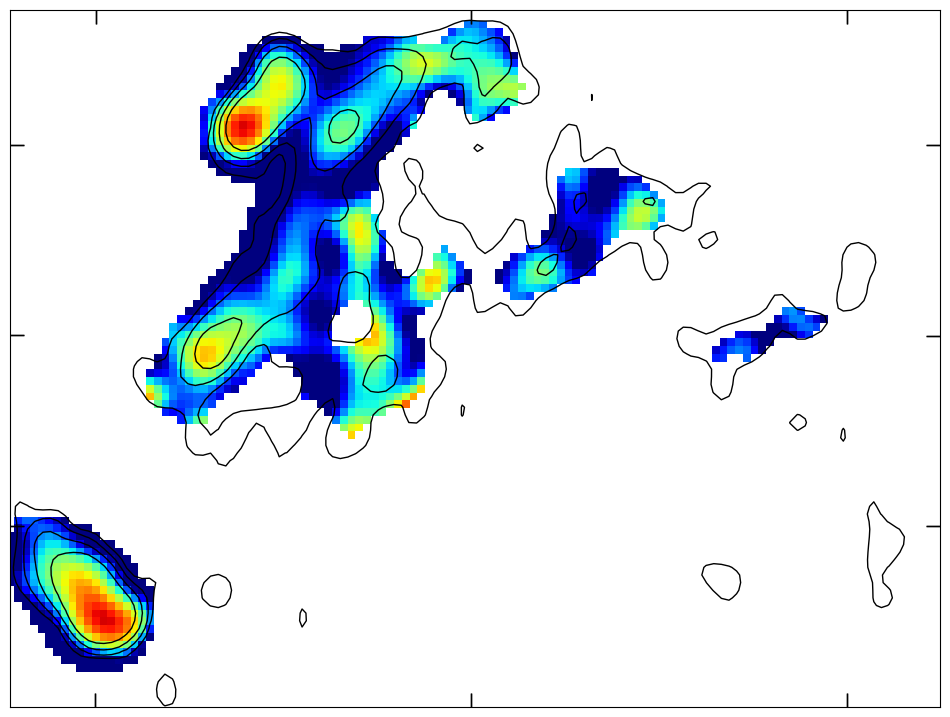}
\includegraphics[height=3.4cm]{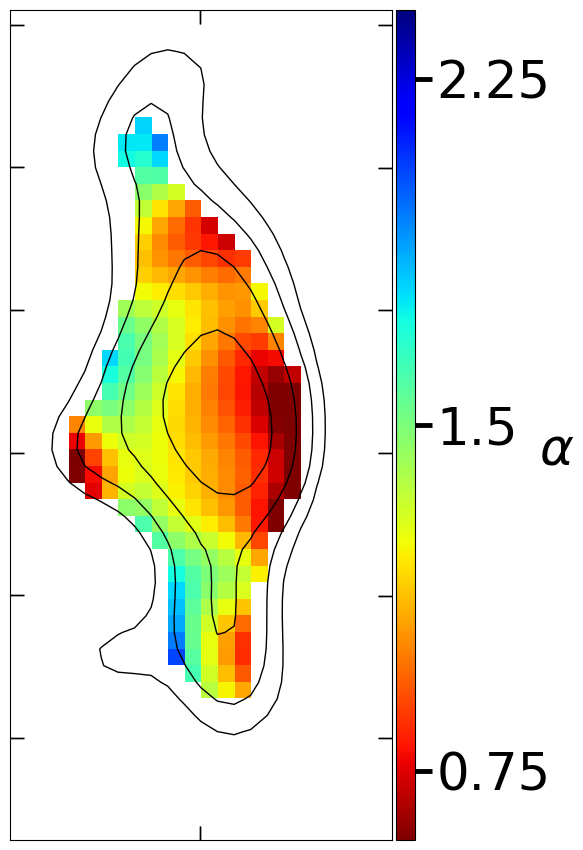}
    \caption{Same as Fig.~\ref{fig:1} but for PSZ2G056.93-55.08 (at 18\arcsec resolution and $\sigma_{RMS} = 11.6 ~\mu \rm{Jy ~ beam^{-1}}$), showing the spectral index map of the region B and C (at $\sim 7\arcsec$) and for the relic in the west. The spectral index uncertainty maps are shown in Fig~\ref{fig:2err}.}
    \label{fig:2}
\end{figure}

\paragraph{PSZ2G172.98-53.55 (A370, Fig.~\ref{fig:3})}
This cluster has been well studied in the optical band as it was the first object for which a gravitational arc was observed (\citealt{Soucail1987}). Its X-ray emission is elongated in the north-south (NS) direction and it has been studied in detail through \emph{Chandra} observations by \cite{Botteon2018-discontinuities}. They found two instances of surface brightness discontinuity whose nature is still uncertain.
Subsequently, \cite{Xie2020} studied the radio emission coming from this object with uGMRT and JVLA, finding a very faint diffuse emission, which they classified as a radio halo. 
More recently, \cite{Knowles2022} and \cite{Duchesne2024-EMU} detected an extended central radio source which was classified as a candidate radio halo.
Using our low-resolution observations, we detected a diffuse source that can be labelled as a radio halo and which is brighter in the southern part of the cluster (Fig.~\ref{fig:3}).
To recover this diffuse emission, we had to degrade the resolution by applying a Gaussian kernel of 200 kpc at the cluster redshift. 
This emission appears to fill a large portion of the cluster, particularly in the southern part.
\begin{figure}[ht!]
    \centering
    \includegraphics[width=\linewidth]{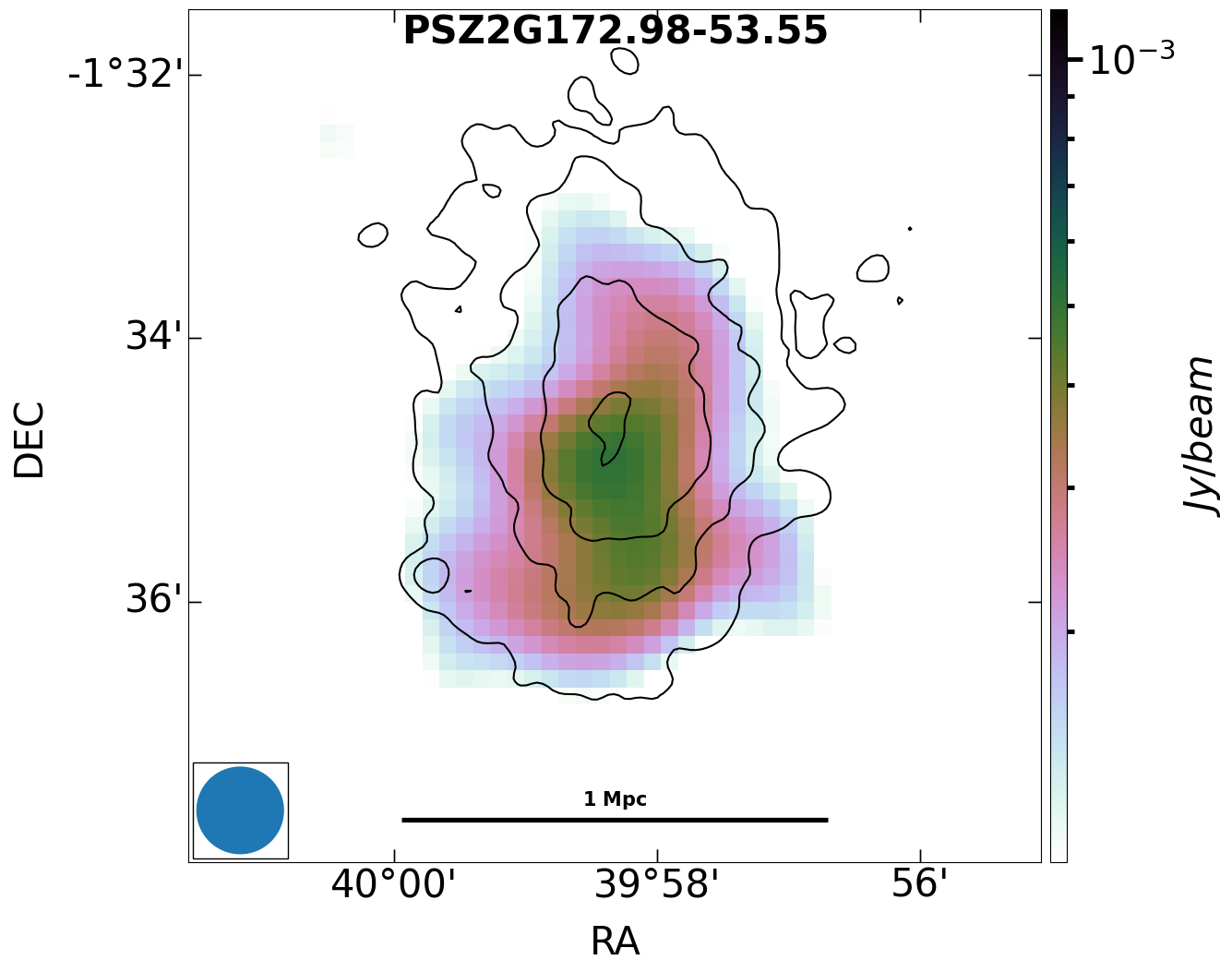}
    \caption{Same as Fig.~\ref{fig:1}, but for PSZ2G172.98-53.55 (at 40\arcsec resolution and $\sigma_{RMS} = 40.8 ~\mu \rm{Jy ~ beam^{-1}}$)}.
    \label{fig:3}
\end{figure}
\paragraph{PSZ2G225.93-19.99 (MACSJ0600.1-2008, Fig.~\ref{fig:5})}
PSZ2G225.93-19.99 is the fourth most perturbed object of CHEX-MATE.
Given its high SZ mass ($M_{500} \sim 10^{15} M_{\odot}$), it was included in the REionisation LensIng Cluster Survey (RELICS, \citealt{Coe2019}), which suggested a complex mass distribution.
More recently, \cite{Furtak2024} performed a detail lensing analysis of this cluster, finding an extended structure
to the north-east and, exploiting \textit{XMM-Newton} data, measuring a cooler temperature in the X-ray peak (south-west) region than the eastern part of the system.
Hee, we present the first radio image of this cluster, where diffuse radio halo emission is detected at high significance (Fig.~\ref{fig:5}). Interestingly, there is a clear offset between the bulk of the radio emission and the X-ray peak. 
\begin{figure}[ht!]
    \centering
    \includegraphics[width=\linewidth]{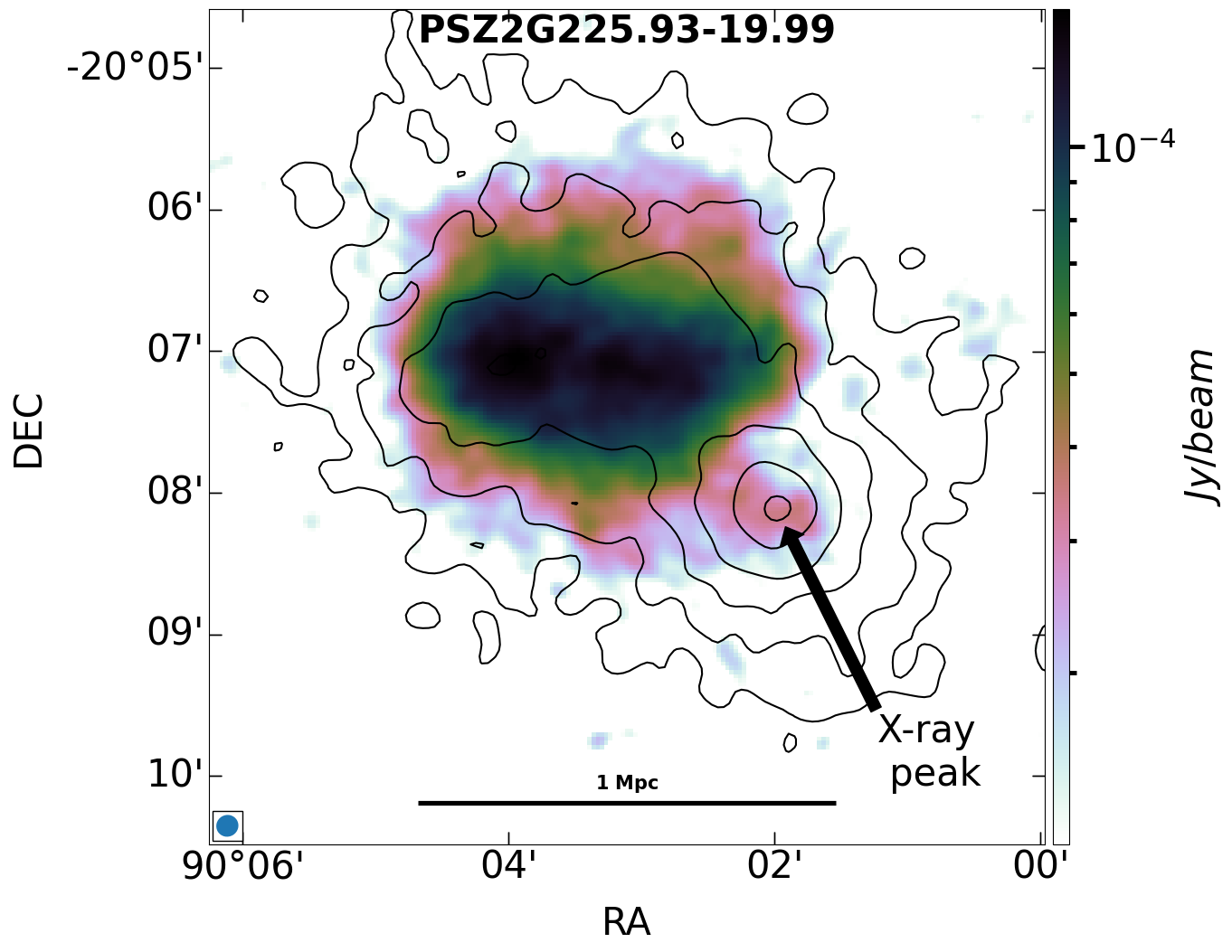}
    \caption{Same as Fig.~\ref{fig:3} but for PSZ2G225.93-19.99 (at 9\arcsec resolution and $\sigma_{RMS} = 4.6 ~\mu \rm{Jy ~ beam^{-1}}$).}
    \label{fig:5}
\end{figure}
\paragraph{PSZ2G239.27-26.01 (MACSJ0553.4-334, Fig.~\ref{fig:6})}
Optical and X-ray data of this cluster suggest that this is a dynamically active cluster observed in a post-core passage.
However, despite agreeing on the ongoing merger process in this system, different studies have reported contrasting merger scenarios, debating whether it is occurring along the line of sight or on the plane of the sky \citep{Mann2012,Ebeling2017,Botteon2018-discontinuities}.  
Dedicated X-ray studies by \cite{Pange2017} and \cite{Botteon2018-discontinuities} found a series of cold fronts and shock fronts depicting a complicated merger scenario, with the primary activity seen in the EW direction, but with the accretion of sub-clusters in other directions given the presence of an extend X-ray tail in the NE. 
On the radio side, \citet[and subsequently \citealt{Wilber2020}]{Bonafede2012} reported the presence of a radio halo emission but no radio relic emission, despite the high mass of the system.
Additionally, \cite{Botteon2018-discontinuities} highlighted how the boundary of the radio halo seems to be confined by the Eastern shock observed in the X-rays.
Here, we also detect the radio halo, which appears bright in our image and which promptly follows  the X-ray emission, supporting the idea of a confined non-thermal plasma (Fig.~\ref{fig:6}). 
We note how the brightest radio regions are co-spatial with the two X-ray peaks.

\begin{figure}[ht!]
    \centering
    \includegraphics[width=\linewidth]{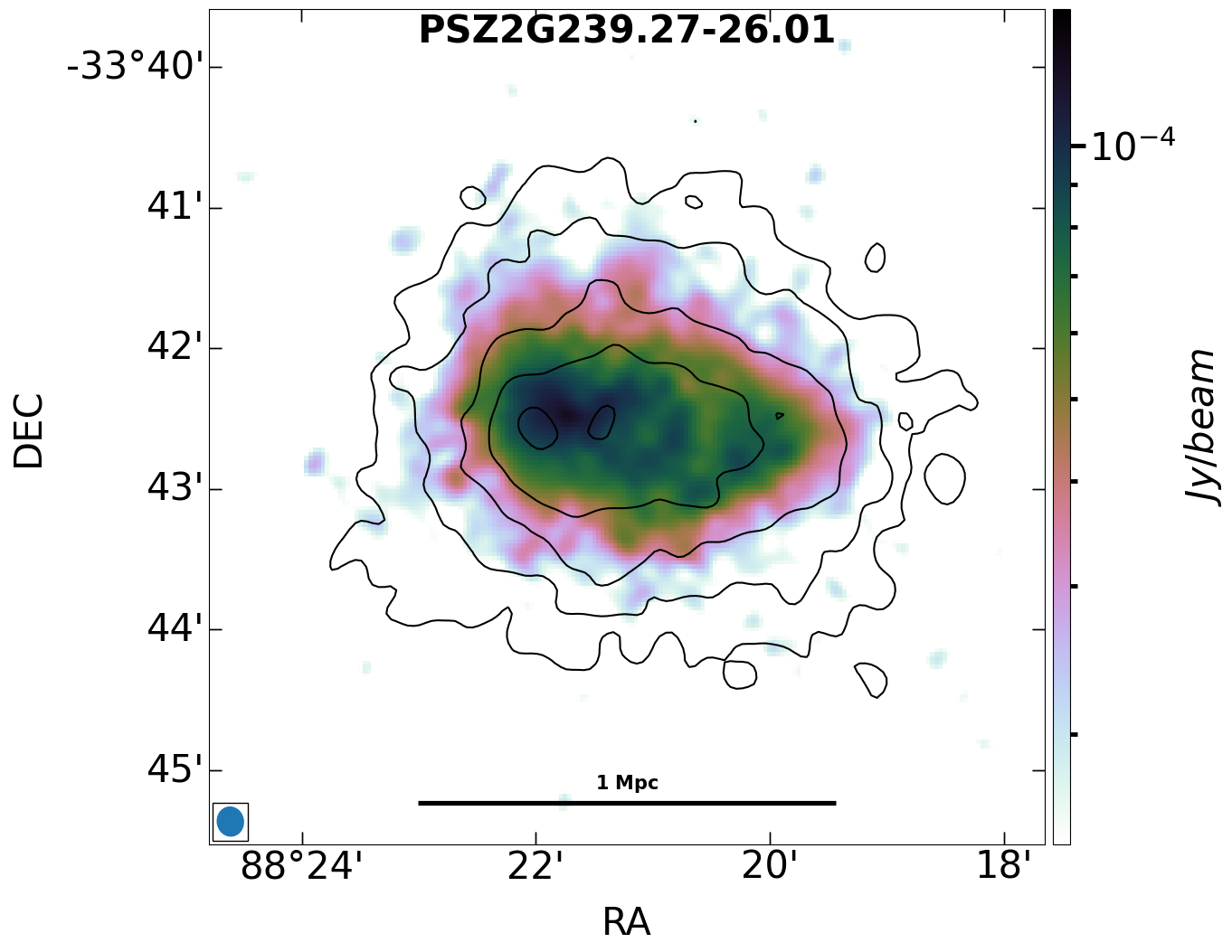}
    \caption{Same as Fig.~\ref{fig:3} but for PSZ2G239.27-26.01 (at 11\arcsec resolution and $\sigma_{RMS} = 5.3 ~\mu \rm{Jy ~ beam^{-1}}$).}
    \label{fig:6}
\end{figure}

\paragraph{PSZ2G243.15-73.84 (MACSJ0159.0-3412, Fig.~\ref{fig:7})}
No dedicated studies were carried out on this target. However, different X-ray works on large samples have classified it as disturbed \citep{Yuan2020} or, in the case of a non-binary classification, C22 has labelled this object as 'mixed'. 
The pilot study of the MERGHERS project (\citealt{Knowles2021})  discovered a radio halo and two radio relics in this system using shallow MeerKAT L-band observations. 
With our deeper observations,  using the spectral radio maps, we can clearly identify the steepening of both relic sources and we are also able to detect a new elongated structure in the east, at $\sim 1$ Mpc from the halo (Fig.~\ref{fig:7}). 
Given its elongation, size, position, and absence of a connection with radio galaxies, we classified this new detection as a candidate relic. This source adds to the number of detection of faint and distant relics made with MeerKAT \cite[e.g.][]{Knowles2022, Botteon2024-A754}.

The overall distribution of the X-ray emission suggests a large scale extension in the EW direction is consistent with the two relics (one of which is found in our study).
In addition, we note a NS extension in the X-ray image on smaller scales, which is consistent with the presence of a second merger and might explain the detection of a relic in the north.
\begin{figure}[ht]

{\includegraphics[width=\linewidth]{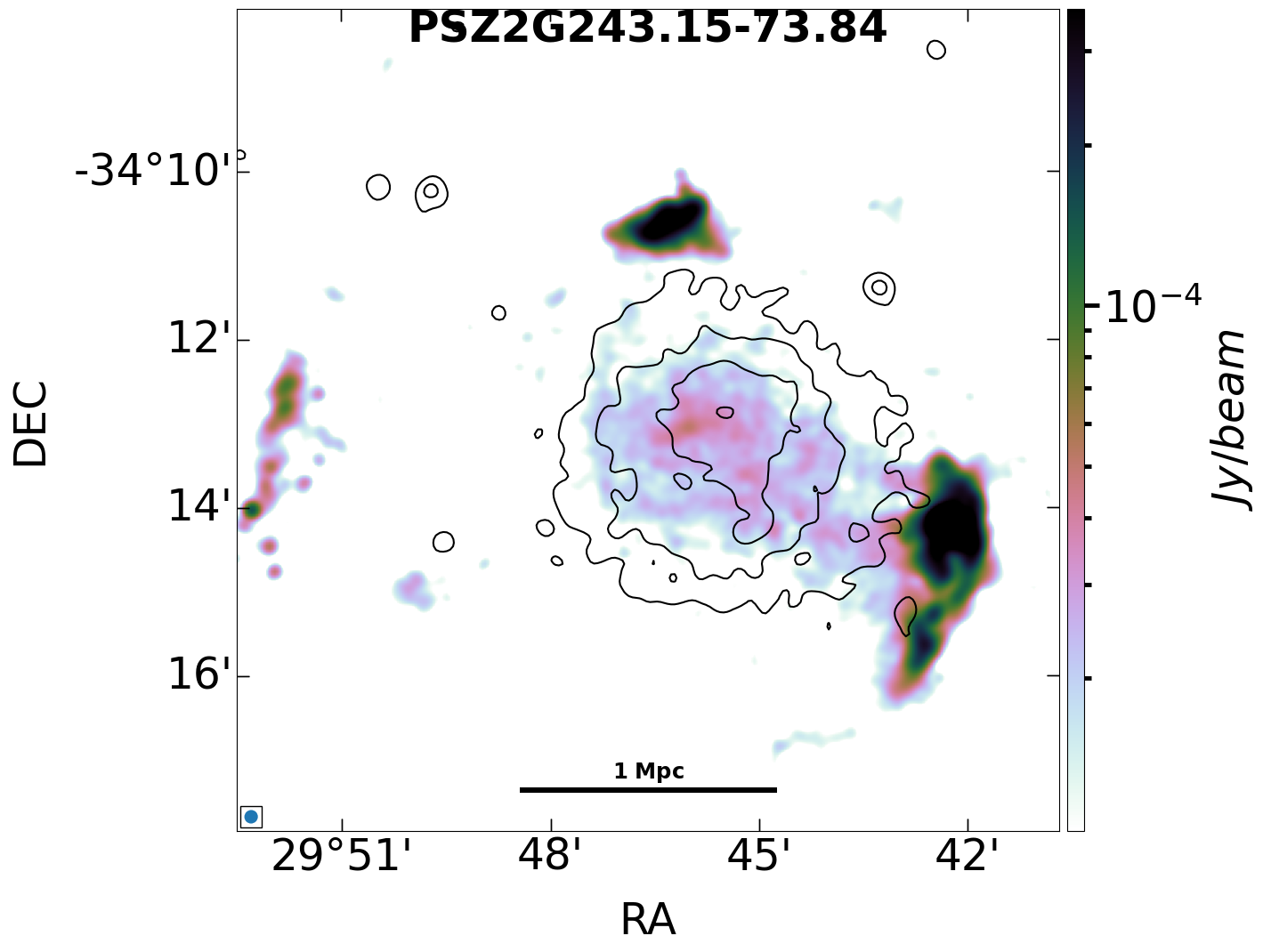}}

    \includegraphics[scale=0.18]{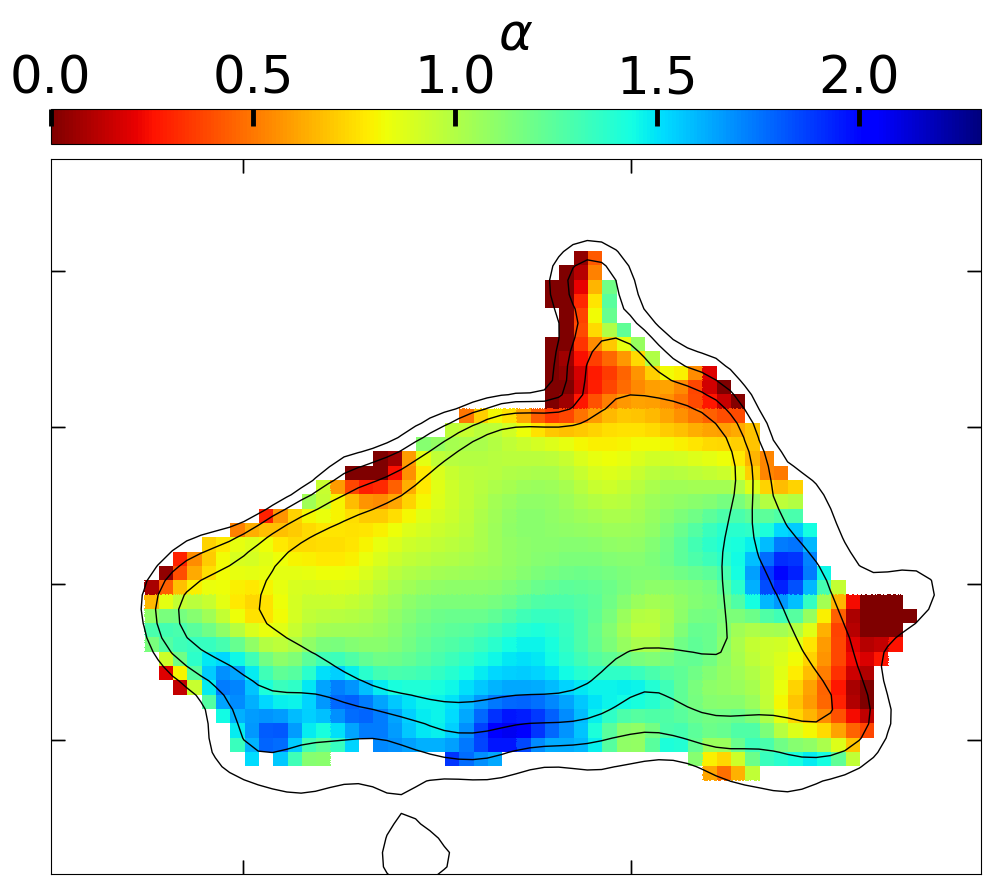}
    \includegraphics[scale=0.17]{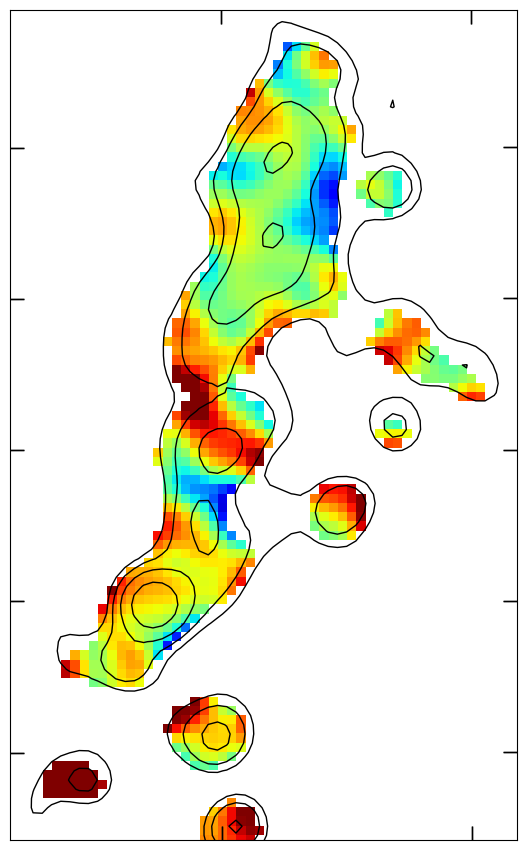}
    \includegraphics[scale=0.17]{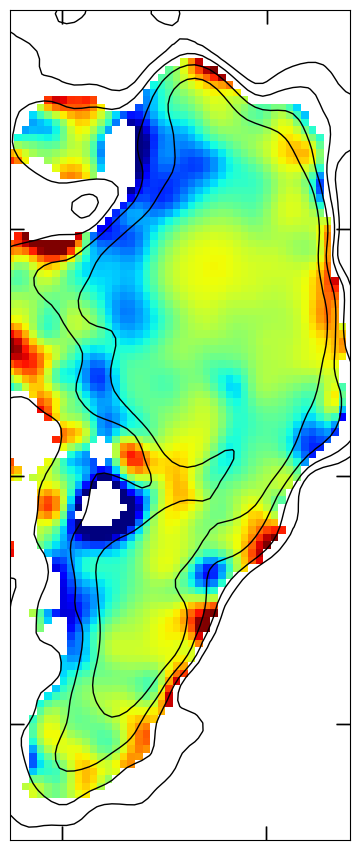} 

    \caption{Same as Fig.~\ref{fig:1} but for PSZ2G243.15-73.84 (at 10\arcsec resolution and $\sigma_{RMS} = 3.7 ~\mu \rm{Jy ~ beam^{-1}}$), showing the spectral index map of the three detected radio relics. The spectral index uncertainty maps are shown in Fig~\ref{fig:7err}.}
    \label{fig:7}
\end{figure}
\paragraph{PSZ2G262.27-35.38 (AS0520, Fig.~\ref{fig:9})}
No detailed studies were conducted on PSZ2G262.27-35.38 (also named ACO S520). C22 classified this object as disturbed, placing it in the first quartile of the most dynamically disturbed CHEX-MATE objects.
{Indeed, it shows a complex and elongated X-ray morphology, with a misalignment between the centroid and the X-ray peak.}
In the radio band, \cite{Knowles2022} identified the central halo, the complex northern source as a radio relic and the southern diffuse emission as a candidate relic.
We confirm the halo detection and definitively classify the southern emission as a relic from the spectral map (Fig.~\ref{fig:9}). We also detected diffuse emission connecting the halo and the southern relic, with no X-ray counterpart.
Regarding the northern source, we argue that particular care must be taken in classifying it. In fact, the misalignment with the main elongation axis of the cluster and the complex spectral properties make it appear different from a classical relic.
As a possible explanation, we propose that the emission consists of a combination of two sources.
In particular, in the high-resolution map (top right panel of Fig.~\ref{fig:9}), it is possible to identify an arc-like structure in the lower part of this source oriented in the NS direction (as the other relic), which displays a mild spectral steepening towards the cluster centre (middle right panel of Fig.~\ref{fig:9}). 
The emission in the upper part, instead, could be associated with a tailed radio galaxy crossing the relic region.
Arguing in favour of this scenario, we also note how the radio relic emission identified here displays a rather uniform emission across its extension, while the one from the radio galaxy shows the typical dimming moving away from the peak.

\begin{figure}[h]

{\includegraphics[width=\linewidth]{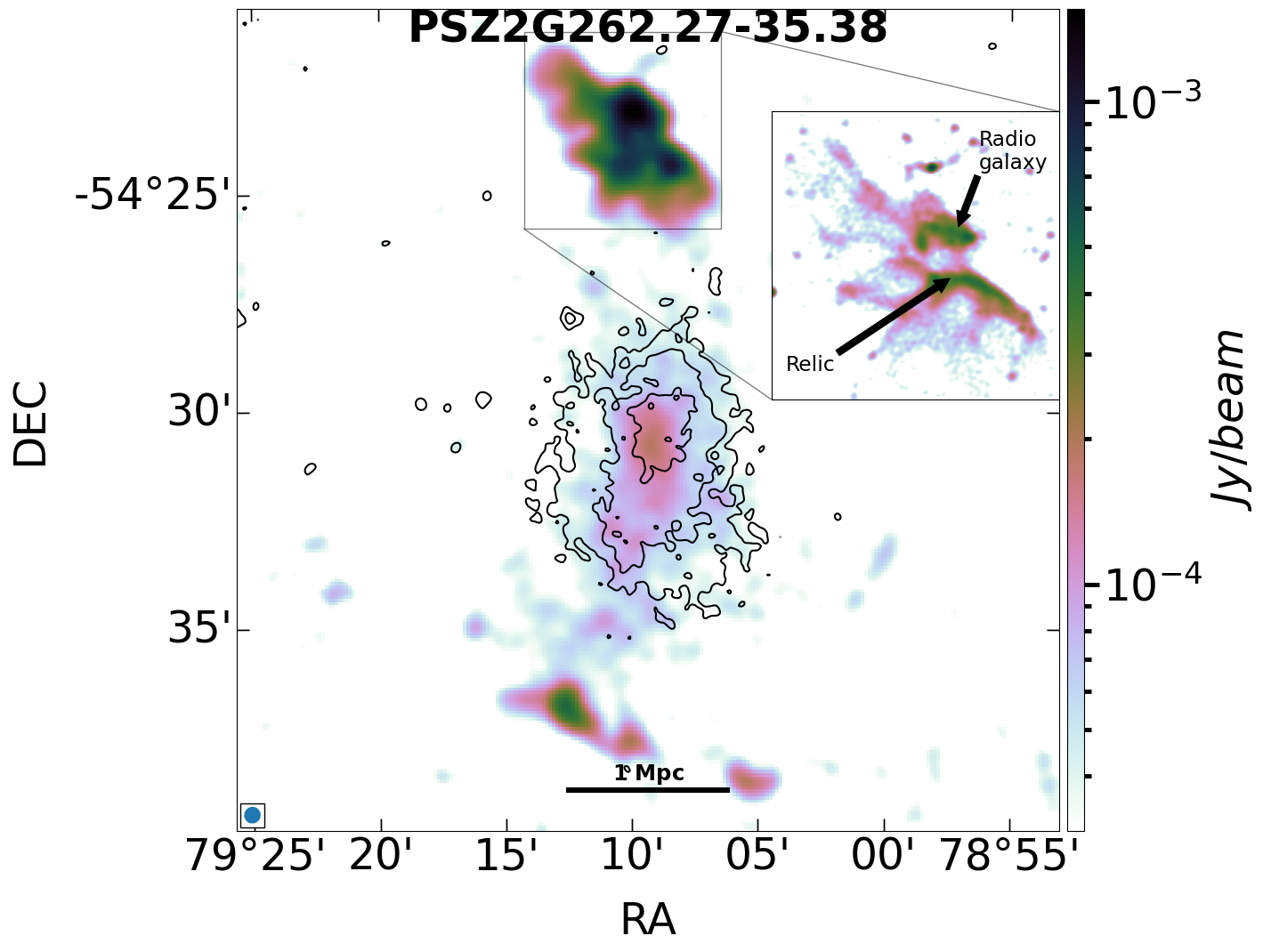}}

\includegraphics[height=2.8cm]{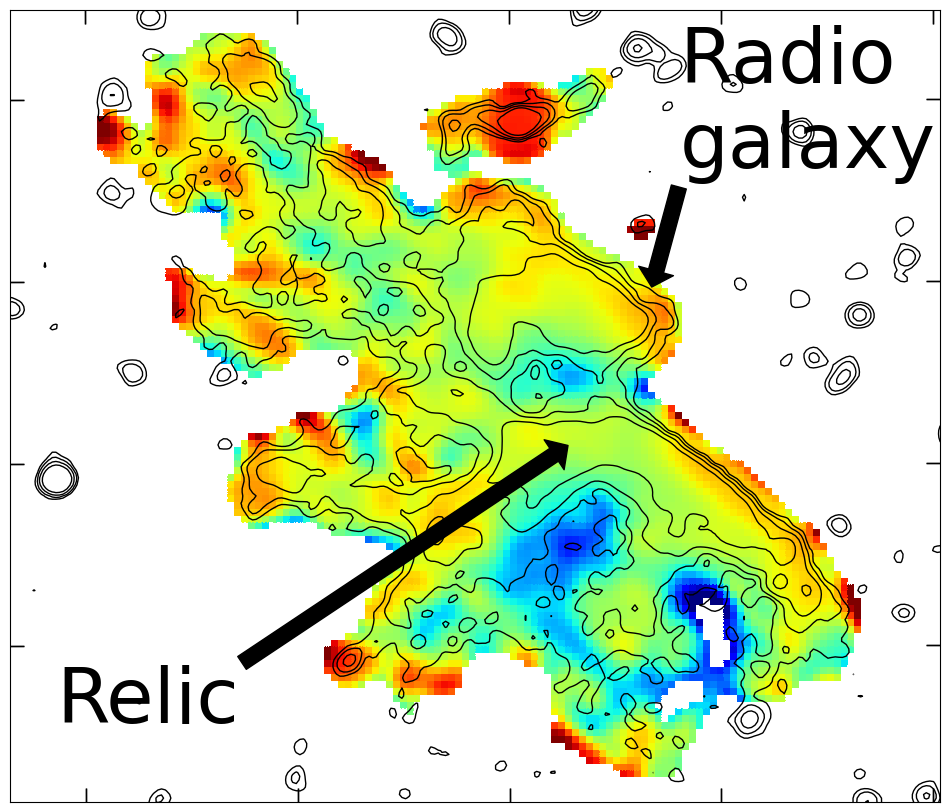}
\includegraphics[height=2.8cm]{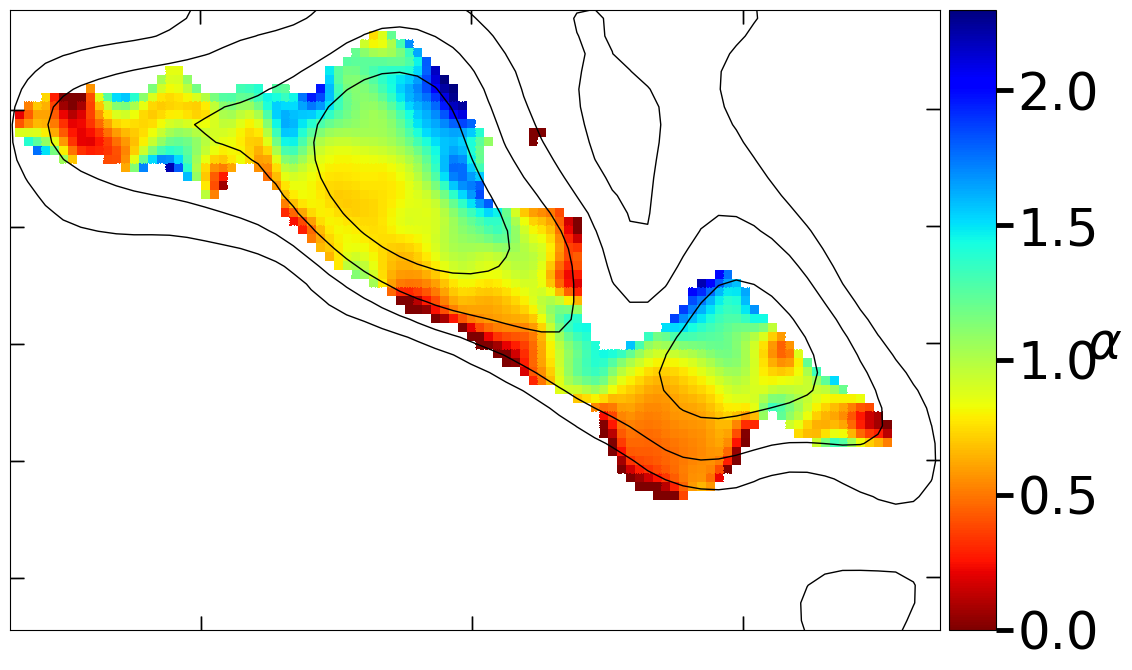} 

\caption{Same as Fig.~\ref{fig:1}, but for PSZ2G262.27-35.38 (at 23\arcsec resolution and $\sigma_{RMS} = 11 ~\mu \rm{Jy ~ beam^{-1}}$), showing the spectral index maps of the source in the north (also highlighted by the high-resolution inset) and the relic in the south. The spectral index uncertainty maps are shown in Fig~\ref{fig:9err}.}
    \label{fig:9}
\end{figure}
\paragraph{PSZ2G277.76-51.74 (Fig.~\ref{fig:11})}
{C22 classified this cluster as a disturbed system.}
In the radio, \cite{Martinez-Aviles2018} investigated its emission with the Australia Telescope Compact Array (ATCA) with two sets of images, the first between 1.1-3.1 GHz at 5" resolution and the second between 1.63-2.13 GHz tapered to 35" resolution, without detecting any diffuse radio emission.

Here, we apply a larger uv-taper when imaging, corresponding to a physical size of 150 kpc at the cluster redshift, in order to clearly detect the fainter SE diffuse emission.
We discovered a large scale halo source, which nicely follows the thermal X-ray emission (Fig.~\ref{fig:11}).
We detected an elongated structure that we classified as a radio relic given its location in the cluster outskirts, the elongated shape of $\sim$Mpc size and, more importantly, since it is not associated with any compact source or radio galaxy emission. 
A high-resolution view of this source is shown in the inset panel of Fig.~\ref{fig:11}, displaying the thin and elongated structure of this source, consistent with that of a radio relic.
However, its spectral index trend is not consistent with being a relic, therefore, we classify it as a candidate radio relic. 
\begin{figure}
    \centering
    \includegraphics[width=\linewidth]{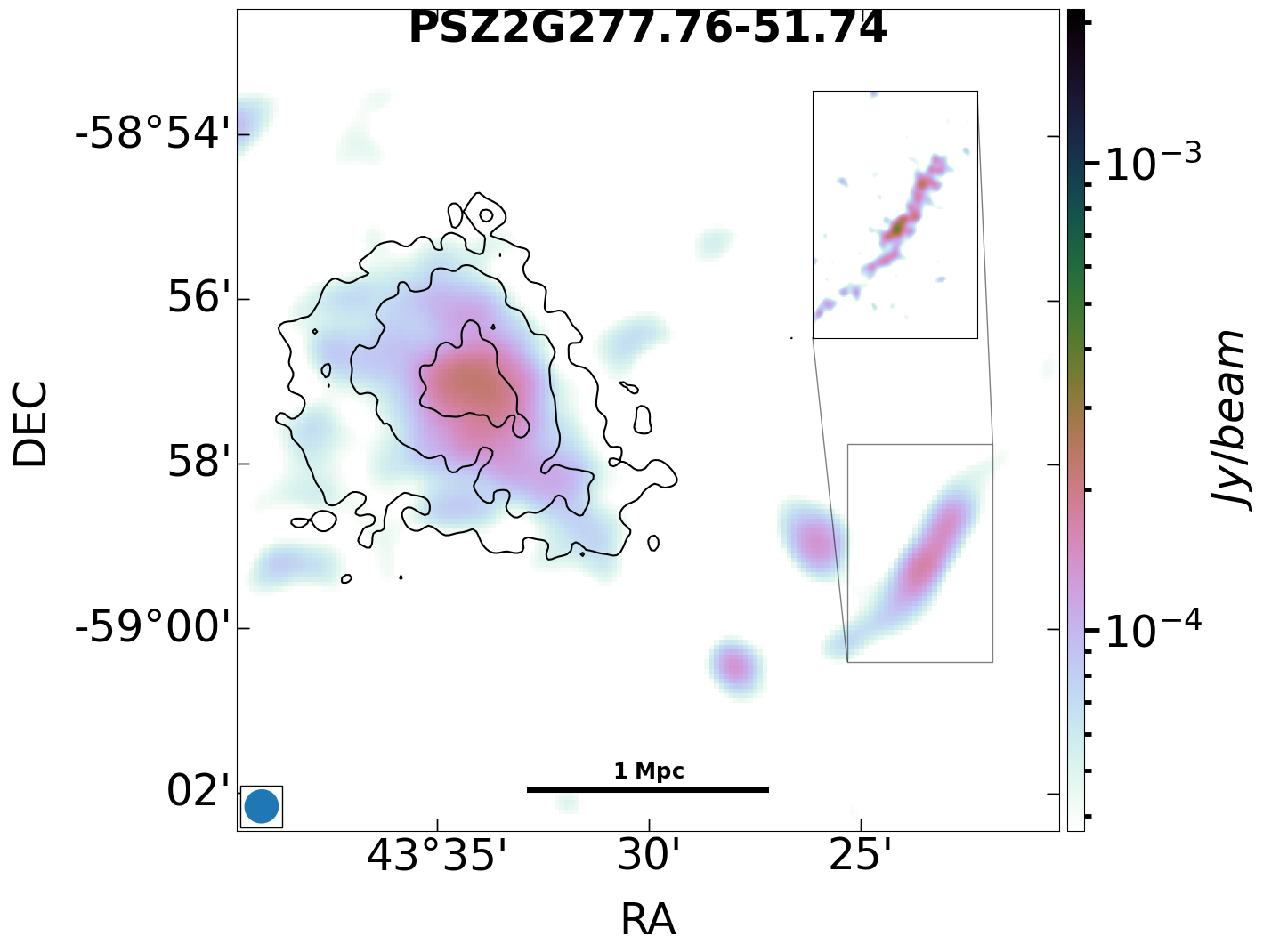}
    \caption{Same as Fig.~\ref{fig:3}, but for PSZ2G277.76-51.74 (at 25\arcsec resolution and $\sigma_{RMS} = 12.6 ~\mu \rm{Jy ~ beam^{-1}}$).}
    \label{fig:11}
\end{figure}
\paragraph{PSZ2G278.58+39.16 (A1300, Fig.~\ref{fig:12})}
It was first surveyed in the X-ray band during the ROSAT All Sky Survey (\citealt{Pierre1994}) and subsequently analysed by \cite{Lemonon1997} and \cite{Pierre1997} in both the optical and X-ray. They found evidence of sub-structures in the thermal emission and suggested that this object is in a post-merging phase with a cluster of similar mass. Further optical and X-ray \emph{XMM-Newton} analyses by \cite{Ziparo2012} detected the presence of filamentary structure in the north, suggesting an ongoing accretion of a group of galaxies. They estimated the elapsed time since the merger of the two cluster progenitors to be $\sim3$ Gyr ago. 
In the radio, the cluster hosts a giant radio halo and a relic located in the south-west (SW) region \citep{Reid1999,Venturi13,TernideGregory2021}.
Our MeerKAT observations recovered a more extended radio emission than previously observed,  allowing us to obtain a resolved spectral index map for the relic (Fig.~\ref{fig:12}).
In particular, we find an extension of the non-thermal component in the northern part, as well as a connection between the radio halo and the relic. 
We also detect a peculiar elongated emission extending south from the relic region, not associated with compact sources.

\begin{figure}[ht!]
    \centering
    \includegraphics[width=\linewidth]{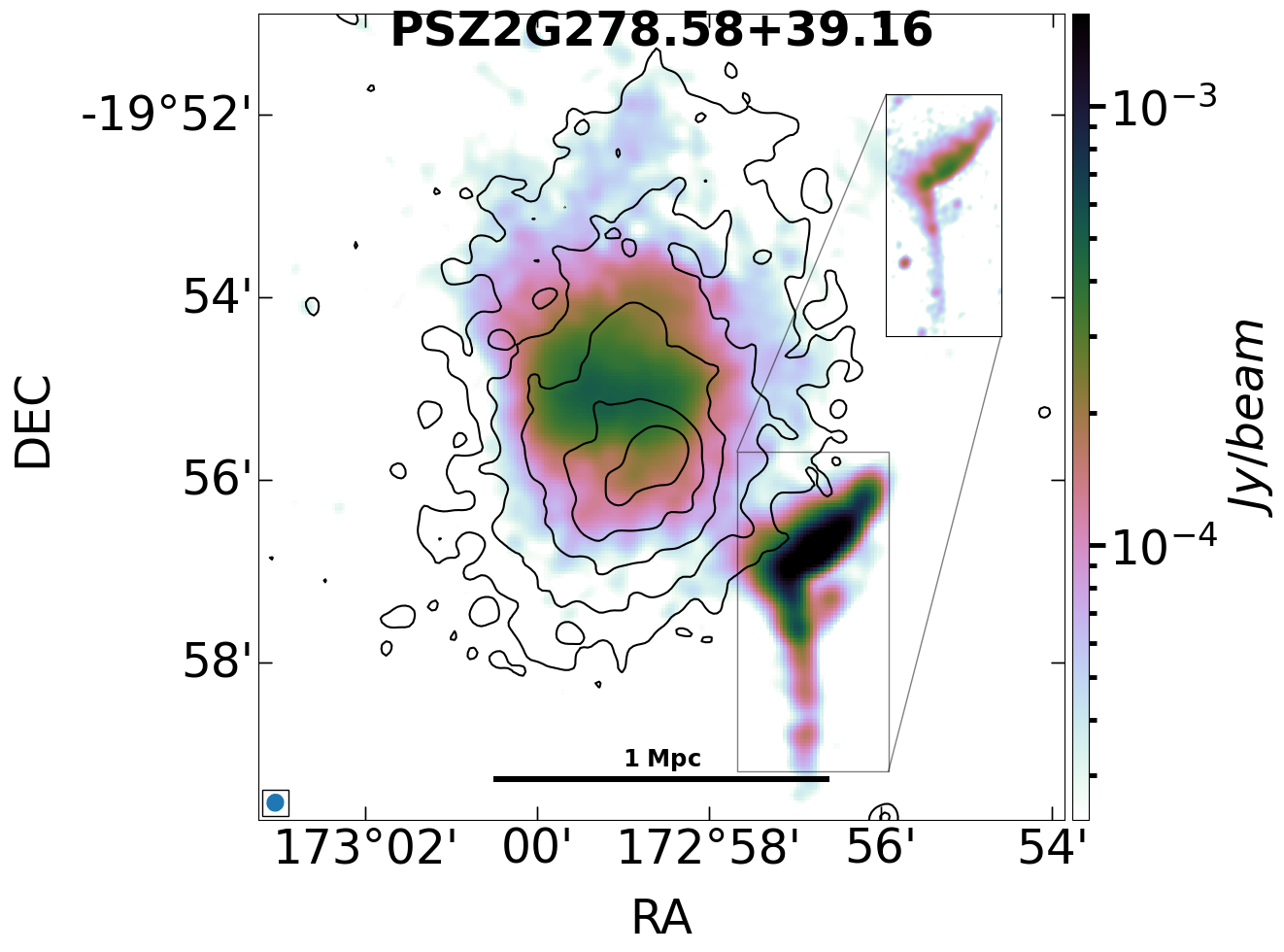}
    \includegraphics[width=.275\linewidth, angle=90]{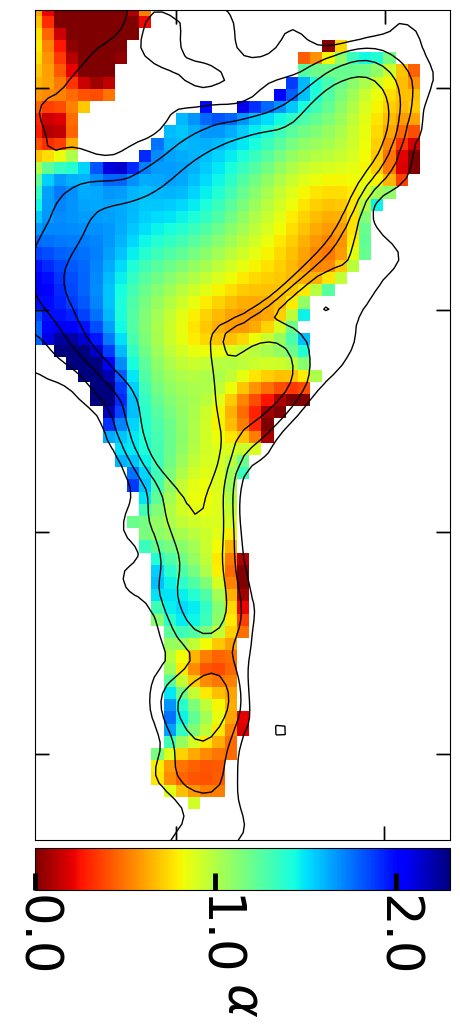}
    \caption{Same as Fig.~\ref{fig:1} but for PSZ2G278.58+39.16 (at 12\arcsec resolution and $\sigma_{RMS} = 8.5 ~\mu \rm{Jy ~ beam^{-1}}$), showing the (rotated) spectral index map of the relic in the south-west. Spectral index uncertainty maps are shown in Fig~\ref{fig:12err}.}
    \label{fig:12}
\end{figure}

\paragraph{PSZ2G286.98+32.90 (Fig.~\ref{fig:13})}
It is the most massive of the whole CHEX-MATE sample, with $M_{500} \sim 1.4 \times 10^{15} ~ M_{\odot}$.
Given its high mass, it has been used in strong and weak lensing analyses, which detected multiple sub-structures and determined it to be a massive, merging cluster (\citealt{Finner2017,Zitrin2017,DAddona2024}).
Recently \cite{Gitti2025} analysed  the central X-ray emission of this target in detail by means of deep \textit{Chandra} observations. They found a cold front and a shock in the NW direction and argued that this object has experienced past dynamical activities that heated the cluster cool core.
This view is further supported by radio features found by \cite{Bagchi2011} and \cite{Bonafede2014}, who detected a giant radio halo with two relics and filamentary structures within their extension.
The MeerKAT high-quality data are capable of recovering at high significance all the known features and to discover new diffuse structures toward the cluster outskirts (Fig.~\ref{fig:13}). 
In particular, we observed a number of filamentary structures beyond the NW relic that were recently associated with a large scale contribution of the diffuse halo emission \citep{Salunkhe2025, Rajpurohit2025}.
\begin{figure}[ht]

\includegraphics[width=\linewidth]{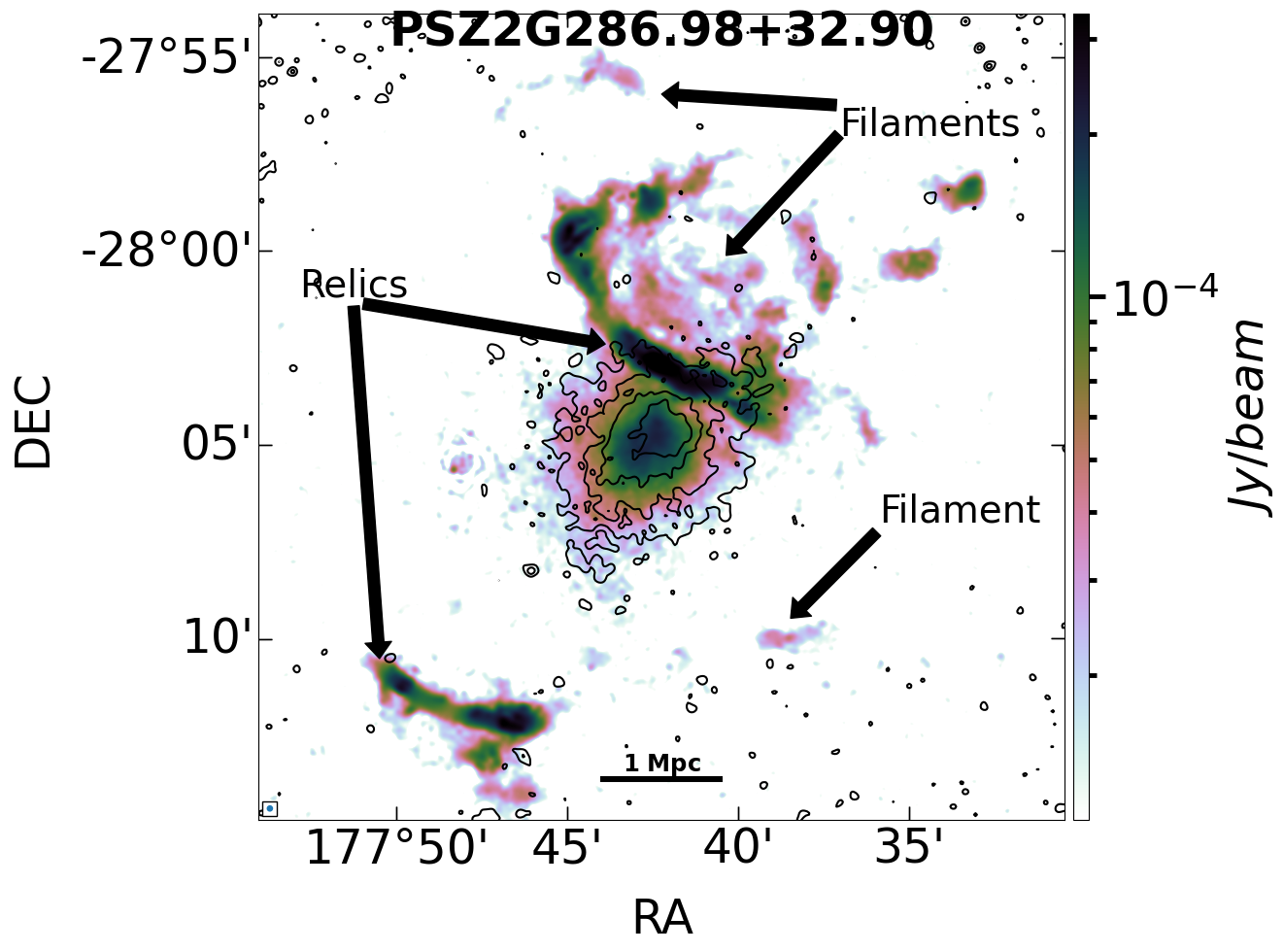}
\includegraphics[width=0.53\linewidth]{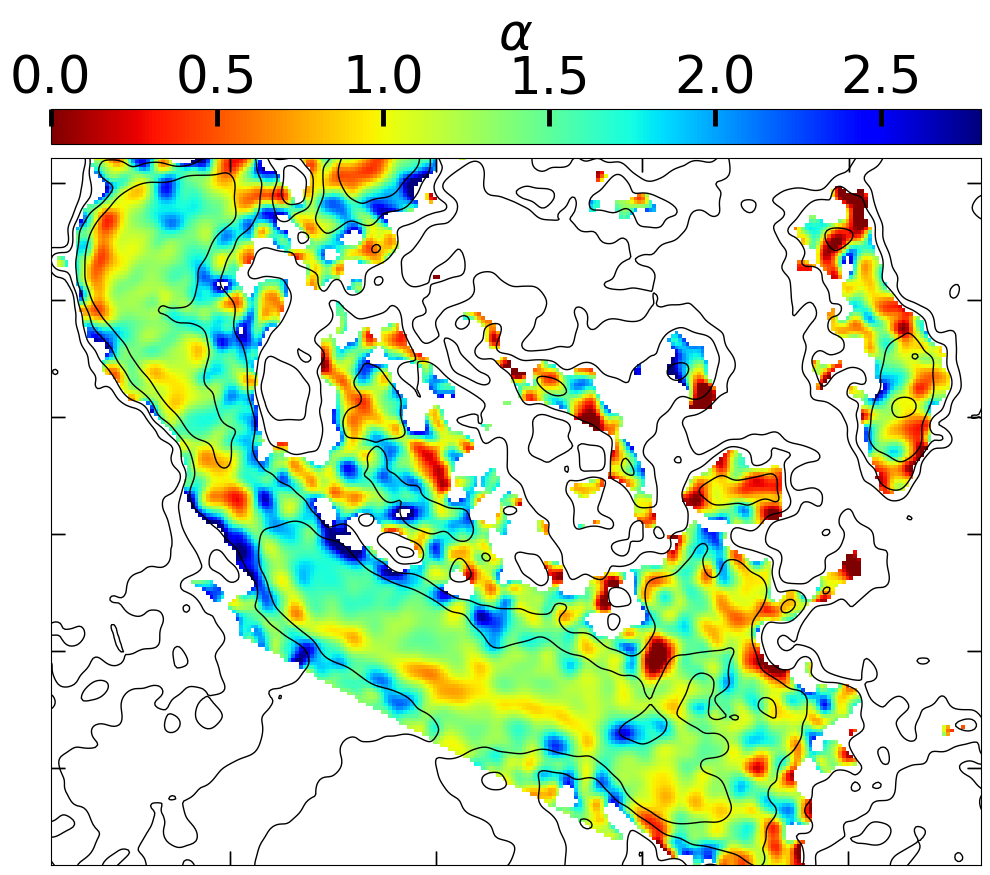}
\includegraphics[width=4cm, height=3.5cm]{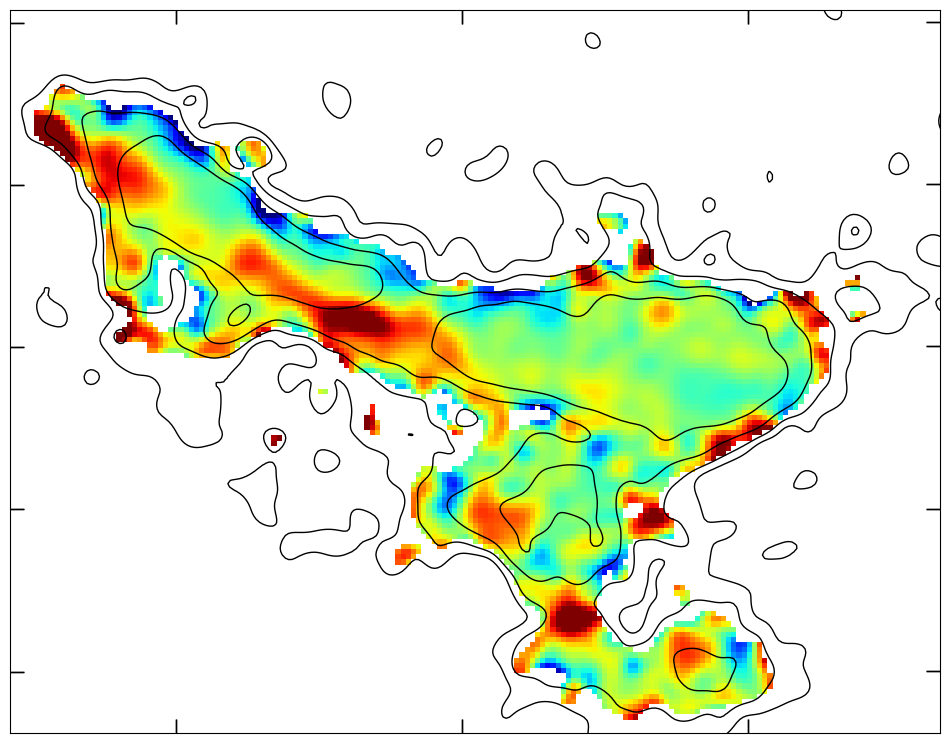} 

\caption{Same as Fig.~\ref{fig:1}, but for PSZ2G286.98+32.90 (at 10\arcsec resolution and $\sigma_{RMS} = 3.9 ~\mu \rm{Jy ~ beam^{-1}}$), showing the spectral index map of the two relics in the NW and SE. The spectral index uncertainty maps are shown in Fig~\ref{fig:13err}.}
    \label{fig:13}
\end{figure}
 
\paragraph{PSZ2G313.33-17.11 (A1689, Fig.~\ref{fig:14})}
This object has been widely studied in several lensing and X-ray analyses. 
In fact, its high mass and round shape allow for a good modelling of the gravitational lensing effects (e.g. \citealt{Limousin2007}) and represent an ideal case for measuring the cluster hydrostatic mass \citep[e.g.][]{Tchernin2015}. 
It has been shown that lensing and X-ray mass estimates are in agreement when considering a triaxial extension of the system (\citealt{Morandi2011}), which seems to be elongated along the line of sight \citep{Peng2009,Sereno2013, Kim2024, Chappuis2025}.
The most likely scenario is that the cluster is experiencing a merger along the line of sight, also showing three aligned groups of galaxies \citep{Girardi1997}.
In the radio, \cite{Vacca2011} detected a radio halo at 1.2 GHz with the VLA, supporting the merger scenario.
With MeerKAT, we detect larger radio emission, which fills the whole northern extension of the thermal emission and shows a peak emission close to the X-ray one (Fig.~\ref{fig:14}). 
We also note that the overall extension of the radio halo towards the NE agrees well with the orientation found by \cite{Vacca2011} and points to the position of the optical sub-structures identified by \cite{Girardi1997}.
The western diffuse spot is the residual emission from a tailed radio galaxy after the subtraction.
\begin{figure}[ht!]
    \centering
    \includegraphics[width=\linewidth]{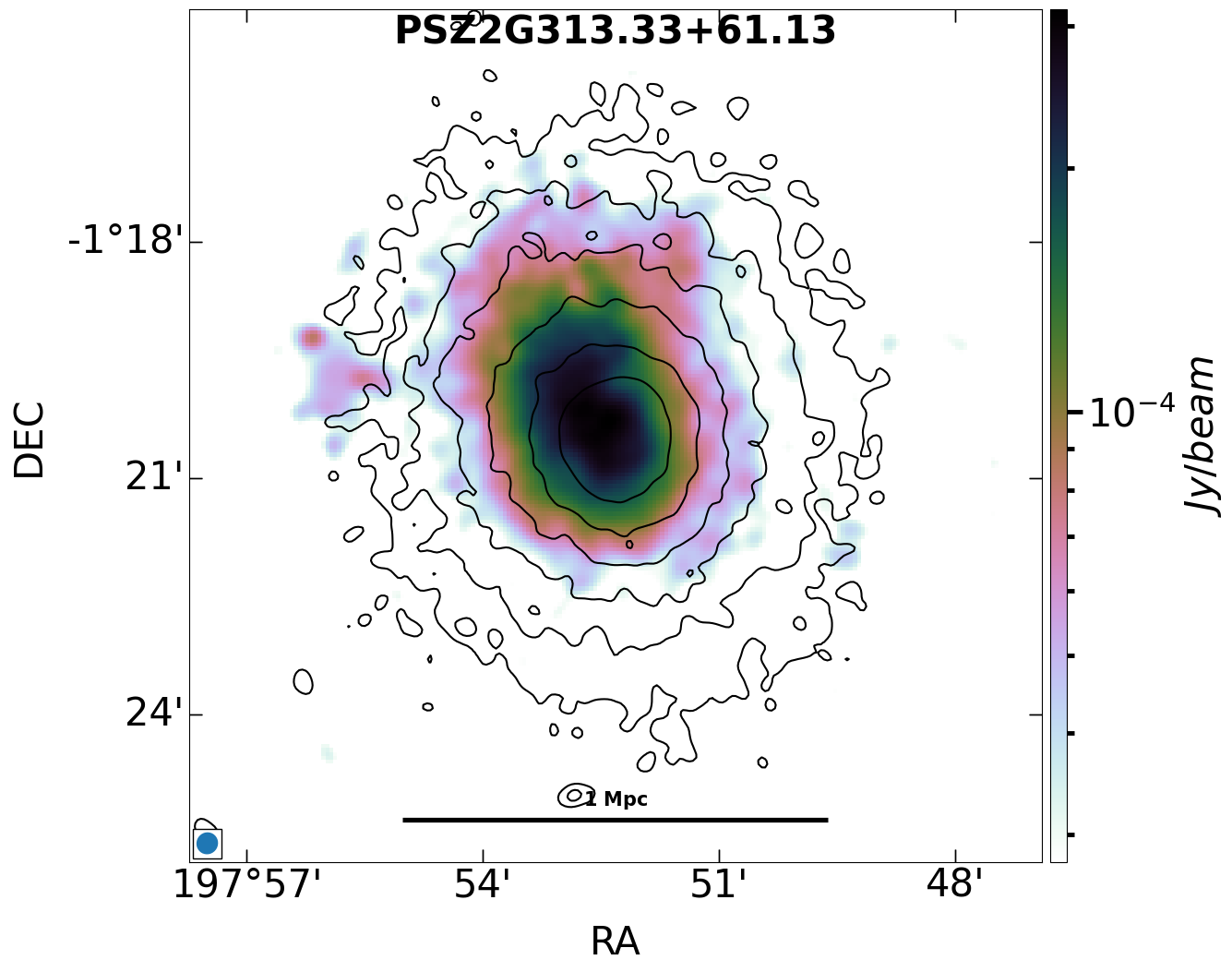}
    \caption{Same as Fig.~\ref{fig:3}, but for PSZ2G313.33+61.13 (at 17\arcsec resolution and $\sigma_{RMS} = 9.9 ~\mu \rm{Jy ~ beam^{-1}}$).}
    \label{fig:14}
\end{figure}
\paragraph{PSZ2G346.61+35.06 (RXCJ1514.9-1523, Fig.~\ref{fig:16})}
This is the third most dynamically disturbed CHEX-MATE cluster and it displays a low X-ray luminosity (\citealt{Bohringer2004}). 
This object has been studied by \cite{Giacintucci2011}, who pointed out the presence of Mpc-scale radio emission, which is fully co-spatial with the thermal one.
In our MeerKAT data, we detected an additional, elongated, diffuse source in the SW, which, given the size, shape, distance from the cluster centre and its alignment with the X-ray elongation, we classified as a radio relic (Fig.~\ref{fig:16}). 
However, we cannot firmly claim its nature using the in-band spectral index map, which displays a patchy distribution of the source with no particular trend visible. 

\begin{figure}[ht!]
    \centering
    \includegraphics[width=\linewidth]{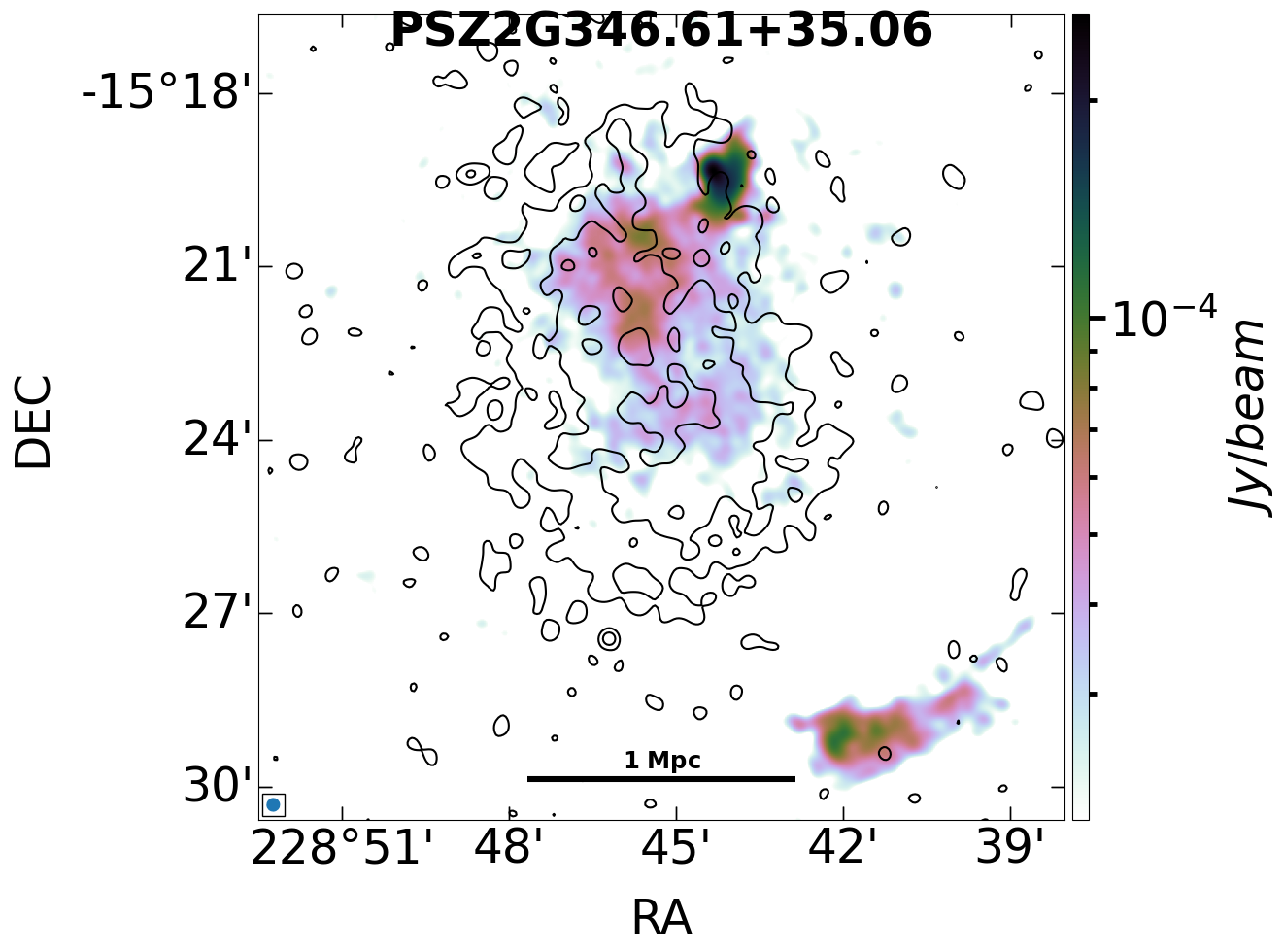} 
    \includegraphics[scale=0.15]{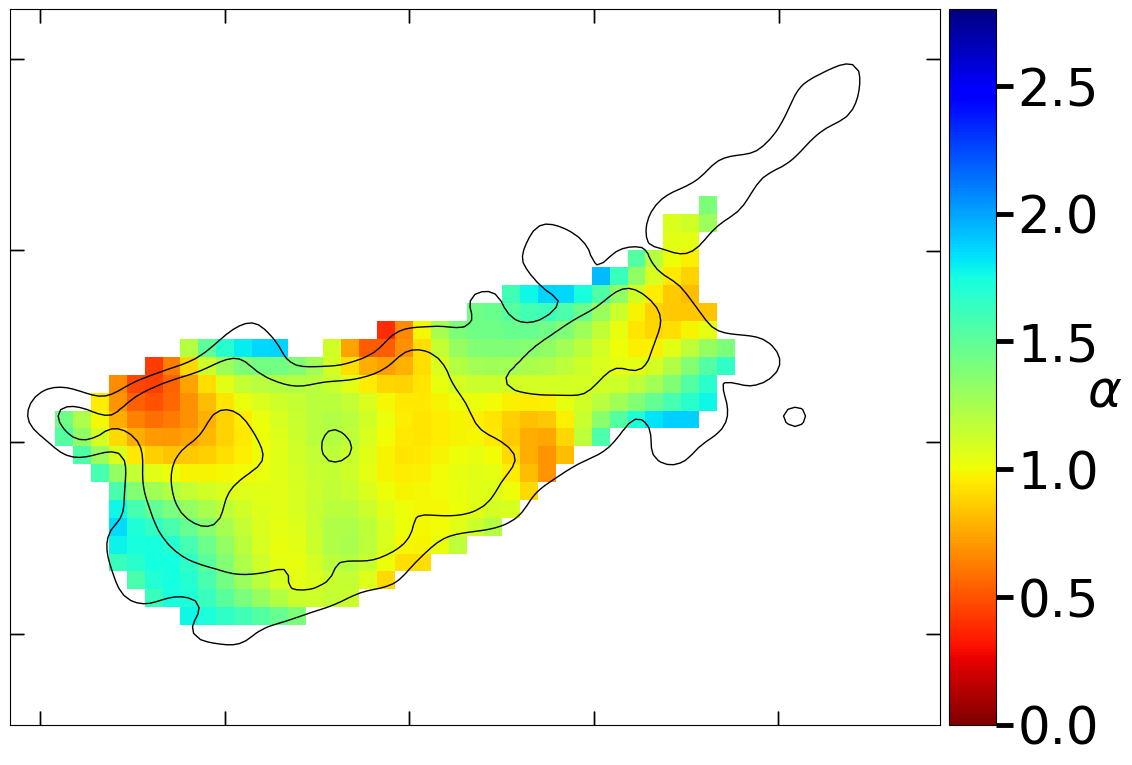}
    \caption{Same as Fig.~\ref{fig:1}, but for PSZ2G346.61+35.06 (at 14\arcsec resolution and $\sigma_{RMS} = 7.2 ~\mu \rm{Jy ~ beam^{-1}}$), showing the spectral index map of the relic in the south-west. The spectral index uncertainty maps are shown in Fig~\ref{fig:16err}.}
    \label{fig:16}
\end{figure}

\section{Radio properties of the halos and relics}

In the following, we perform a systematic analysis of the non-thermal emission of the sources presented in Sect.~\ref{sec:sample} and Table~\ref{tab:radio_info}, comparing our data with the literature results and discussing the implications of our findings.

{Thanks to new MeerKAT observations, we were able to detect and classify several (new) radio sources. We detected two new radio halos and one relic. We also classified one radio halo and two radio relics previously labelled as candidates.}
\subsection{Radio halos}\label{sec:radio_halos}
We detect radio halo emission in every target of the sample, regardless of the cluster redshift.
This suggests that when studying clusters in this high mass range and with such sensitive observations, radio halos can be found in the majority of the systems. 
This is in line with the results obtained by \cite{Cuciti21-2}, who reported a high fraction of halos in high-mass and high-redshift systems, and by \cite{DiGennaro2021}, who detected powerful radio halos at $z>0.6$.

A number of studies have investigated the presence of scaling relations between cluster properties and radio halo power (\citealt{Liang2000, Cassano06, Basu2012, Kale2015, Balboni2025}). The most important is certainly the relation between the radio halo power ($P_{RH}$) and the cluster mass (\citealt{Cassano2007, Cassano2010, Cassano2013, vanWeeren2021,Cuciti21-2,Duchesne2021-EoR,George2021,Cuciti2023}). In fact, it directly points to an underlying physical relation among the two quantities, which is expected in the turbulent re-acceleration scenario for radio halo formation (e.g. \citealt{Cassano2007, Cassano2023}).
The detection of several radio halo sources in our sample allows us to explore the $P_{RH}-M_{500}$ correlation for CHEX-MATE targets observed with MeerKAT. 
To do so, we took the MMF3 cluster mass estimates from the Planck catalogue and measured the radio halo power using the Halo-Flux Density CAlculator (Halo-FDCA, \citealt{HFDCA-2021}) package. 
Halo-FDCA allows us to fit the 2D surface brightness emission of the radio halo through an exponential model with different shapes. We chose the rotated elliptical one to recover possible halo elongations, expressed as
\begin{equation}
    I(r) = I_0 \exp\left(-\sqrt{\left(\frac{r_x}{r_1}\right)^2 + \left(\frac{r_y}{r_2}\right)^2}\right),
\end{equation}
where $r_1$ and $r_2$ are the e-folding radii along the two, rotated (by an angle $\theta$ on the plane of the sky) axis of the ellipse, $r_x$ and $r_y$ (see \citealt{HFDCA-2021} for further details). As commonly done in the literature (e.g. \citealt{Murgia2009,Botteon2022, Duchesne2024-EMU}), the derived integrated radio flux density and power are obtained by analytically integrating the best-fit exponential model up to three times the e-folding radii. 
To limit the contamination from point sources, in the $15"$ resolution images of the MGCLS, we adopted the approach used by \cite{Botteon2023}. Specifically, we manually masked point sources falling within the halo extension and replaced the pixels in the selected regions with values interpolated from neighbouring pixels.
As also discussed  by \cite{Botteon2023}, this approach works well for small discrete sources, whereas for extended ones, it creates large artefacts which were masked out from the subsequent analyses (see Appendix~\ref{appendix:mgcls_imgs}).
In Table~\ref{tab:RH_prop}, we list the best-fit results of the halo exponential fit. \par
In Fig.~\ref{fig:P-M+lotssRH}, we present the $P_{RH}-M_{500}$ relation for the CHEX-MATE sub-sample considered (coloured squares) and compare it with literature results (black points).
Specifically, we report the $P_{RH}-M_{500}$ relation obtained at 150 MHz by \cite{Cuciti2023} for a large sample of Planck clusters covered by the LOFAR Two-mete Sky Survey 2nd Data Release \citep[LoTSS DR2,][]{Botteon2022}. 
The authors analysed 61 radio halos detected by the LoTSS DR2 \citep{Shimwell2022}, which lie above the 50\% completeness line of the Planck sample \citep{Planck2016}\footnote{We note that the LoTSS DR2 masses are taken from Planck measurements as the ones of our clusters. The only difference is that \cite{Cuciti2023} used the masses obtained by combining different techniques, while CHEX-MATE relies on those estimated using the MMF3 algorithm. The differences between the two estimates have no impact on the analyses conducted here ($\lesssim5 - 10\%$).}.
To perform such a comparison, we extrapolated the flux obtained at 150 MHz to the 1.28 GHz of our MeerKAT observations. 
We used a single spectral index value of 1.3 for all the clusters and we accounted (within the errors) for a variation of $\alpha$ between 1.1 and 1.5 (as usually observed in halos; e.g. \citealt{Feretti2012,vanWeeren19}.

As expected, we find that the clusters in our sample show a positive correlation between $P_{RH}$ and $M_{500}$, with a Spearman correlation coefficient $r_S \sim 0.57$ (p-value$\sim0.01$).
Figure~\ref{fig:P-M+lotssRH} also shows the clear mass selection of our targets when compared with a more complete sample.
We also report two best-fit regression lines of the $P_{RH}-M_{500}$ relation found by \citet[at 1.4 GHz]{Cuciti21-2} and \citet[at 150 MHz]{Cuciti2023}, using the LIRA Bayesian regression method \citep{Sereno2016}. We rescaled the best-fit lines using $\alpha=1.3$.
We see how our cluster sample fits rather well in the high-mass range of the $P_{RH}-M_{500}$ found at low frequencies, both in terms of global scatter and relation slope.\par
 {Assuming an ellipsoidal volume for the entire radio halo emission, with semi-axes $3r_1$ and $3r_2$, we can divide $P_{RH}$ by the halo volume to calculate the average radio emissivity of each target ($\langle \varepsilon_{RH} \rangle$)}. 
As  for the radio power, we observe a trend between the mass and $\langle \varepsilon_{RH} \rangle$ with $r_S \sim 0.48$ (p-value $ \sim 0.03$; Fig.~\ref{fig:emR-M}). 
We also report no relation between the radio halo size and the cluster mass ($r_S \sim -0.08$ with p-value $\sim0.72$).
Although caution must be taken to exclude correlations with the mass when working on such a limited mass range, this result suggests that $\langle \varepsilon_{RH} \rangle$ is the physical quantity responsible for the $P_{RH}-M_{500}$ correlation.  
 {This is consistent with models in which radio halos trace turbulent regions in the ICM driven by mergers, where the synchrotron emissivity, originating from the energy cascade across multiple spatial scales, depends on the turbulent injection rate that scales with the cluster mass \citep[e.g.][]{Cassano2005, Gaspari2014}.}
We also note that evidence for the radio emissivity being the driver of the $P_{RH}-M_{500}$ correlation has also been pointed out at lower frequencies by \cite{Balboni2025} for LoTSS DR2 targets.\\ 
\begin{figure}
    \centering
    \includegraphics[width=\linewidth]{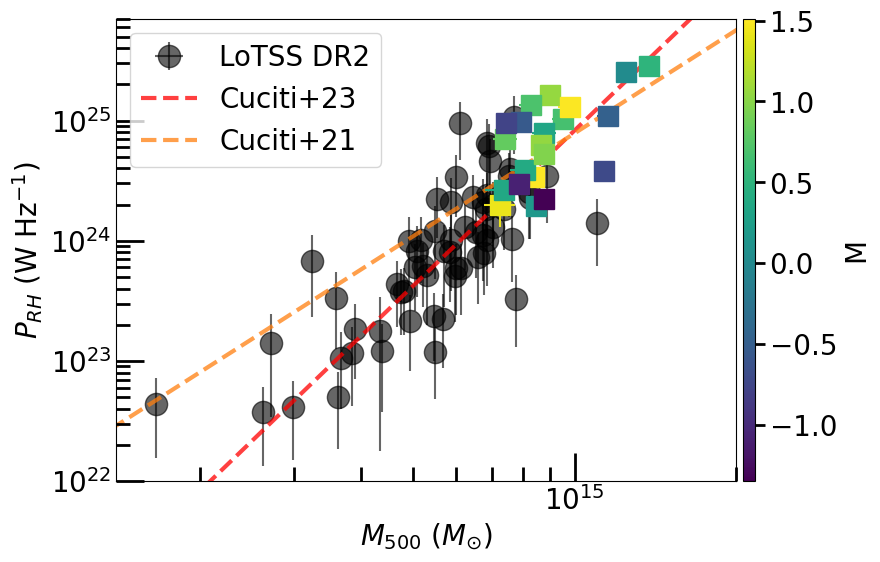}
    \caption{
    Radio power-mass relation, at 1.28 GHz, for the targets considered in this work (errorbars fall within point size) including also the literature results from \cite{Cuciti21-2} at GHz frequencies and \cite{Cuciti2023} at low frequencies, alongside the best-fit regression lines they obtained. CHEX-MATE clusters are colour-coded by their M parameter, which is a proxy for the cluster dynamical state (C22). }    
    \label{fig:P-M+lotssRH}
\end{figure}
\begin{figure}
    \centering
    \includegraphics[width=\linewidth]{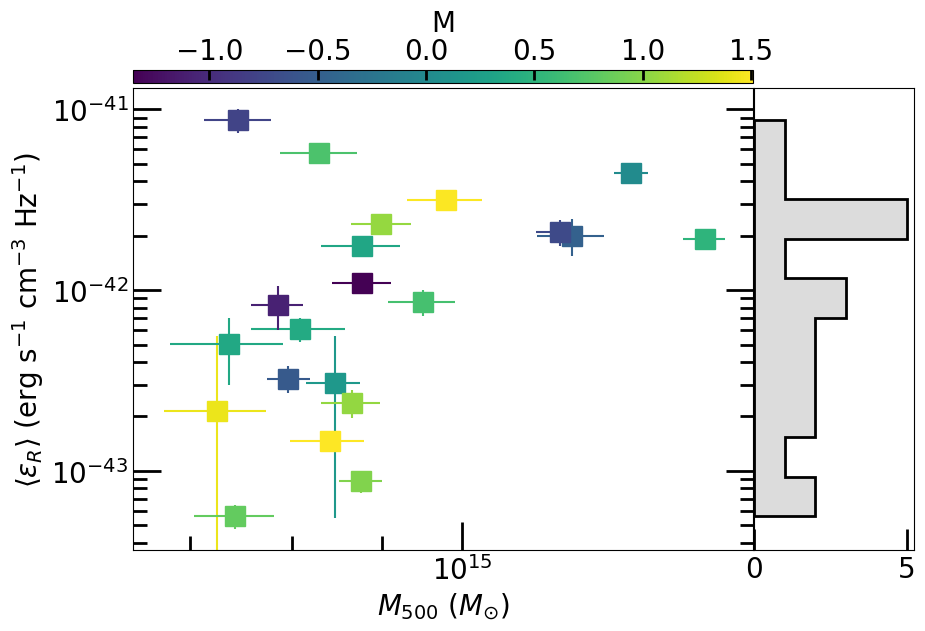}
    \caption{Average radio halo emissivity-mass relation, at 1.28 GHz, of the considered clusters, colour-coded by the M parameter.}
    \label{fig:emR-M}
\end{figure}

\subsection{Radio relics}\label{sec:radio_relics}
As for the halos, radio relic have also been found to display scaling relations with the cluster properties. Correlations between the relic power and the cluster mass or the relic size and its distance from the cluster centre have indeed been observed in several cases (e.g. \citealt{vanWeeren2009, Feretti2012, Bonafede2012, deGasperin2014,Jones2023, Stroe2025}). Alongside the observational results, numerical studies were conducted on several aspects of radio relics, such as their occurrence in clusters or the presence of correlation among quantities (e.g. \citealt{Bruggen2003,Vazza2012, Nuza2017, Bruggen2020}).

In our sample, we detected 20 radio relics and candidate radio relics, with clusters hosting one or more of these sources. 
We derived the relic characteristic quantities following \cite{Jones2023}. 
The radio relic total power ($P_{RR})$ is computed as $P_{RR} = 4 \pi S_{\nu} D_L^2 (1+z)^{\alpha-1}$, where $D_L$ is the luminosity distance, $S_{\nu}$ the flux coming from a region encompassing the relic emission above 3$\sigma_{\rm RMS}$. and $\alpha$ is assumed to be 1.15.
Motivated by the increasing complexity of the relic morphology when observed at high resolution,  to derive the relic position, we computed the flux-weighted position average of the brightest 10\% relic pixels and used this value as the relic centre \citep[see][]{Jones2023}. 
This was the position considered to derive the distance from the cluster centre ($D_{cc-RR}$), assuming the peak of the X-ray emission from the ICM for the latter. 
Given the possible presence of projection effects, this measurement must be considered a lower limit of such a distance.
Following \cite{Jones2023}, we assumed that the offset between the real cluster-relic distance and the observed one is at most $30^{\degree}$, and considered this value as an upper limit of $D_{cc-RR}$. 
Eventually, we derived $D_{cc-RR}$ as the mean value lying between the upper and lower limits, which are then taken as upper and lower error estimates, respectively.
The last quantity we measured was the largest linear size (LLS) of the radio relics. This was simply computed as the maximum distance between two pixels above the 3$\sigma_{\rm RMS}$ level within the relic extension. The error associated to the LLS is set to one beam width.
A summary of all the derived quantities is listed in Table~\ref{tab:RR_prop}.
We then investigated possible correlations among the derived quantities as
previously found in the literature.

\subsubsection{Relic power-mass relation}
In Fig.~\ref{fig:P-M+lotssRR}, we present the $P_{RR}-M_{500}$ relation of our sample (including both candidates and confirmed radio relics) and the results presented by \cite{Jones2023} using LoTSS DR2 data. 
 {Following on the procedure described in Sect.~\ref{sec:radio_halos}, we rescaled the LoTSS DR2 relic luminosities at 150 MHz to 1.28 GHz, assuming an $\alpha=1.15$ and accounting for a variation of it between 1 and 1.3 within the errors.} 
We also report three different regression lines for the $P_{RR}-M_{500}$ relation of radio relics, from both observational and numerical simulation studies. 
 {Specifically, on the observational side, we considered the work made by \cite{Jones2023} and \cite{Stroe2025}, using the BCES orthogonal regression method \citep{Akritas1996}.}
The former, used deep, low-frequency observations obtained by the LoTSS DR2 for a large sample of clusters and performed a systematic study of the radio relics found in those systems.
The latter, instead, updated the lists of all the double radio relic systems known so far and studied their scaling properties. 
On the simulation side, we reported the best-fit $P_{RR}-M_{500}$ relation found by \cite{Lee2024}, who presented an overview of a large number of radio relics in massive cluster mergers identified in the new TNG-Cluster simulation.
\begin{figure}
    \centering
    \includegraphics[width=\linewidth]{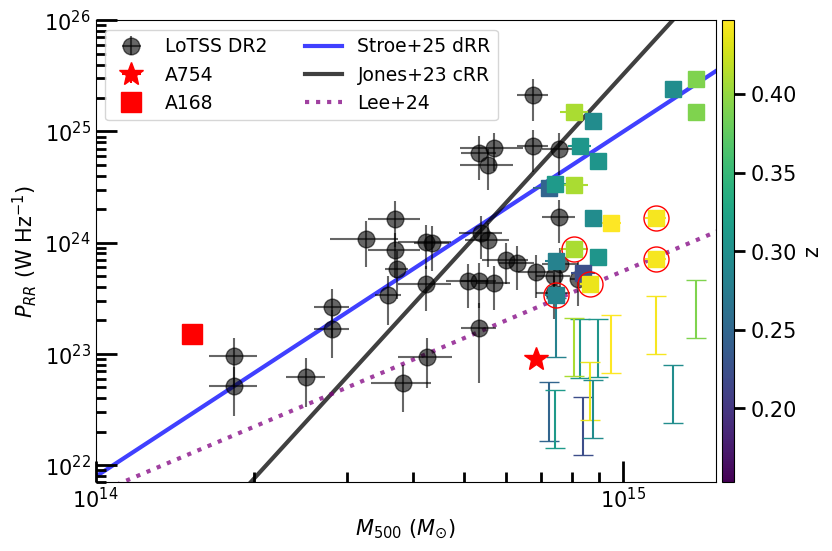}
    \caption{Radio relic power - mass relation of our sample at 1.28 GHz (squares with red circles for candidate relics) compared with literature results.
    The black points represent relics from the LoTSS-DR2 described in \cite{Jones2023}, while the red square and the red star represent the relics in A168 \citep{Dwarakanath2018} and A754 \citep[error bars fall below point size;][]{Botteon2024-A754}. 
    The best-fit relations by both observations (\citealt{Jones2023} and \citealt{Stroe2025}, dashed lines) and simulations (\citealt{Lee2024}, dotted lines) are also reported. The coloured bars indicate the detection thresholds for each of our observations where we detected relic emission as explained in the main text.}    
    \label{fig:P-M+lotssRR}
\end{figure}

We did not find any signs of correlation between $P_{RR}$ and $M_{500}$ in our clusters, possibly due to the limited mass range explored.

When compared to other samples, our targets span a wide range of the relic radio powers, indicating how the high sensitivity of MeerKAT is capable of recovering faint diffuse sources. 
The ability to recover faint relic emissions will be a key feature for the understanding of these sources with the current and next generation of radio telescopes. 
In fact, numerical simulations predict the presence of a large number of low-power radio relics in both high and low-mass systems \citep{Bruggen2020, Lee2024}, with the mass of the hosting cluster being related only to the maximum power of a relic source \citep{Nuza2017,Jones2023, Lee2024}.
Simulations show that the relic radio power is also affected by other factors such as the merger phase, the mass ratio, and the cluster's dynamical history. 
This is aptly explained by \cite{Lee2024}, who exploited numerical simulations to show how a plethora of radio relic powers can originate in high-mass systems, depending on where and when the cluster merger occurs. 
 {However, it is also important to notice that particle acceleration at weak Mach number shocks ($<2$) is poorly understood \citep[e.g.][]{Guo2014} and that the number of faint radio relics may be over-predicted by numerical simulations. Therefore, it is crucial to obtain deep radio observations of (massive) galaxy clusters to probe the presence of faint radio relics.}\par
\cite{Jones2023}  showed how low-frequency observations are capable of detecting new faint radio relics, especially for low-mass and nearby clusters.
Here, using MeerKAT observations, we are starting to explore an analogous range of low-power radio relics but at higher frequencies in massive clusters. 
 {This is better expressed by the detection thresholds of the observations used here and shown in Fig.~\ref{fig:P-M+lotssRR} with coloured bars. 
Following \cite{Jones2023}, we computed these limits by considering the power emitted by a region of size $300 ~ {\rm kpc} \times 100 ~{\rm kpc}$, for the lower limit, and $1000 ~ {\rm kpc} \times 100 ~{\rm kpc}$, for the upper limit, and assuming an average surface brightness of twice the image noise level ($2\sigma_{RMS}$)}. 
We see how the detection thresholds extend even below $10^{23} ~{\rm W ~ Hz^{-1}}$, a region where simulations are predicting a large number of radio relics even at high masses \citep[e.g.][]{Nuza2017,Lee2024}.
 {In this respect, we note that despite the wide range of $P_{RR}$ detected, in the considered clusters, we did not recover a significant fraction of radio relics with $P_{RR} < 10^{24} ~ {\rm W ~ Hz^{-1}}$}.
 {As a reference, we report two low-power radio relics observed at GHz frequencies by \citet[A168]{Dwarakanath2018} and \citet[Abell 754]{Botteon2024-A754} after rescaling the emissions to our 1.28 GHz observing frequency using their spectral index. 
Both sources are amongst the faintest relics observed so far and their detections have been possible thanks to deep MeerKAT and VLA observations at GHz frequencies. 
From Fig.~\ref{fig:P-M+lotssRR}, we can see how, in the lower redshift regime, our MeerKAT L-band observations would be entirely capable of recovering both GHz emissions. Noticeably, the two radio relics could also be detected in most of the higher redshift cluster observations when considering the lower-sensitivity detection limits.}
Hence, current MeerKAT observations are opening a new window of analyses for the radio relic at GHz frequencies, allowing us to detect the faint end of these sources and test numerical simulation predictions.

Additionally, the detection of low-power radio relics will also enable to test DSA models for radio relic production. 
Given the low Mach numbers of shock waves in clusters, such a mechanism usually requires an unphysically high particle acceleration efficiency to match the observed radio relic luminosities.
In the case of faint radio relics, instead, it is possible to find reasonable conditions in which these models reproduce the observed radio luminosities. In addition, it would also be possible to derive constraints on the shock acceleration efficiency and magnetic fields \citep[e.g][]{Locatelli2020, Rajpurohit2024,Botteon2024-A754}.\\
\subsubsection{LLS $D_{cc-RR}$ relation}\label{sec:lls-dist}
 {In Fig.~\ref{fig:LLS-dist}, we report the LLS-$D_{cc-RR}$ relation for our radio relics and candidate radio relics (red circles),  including the measurements made by \cite{Jones2023}.
Following the formalism of \cite{Lee2024}, we removed the mass dependence by rescaling LLS and $D_{cc-RR}$ for $R_{500}$. 
We then focused only on the impact of the timing of the merger on the luminosity of the relics by rescaling the latter for the best-fit value of the $P_{RR}-M_{500}$ relation found by \citet[$P_{RR}/P_{fit}$, where $P_{fit} \propto M_{500}^{3.1}$]{Stroe2025}.}
\begin{figure}
    \centering
    \includegraphics[width=0.95\linewidth]{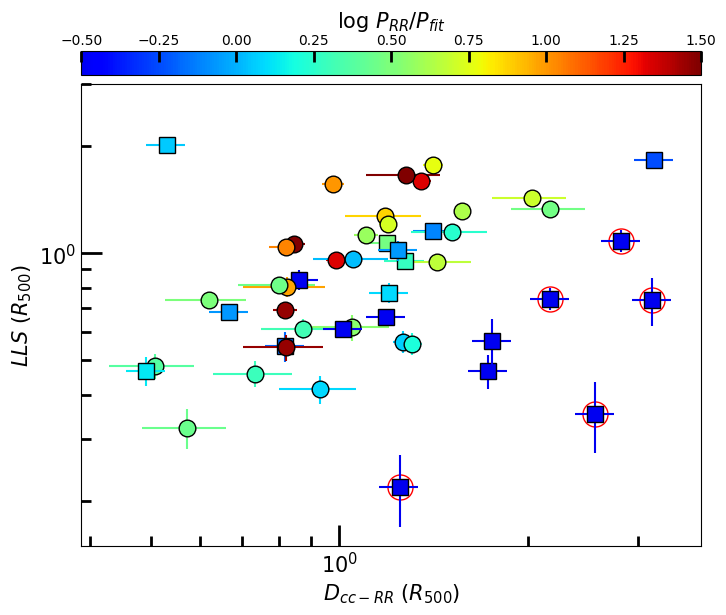}
    \caption{Relic LLS as a function of $D_{cc-RR}$, both rescaled for the cluster $R_{500}$, and colour-coded by $P_{RR}/P_{fit}$. Squares are the targets presented in this work (red circles indicate candidate radio relics) while circles are taken from \cite{Jones2023}.}
    \label{fig:LLS-dist}
\end{figure}
Apart from the outlier PSZ2G286.98+32.90-NW seen in the top-left part of the plot, the relics presented here using MeerKAT observations (squares), display a weak ($r_S\sim 0.4$) correlation in the LLS-$D_{cc-RR}$ plane, with a scatter that increases at higher $D_{cc-RR}$, mainly due to candidate relic sources.
\cite{Lee2024} highlighted that when studying a large number of simulated relics, such a relation is hardly found and that it is possible to identify a dependence on the $P_{RR}/P_{fit}$ in the LLS-$D_{cc-RR}$ plane.
 {Specifically, for a given LLS, relics with high (low) $D_{cc-RR}$ present low (high) values of $P_{RR}/P_{fit}$. In addition, they show how this trend becomes more evident considering smaller LLS.
When looking for such a trend in Fig.~\ref{fig:LLS-dist}, we do not recover any unambiguous correlations among the involved quantities. 
However, we do note how low values of $P_{RR}/P_{fit}$ are mostly observed for high $D_{cc-RR}$; however, when considering lower $D_{cc-RR}$, higher values of such ratio are also present (see also Fig. 12 of \citealt{Lee2024}).}

These findings, considered from another perspective, show that for a given LLS the radio relic power can vary by almost two orders of magnitude depending on the relic distance. 
The relic position can be seen as a proxy of the relic evolutionary stage, as these sources would evolve once the merger happens.
In particular, the relic is expected to be faint when the shock forms close to the cluster centre. Then, it progressively increases its luminosity moving toward the outskirts of the system, peaking at $\sim 1$ Mpc, and fading at the periphery of the cluster \citep[e.g.][]{Skillman2011, Vazza2012}. 
 {Hence, despite still being a qualitative comparison, the  trend suggested by $P_{RR}/P_{fit}$ in Fig.~\ref{fig:LLS-dist} aligns with the idea that the measured radio power also depends on the phase of the relic evolution, which may be trace by $D_{cc-RR}$. }
\section{Summary and conclusions}\label{sec:sum&concl}

In this work, we present MeerKAT L-band observations for a sample of galaxy clusters in the southern sky, combining archival and proprietary data.
These targets are part of the CHEX-MATE sample (\citealt{Arnaud21}), which provides deep X-ray observations of the thermal counterpart. 
Given their high mass and (mostly) disturbed dynamical state, they are all deemed as ideal candidates to host diffuse radio sources.
Thanks also to the new calibration strategy presented in \cite{Botteon2024-A754}, the depth reached in the radio images has enabled the detection of extended radio emission in all of the considered targets.

We performed a systematic study of the detected halos and relics, comparing our results with literature works and numerical simulations, and highlighting how current and forthcoming radio surveys will improve our view of non-thermal cluster phenomena. Our results can be summarised as follows:
\begin{itemize}
    \item The new MeerKAT observations of massive and dynamically active CHEX-MATE clusters have largely satisfied the goal of detecting (faint) diffuse radio sources in the considered redshift range. This further probes the issue of how X-ray morphological parameters
    can be exploited to search for radio diffuse emission in unobserved objects. 
    We report five new sources among relics and halos, alongside a few new sources associated with radio galaxies. Remarkably, all of the studied clusters exhibited the presence of diffuse radio emission, highlighting the role of cluster mass and dynamics in producing this type of emission (see Table~\ref{tab:radio_info}).\\
    \item In this work, we considered 21 radio halos, 2 of which have been classified for the first time. The sample allowed us to study the well-known scaling relation between the radio halo power and the cluster mass. We found a strong correlation between $P_{RH}$ and $M_{500}$, in line with the scenario that cluster mergers are responsible for halo emission. In addition, the $P_{RH}-M_{500}$ relation of our sample is in good agreement with the results obtained by \cite{Cuciti2023}, who explored a wider range of masses. 
    \item We investigated possible scaling relations between the halo average emissivity or the cluster size with the mass. We reported a positive $\varepsilon_{RH}-M_{500}$ relation ($r_S\sim0.48$). This result suggests how, in the considered high-mass range, the $\varepsilon_{RH}$ is the real driver of the $P_{RH}-M_{500}$ relation, leaving no role for $R_H$. Future studies on broader samples will shed light on this and establish whether it is a general property or whether it depends on cluster-specific characteristics.
    \item We report 20 radio relics and candidate radio relics, 3 of which are classified here for the first time. We show how, thanks to high-sensitivity MeerKAT observations, it is now possible to explore a low-surface brightness population of radio relics and carrying out new tests for DSA models.
    \item We investigated scaling relations for relic sources. We did not find a $P_{RR}-M_{500}$ relation in the limited mass range explored. However, the selected mass range of our sample allowed us to highlight the variety of $P_{RR}$ that can be found in systems with similar masses when the study is no longer limited by the sensitivity.
    \item  {Finally, we have reproduced the work carried out by \cite{Lee2024} for simulated relics on our sample, searching for possible dependences of $P_{RR}/P_{fit}$ on $D_{cc_RR}$ or LLS. Once the possible mass dependences were removed from the LLS and $D_{cc-RR}$, no distinct correlation between $D_{cc_RR}$ or LLS with $P_{RR}/P_{fit}$ has been found. However, we found that $P_{RR}$ can vary by almost two orders of magnitude for a given relic size, while also depending on its relative position to the cluster centre.
    This suggests a dependence of $P_{RR}$ on the evolutionary stage of the relic, traced by $D_{cc-RR}$, which can be investigated in analyses that combine observations and simulations, in a similar way to what is proposed in Sect.~\ref{sec:lls-dist}.}
\end{itemize}

\begin{acknowledgements}
We would like to thank the anonymous referee for their helpful suggestions, which have improved the presentation of this manuscript.
The MeerKAT telescope is operated by the South African Radio Astronomy Observatory, which is a facility of the National Research Foundation, an agency of the Department of Science and Innovation.
LOFAR data products were provided by the LOFAR Surveys Key Science project
(LSKSP; https://lofar-surveys.org/) and were derived from observations with the International LOFAR Telescope (ILT). LOFAR \citep{LOFAR2013} is the Low Frequency Array designed and constructed by ASTRON. It has observing, data processing, and data storage facilities in several countries, that are owned by various parties (each with their own funding sources), and that are collectively operated by the ILT foundation under a joint scientific policy. The efforts of the LSKSP have benefited from funding from the European Research Council, NOVA, NWO, CNRS-INSU, the SURF Co-operative, the UK Science and Technology Funding Council and the Jülich Supercomputing Centre.\\
This research was supported by the International Space Science Institute (ISSI) in Bern, through ISSI International Team project \#565 ({\it Multi-Wavelength Studies of the Culmination of Structure Formation in the Universe}).
MB, FG, AB, MR acknowledge the financial contribution from the INAF GO grant 1.05.24.02.10 {\it Extended Radio Emission in Galaxy Clusters at deep focus with MeerKAT}. 
AB acknowledges support from the ERC CoG $\vec{B}$ELOVED, GA N.101169773.
SE and MR acknowledge the financial contribution from the contracts Prin-MUR 2022 supported by Next Generation EU (M4.C2.1.1, n.20227RNLY3 {\it The concordance cosmological model: stress-tests with galaxy clusters}), and from the European Union’s Horizon 2020 Programme under the AHEAD2020 project (grant agreement n. 871158).
LL acknowledges the financial contribution from the INAF grant 1.05.12.04.01.
JS was supported by NASA Astrophysics Data Analysis Program (ADAP) Grant 80NSSC21K1571.
MS acknowledges the financial contributions from contract INAF mainstream project 1.05.01.86.10 and INAF Theory Grant 2023: Gravitational lensing detection of matter distribution at galaxy cluster boundaries and beyond (1.05.23.06.17).
EP acknowledges support from CNES, the French national space agency, and from ANR the French Agence Nationale de la Recherche, under grant ANR-22-CE31-0010.
M.G. acknowledges support from the ERC Consolidator Grant \textit{BlackHoleWeather} (101086804). R.C. acknowledges financial support from the INAF grant 2023 ``Testing the origin of giant radio halos with joint LOFAR-uGMRT observations'' (1.05.23.05.11).\\
We acknowledge the developers of the following Python packages which were used in this work: \textsc{ASTROPY} \citep{astropy:2013,astropy:2018,astropy:2022}, \textsc{MATPLOTLIB} \citep{matplotlib}, \textsc{SCIPY} \citep{SciPy-NMeth}, \textsc{NUMPY} \citep{numpy} and \textsc{PANDAS} \citep{pandas-mckinney,pandas}. 
\end{acknowledgements}

\bibliographystyle{aa}

\bibliography{biblio.bib} 

\begin{appendix}

\section{Spectral index uncertainty maps}\label{appendix:spidx_err}

In the following, we report the uncertainty maps for all the spectral index maps showed in Sect.~\ref{sec:sample} following their order of appearance.
\begin{figure}[ht!]
    \includegraphics[scale=0.115]{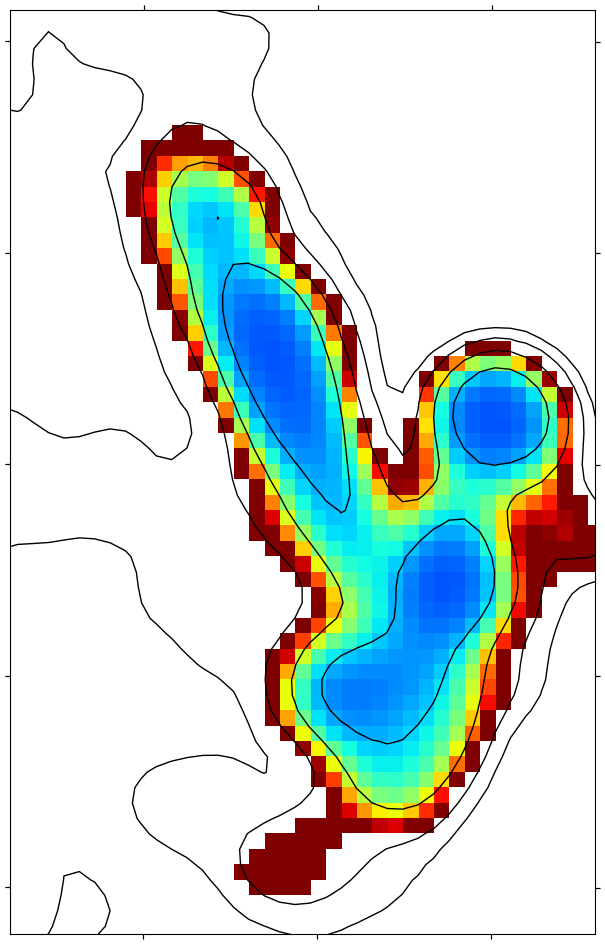}
    \includegraphics[scale=0.2]{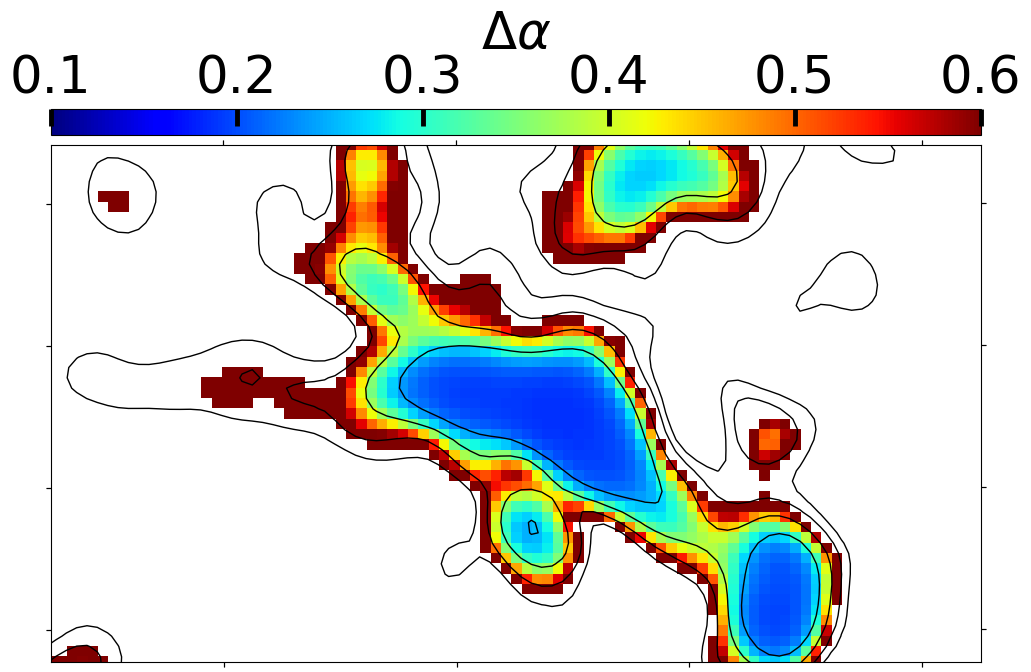}
    \caption{Spectral index uncertainty maps for Fig.~\ref{fig:1}.}
    \label{fig:1err}
\end{figure}
\begin{figure}[ht!]
\includegraphics[height=3.4cm]{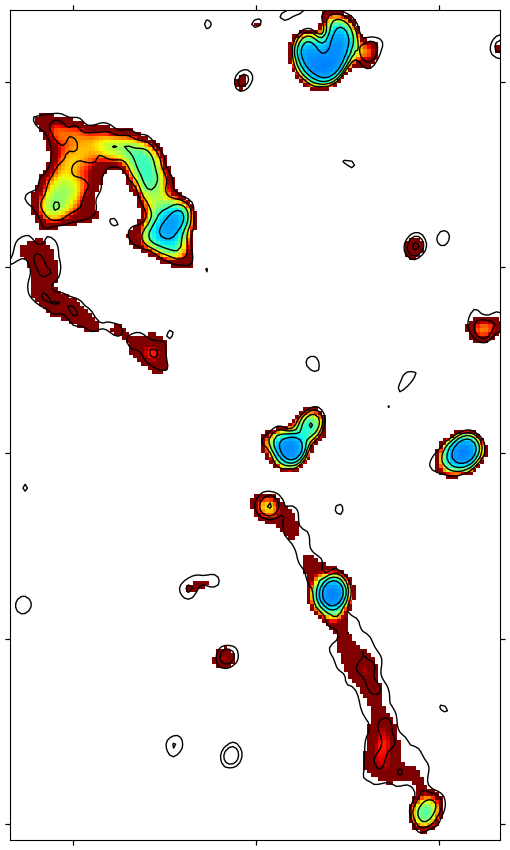}
\includegraphics[height=3.4cm]{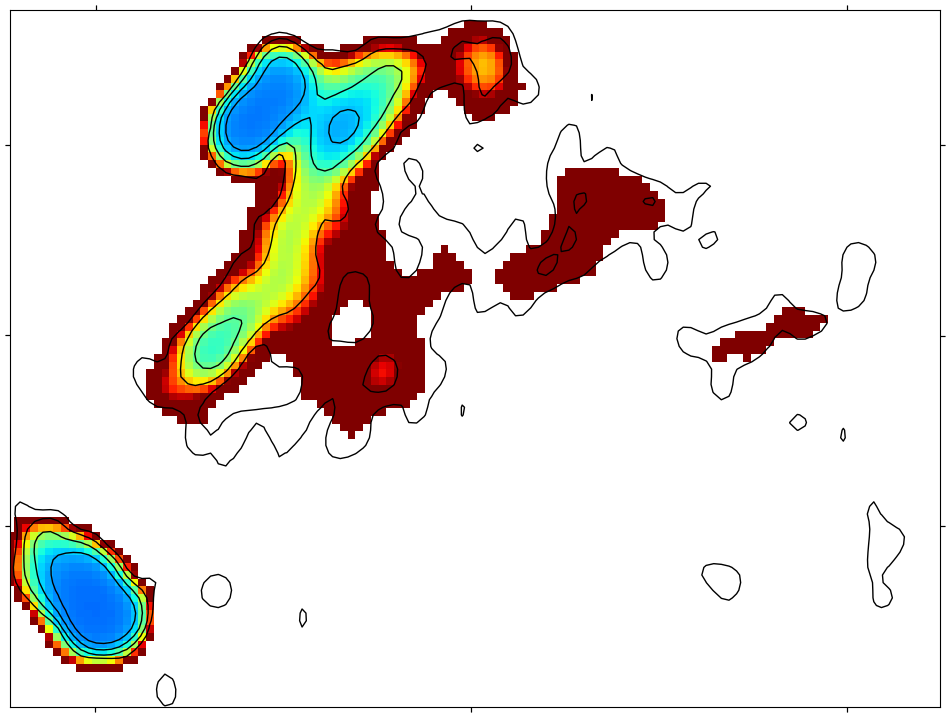}
\includegraphics[height=3.4cm]{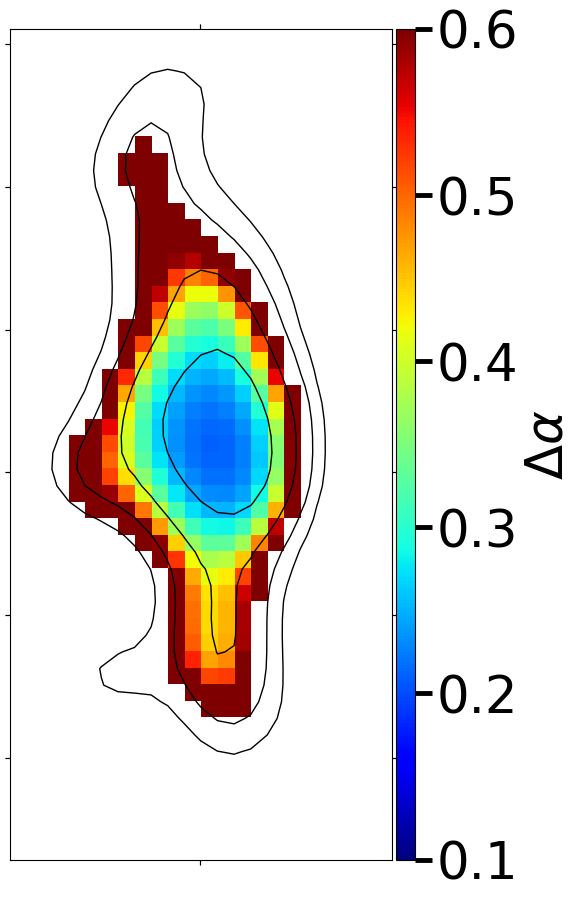}
    \caption{Spectral index uncertainty maps for Fig.~\ref{fig:2}.}
    \label{fig:2err}
\end{figure}
\begin{figure}[ht!]
    \includegraphics[scale=0.18]{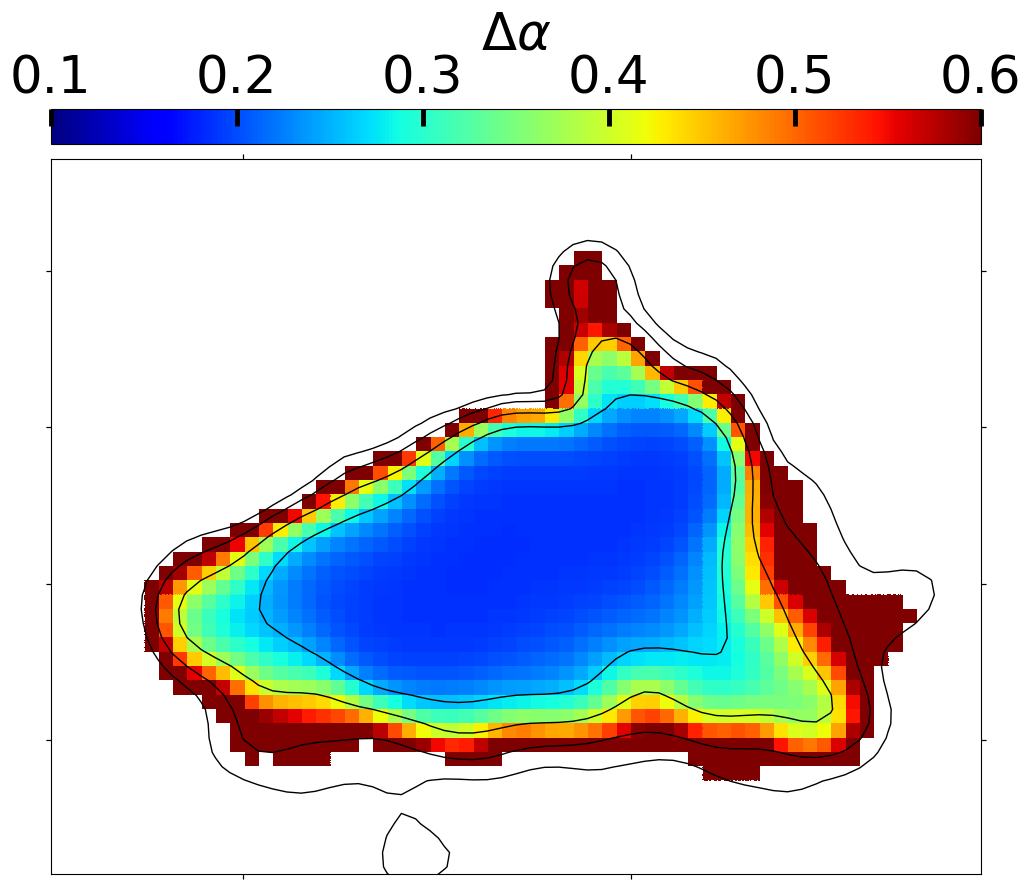}
    \includegraphics[scale=0.17]{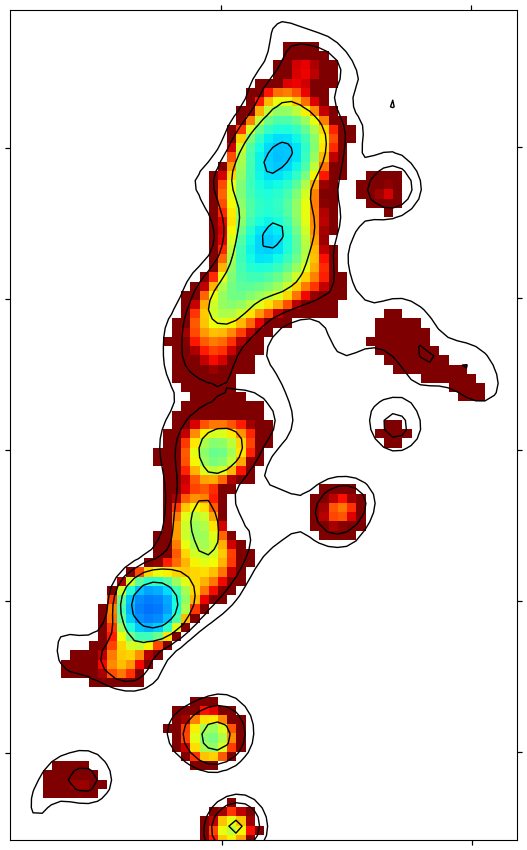}
    \includegraphics[scale=0.17]{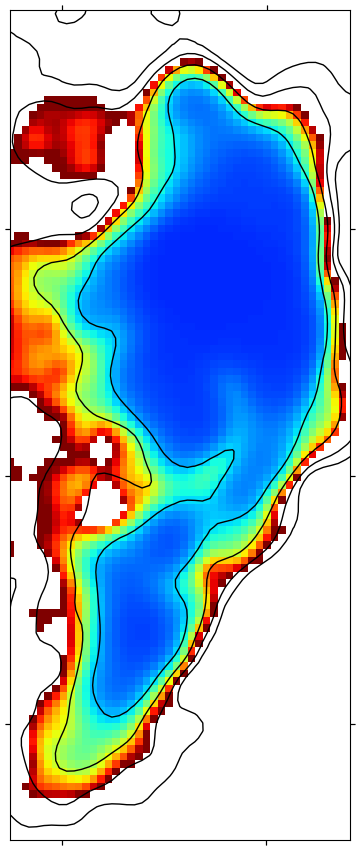} 
    \caption{Spectral index uncertainty maps for Fig.~\ref{fig:7}.}
    \label{fig:7err}
\end{figure}
\begin{figure}[ht!]
\includegraphics[height=2.8cm]{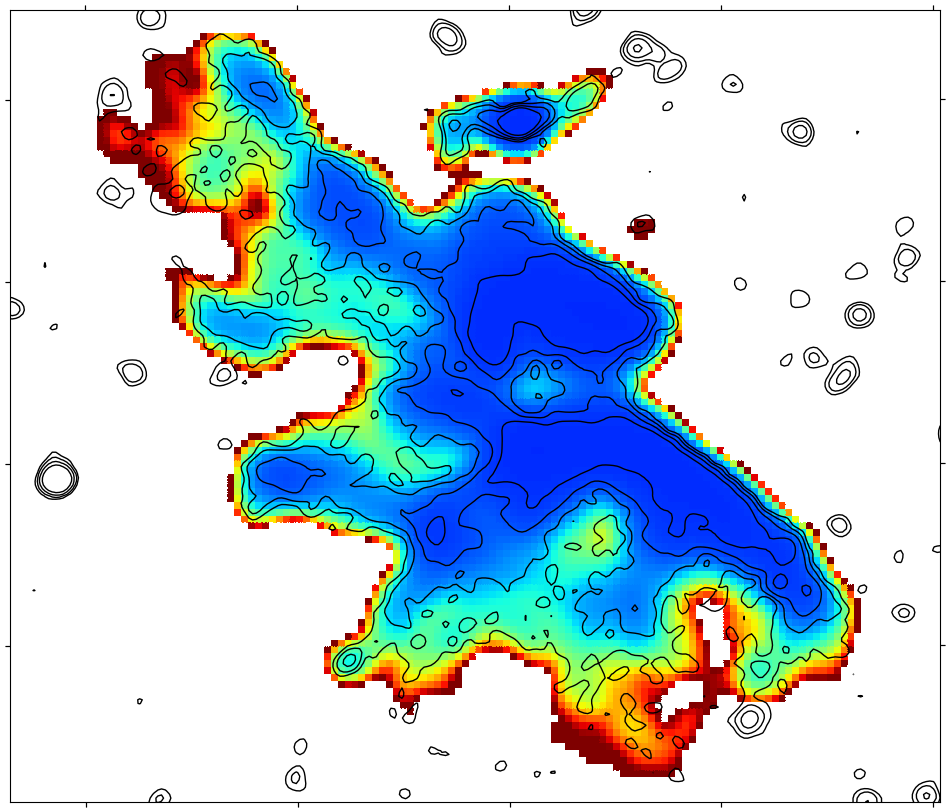}
\includegraphics[height=2.8cm]{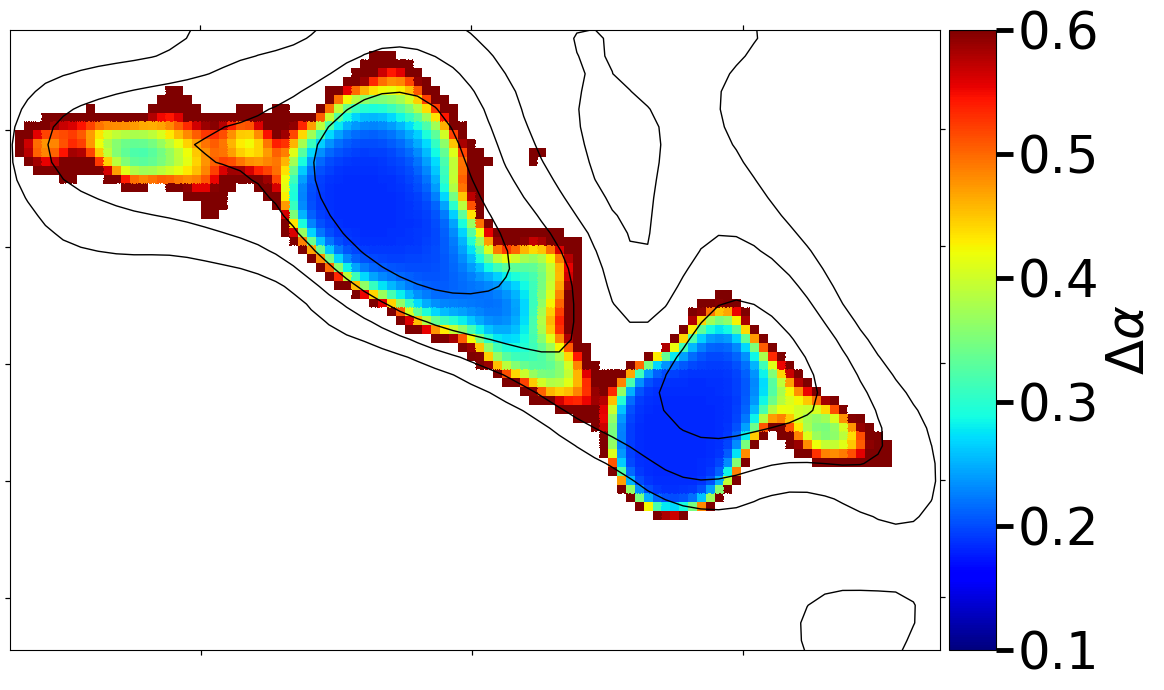} 
\caption{Spectral index uncertainty maps for Fig.~\ref{fig:9}.}
    \label{fig:9err}
\end{figure}
\begin{figure}[ht!]
\includegraphics[width=.275\linewidth, angle=90]{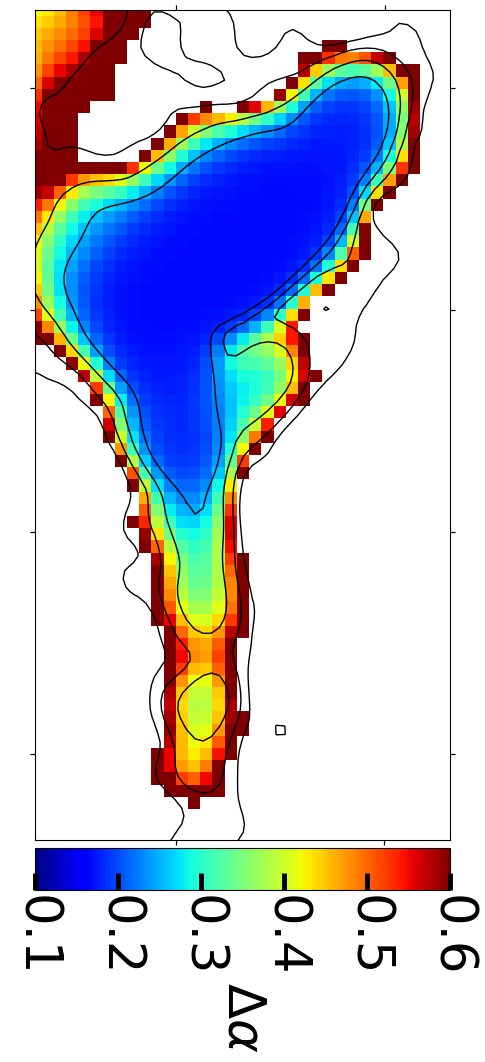}
    \caption{Spectral index uncertainty maps for Fig.~\ref{fig:12}.}
    \label{fig:12err}
\end{figure}
\begin{figure}[ht!]
\includegraphics[width=0.53\linewidth]{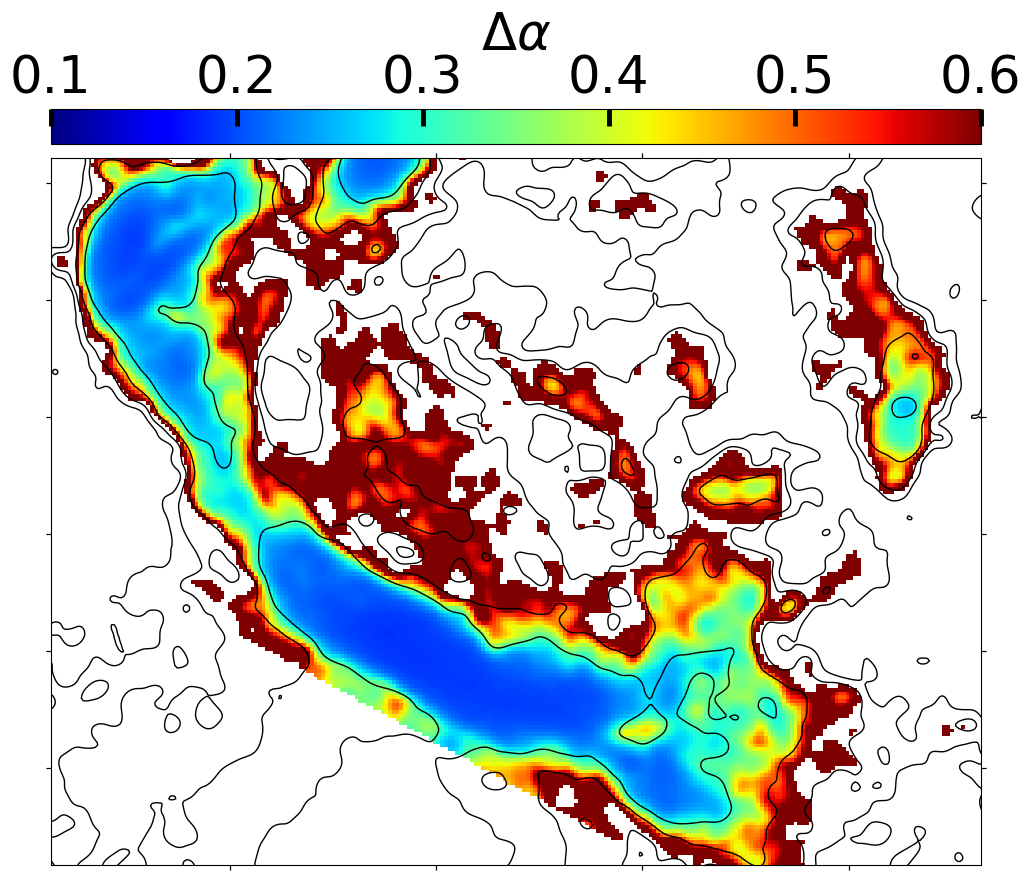}
\includegraphics[width=4cm, height=3.5cm]{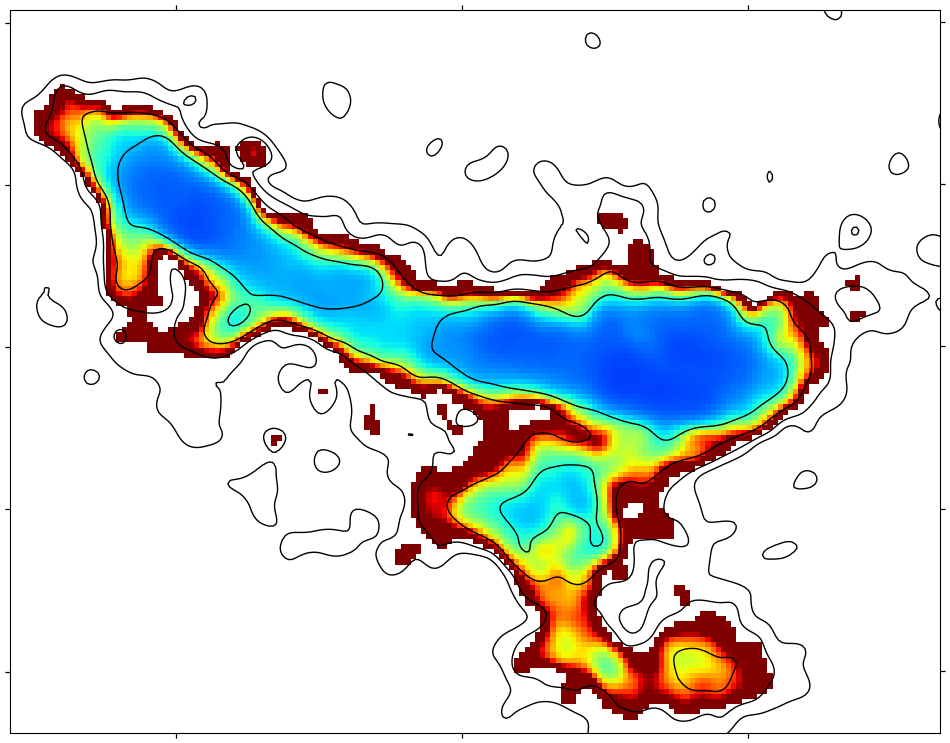} 
\caption{Spectral index uncertainty maps for Fig.~\ref{fig:13}.}
    \label{fig:13err}
\end{figure}
\begin{figure}[ht!]
    \includegraphics[scale=0.15]{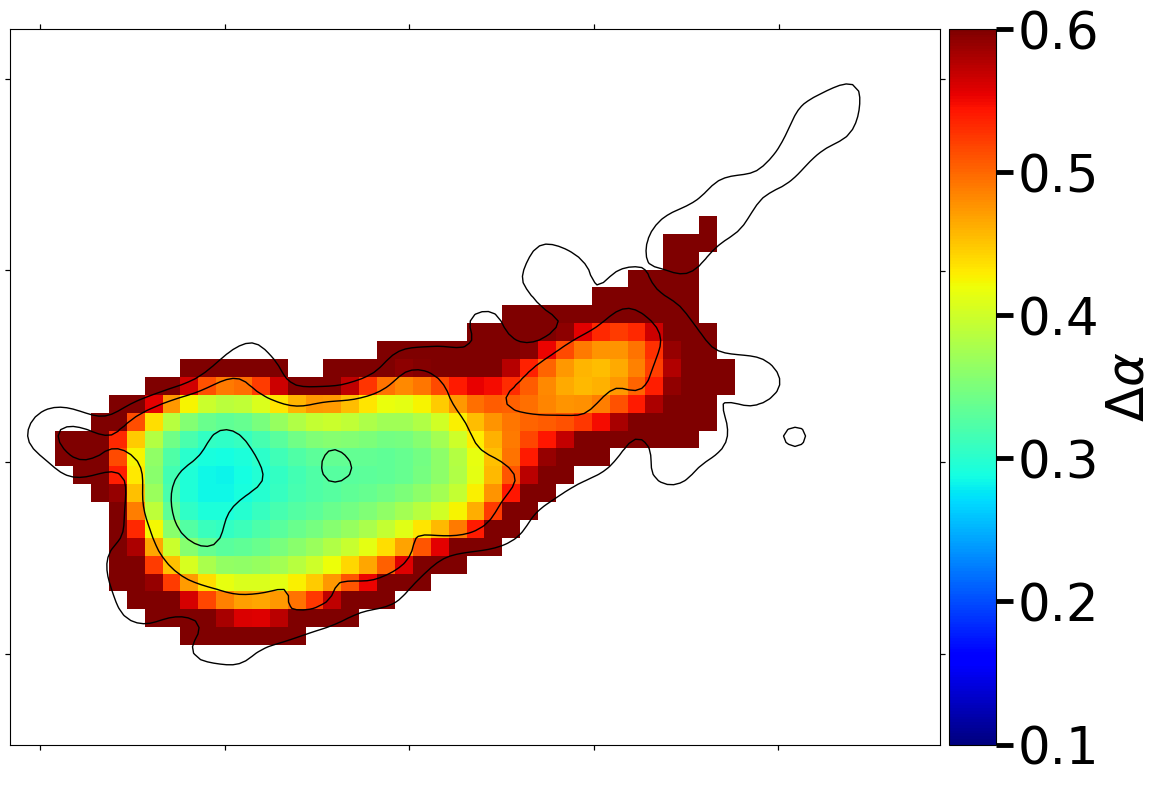}
    \caption{Spectral index uncertainty maps for Fig.~\ref{fig:16}.}
    \label{fig:16err}
\end{figure}
\section{MGCLS images}\label{appendix:mgcls_imgs}

Here, we report the images used in the paper to derive the radio halos and radio relics properties of the nine MGCLS clusters, once PSZ2G106.87-83.23 has been removed due to residual calibration artefacts.
\begin{figure*}[ht!]
    \centering
    \includegraphics[width=6cm, height=5cm]{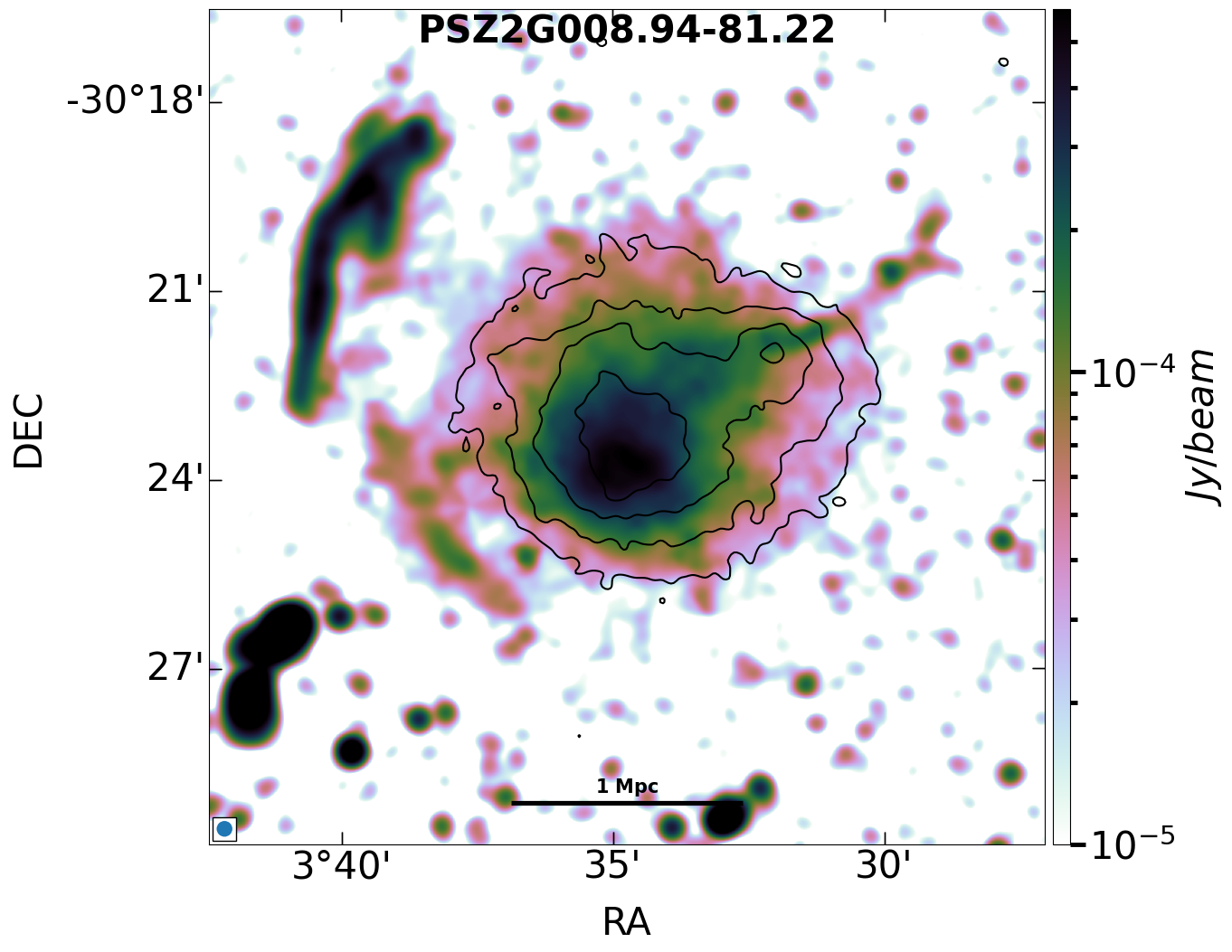}
    \includegraphics[width=6.15cm, height=5.cm]{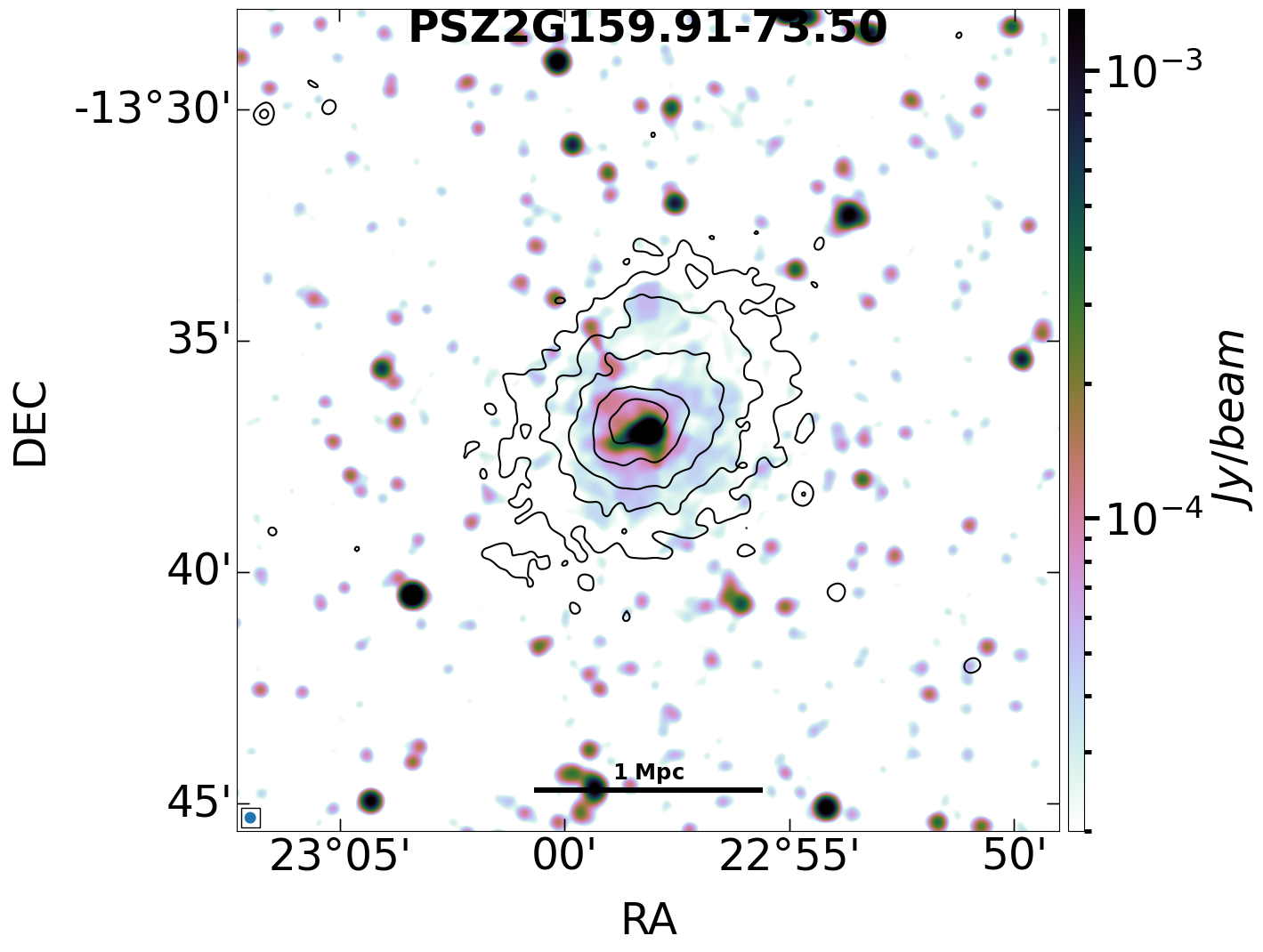}
    \includegraphics[width=6cm, height=5cm]{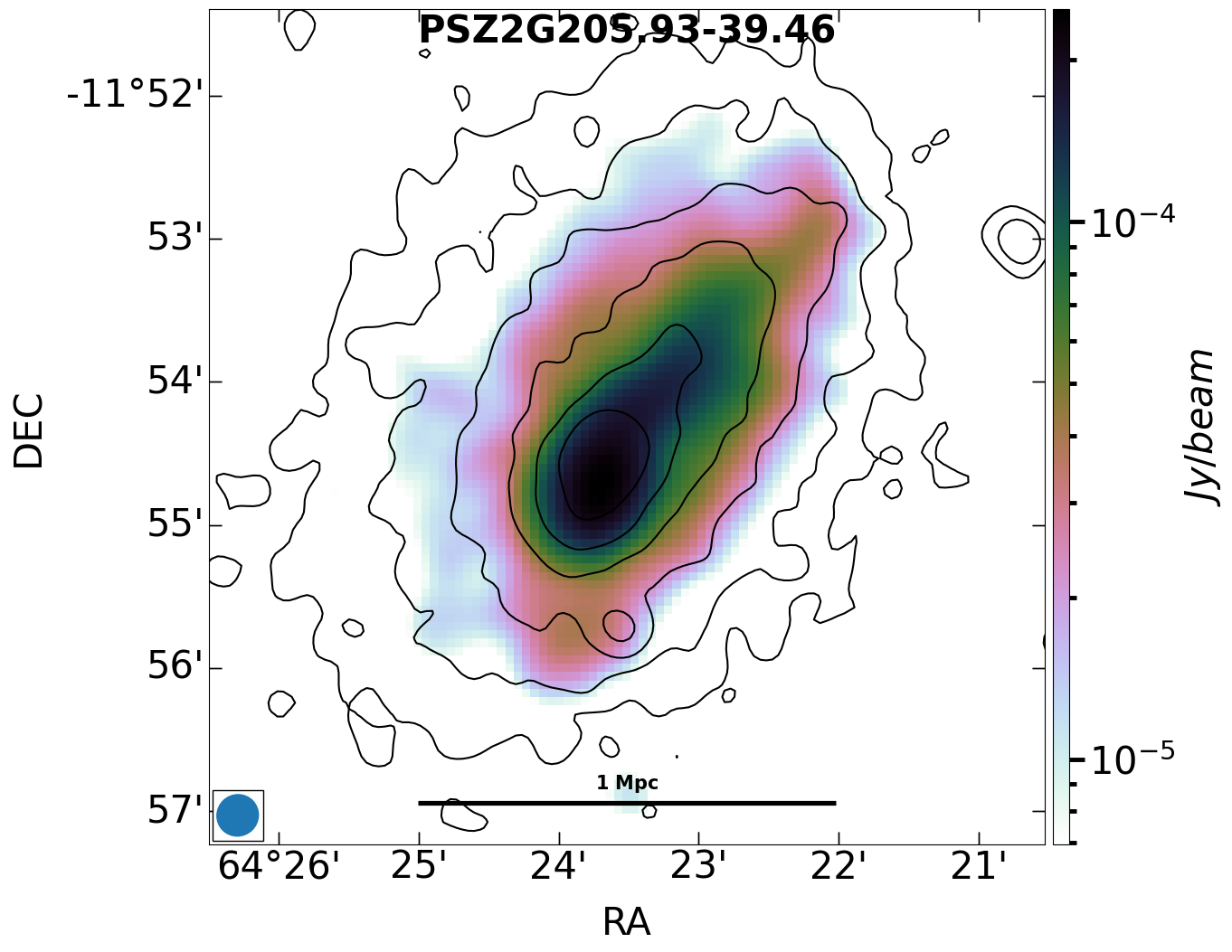}
    \includegraphics[width=6cm, height=5cm]{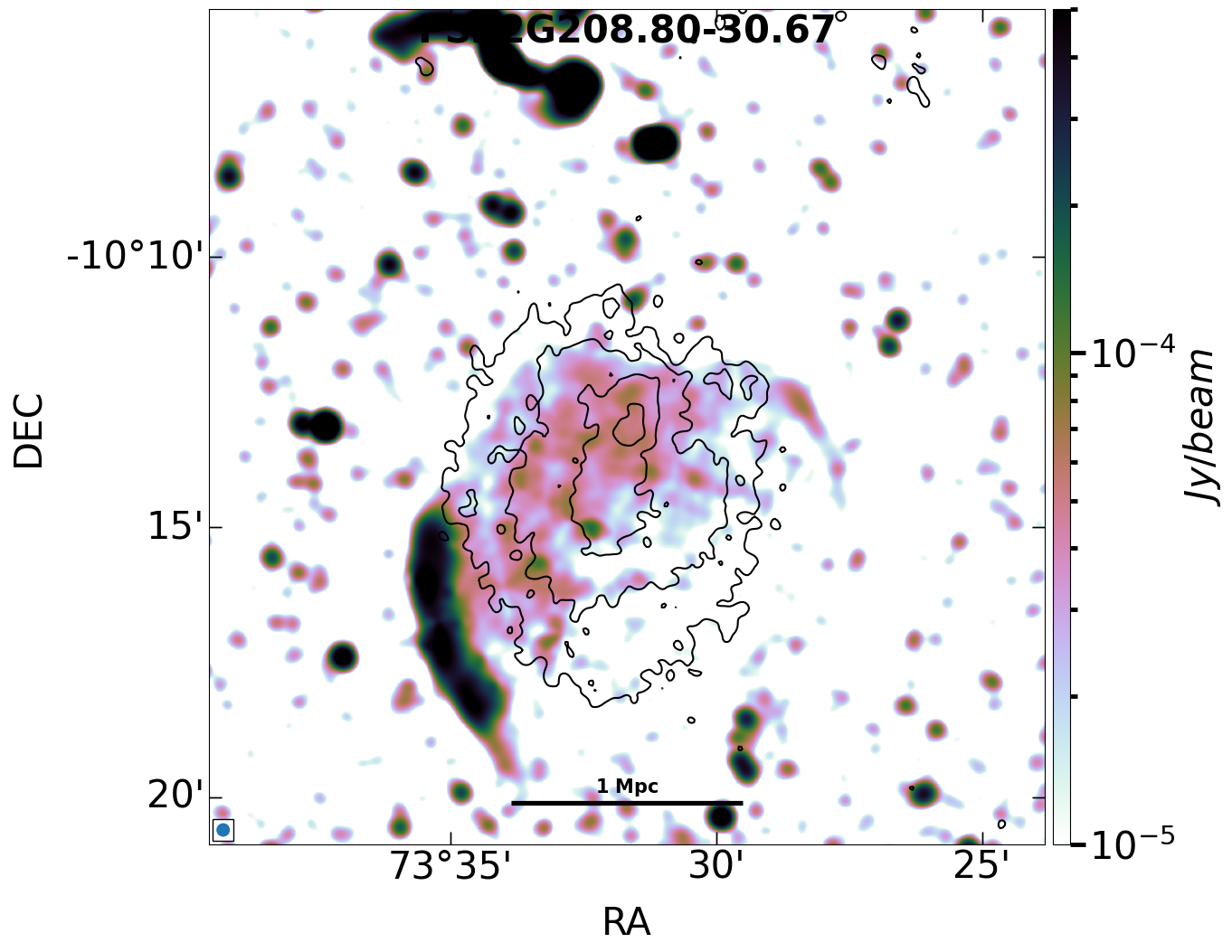}
    \includegraphics[width=6cm, height=5cm]{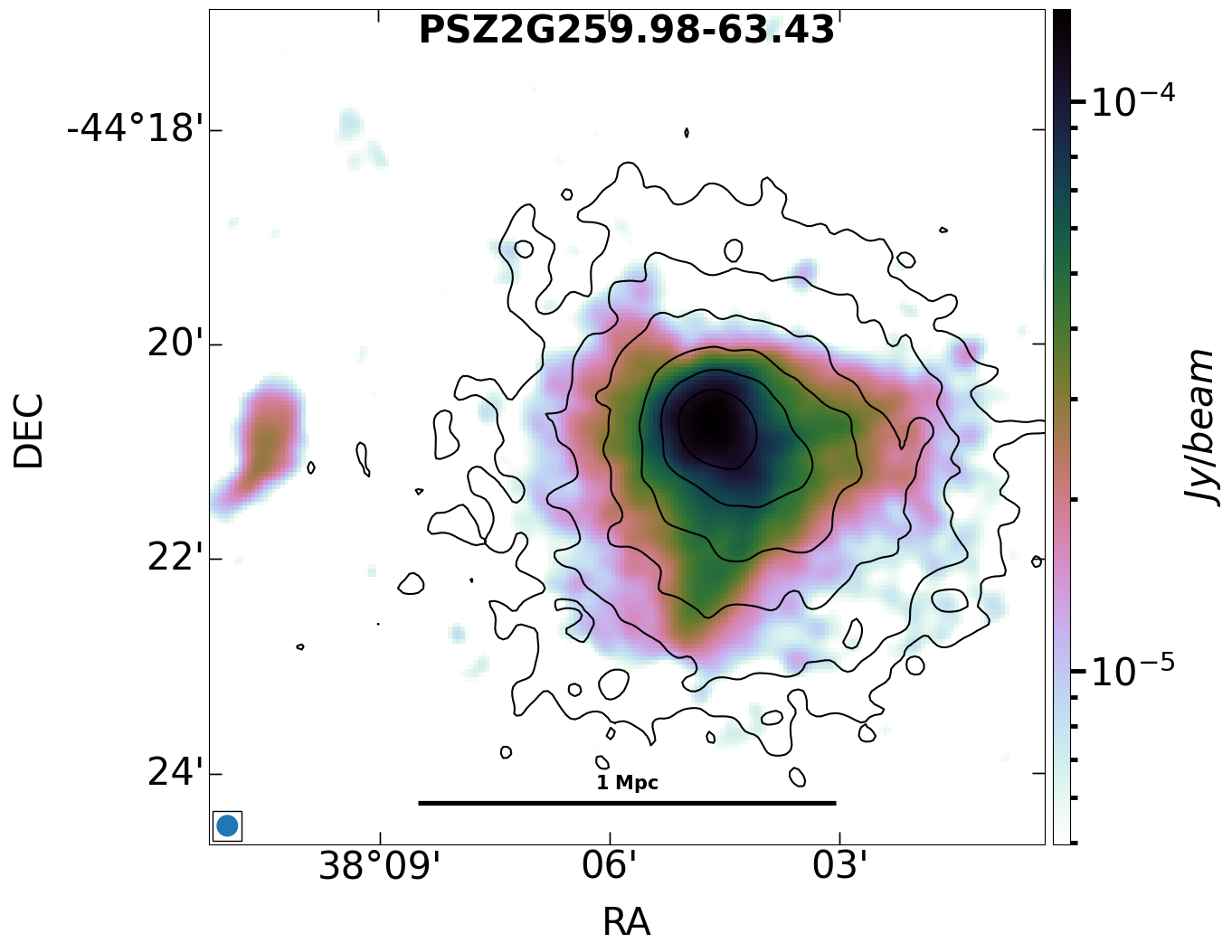}
    \includegraphics[width=6cm, height=5cm]{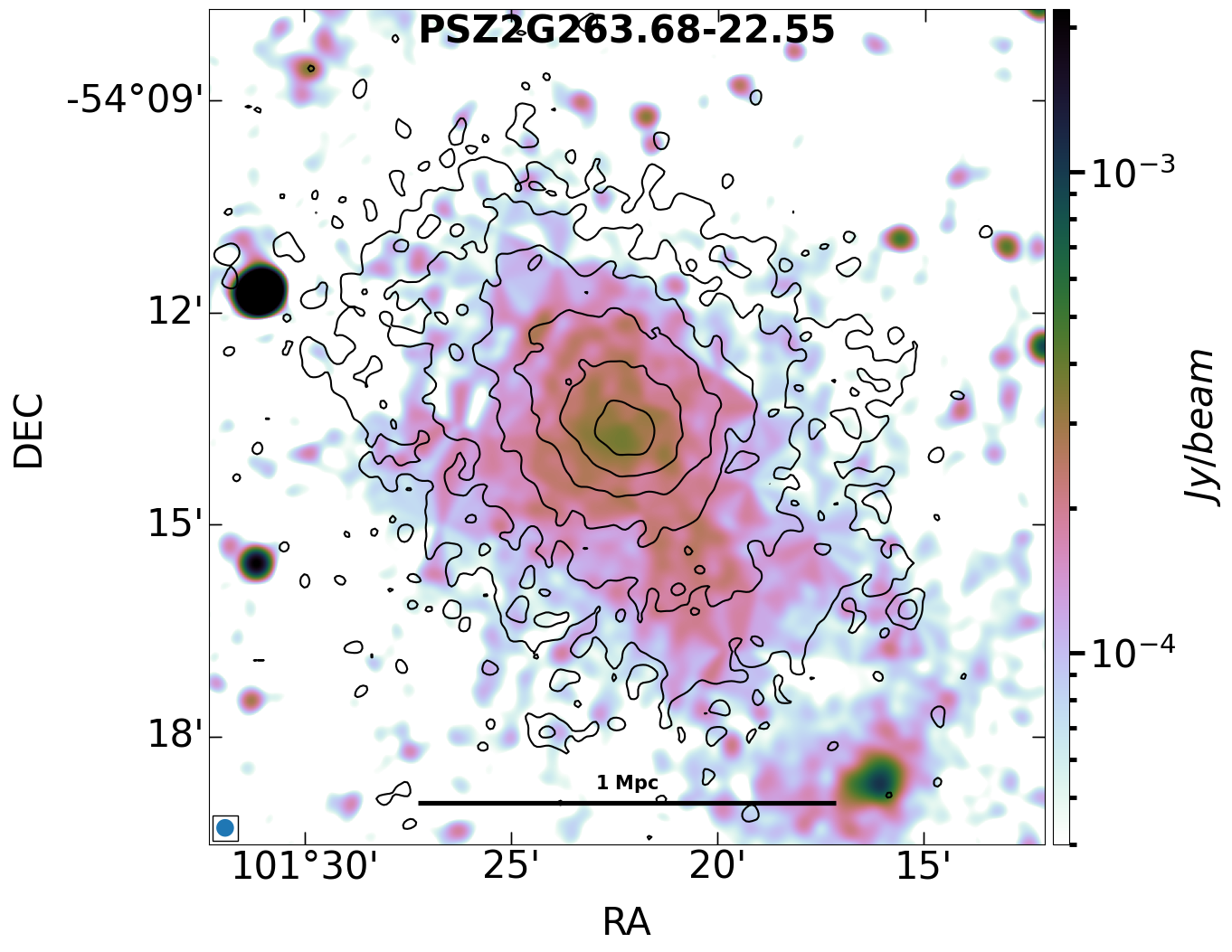}
    \includegraphics[width=6cm, height=5cm]{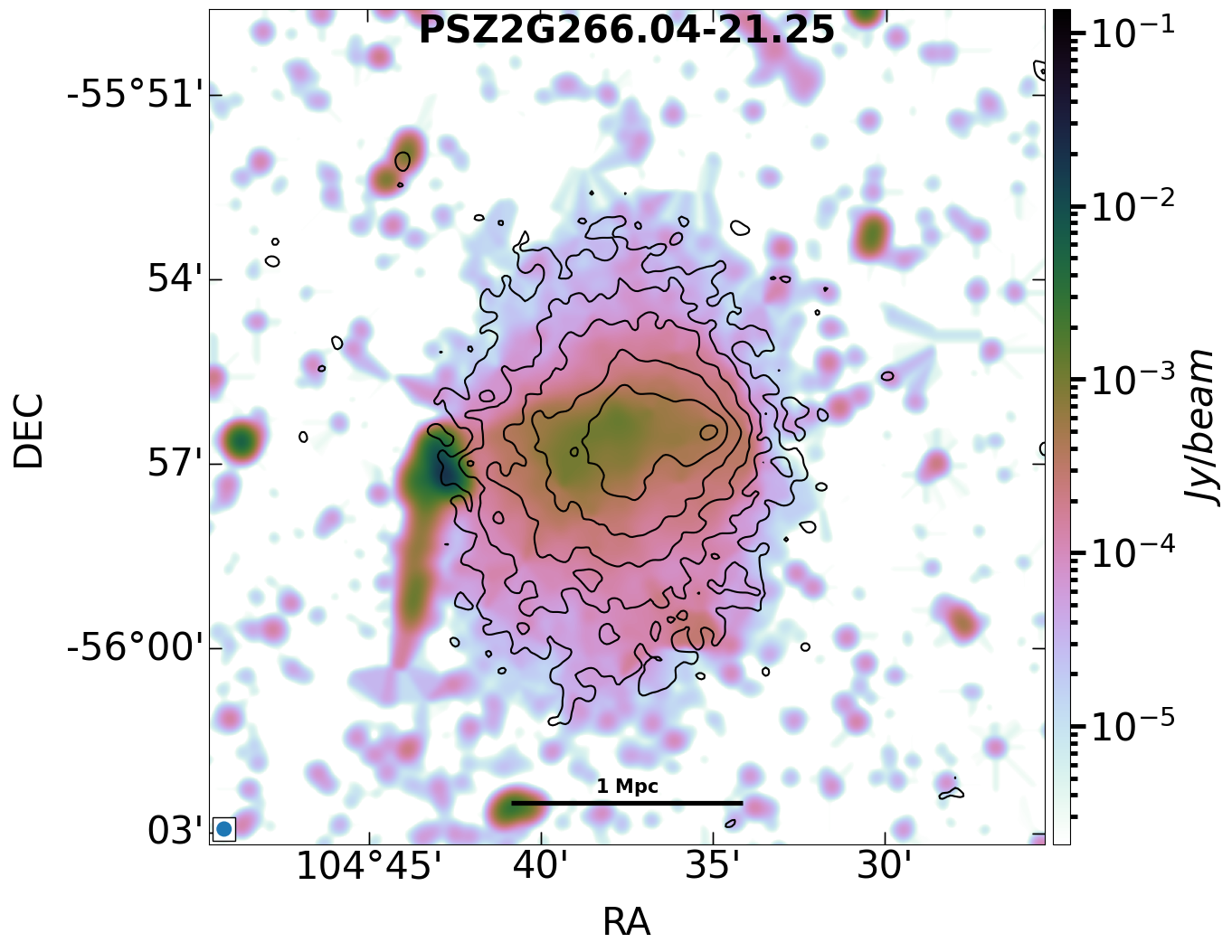}
    \includegraphics[width=6cm, height=5cm]{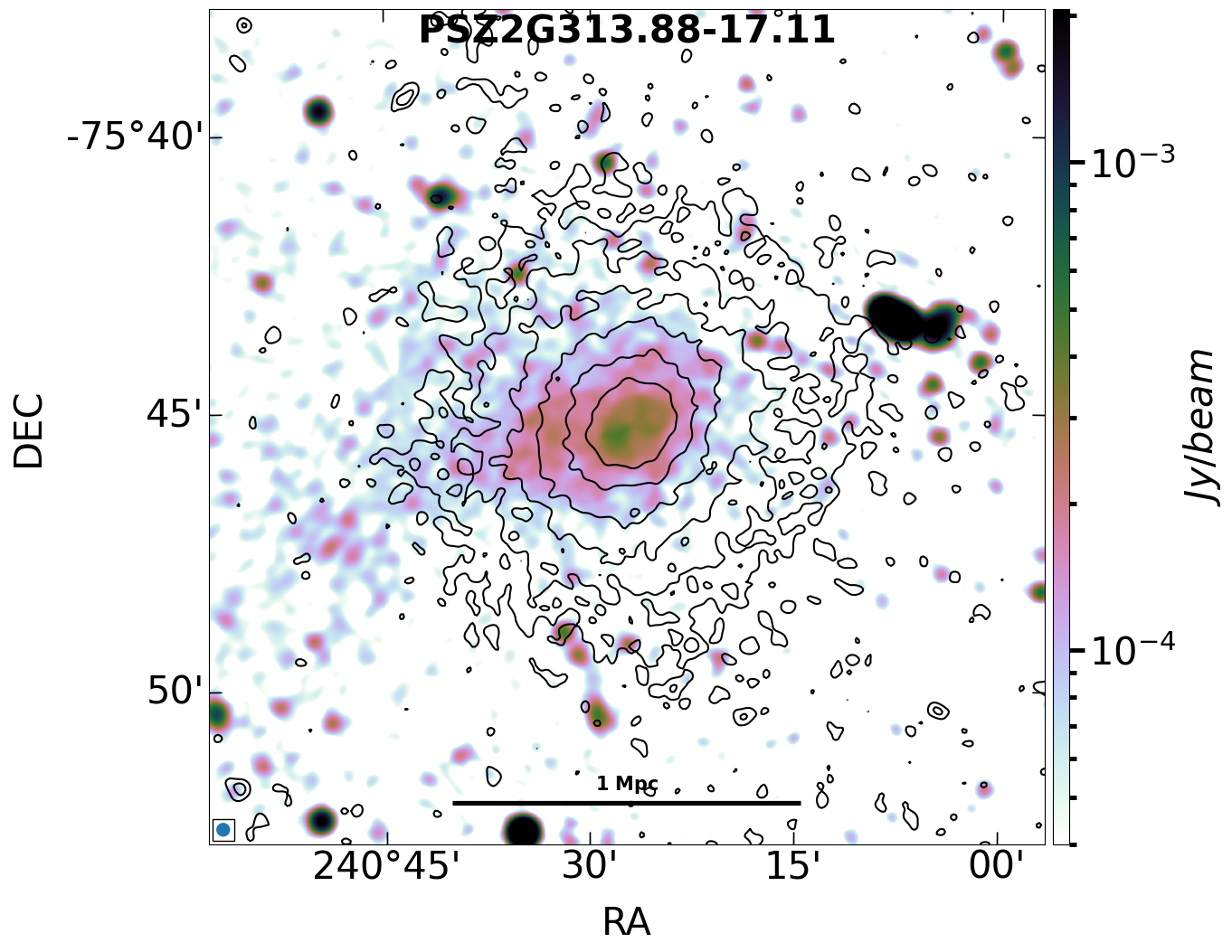}
    \includegraphics[width=6cm, height=5cm]{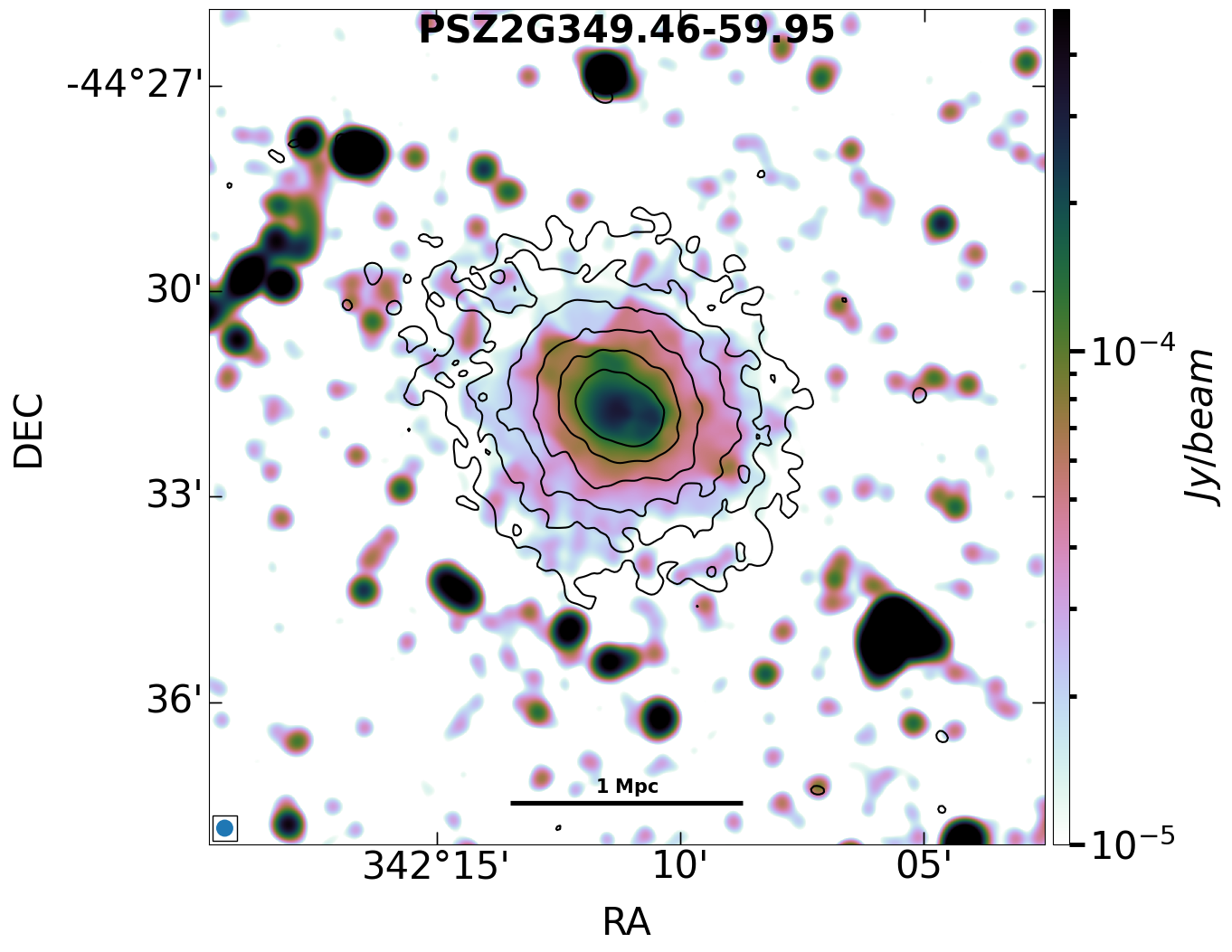}
    \caption{L-band image of MGCLS clusters considered with discrete sources within the cluster extension replaced by interpolated values.}
    \label{fig:MGCLS-clusters}
\end{figure*}

\section{Tables}
\begin{table*}
    \centering
    \caption{General information about the presented clusters. }
    \label{tab:targets_info}
    \begin{tabular}{lcccccccc}
    \toprule
    Name & $z$ & \thead{$R_{500}$ \\ (arcmin) } & \thead{$M_{500}$ \\ $(10^{14} {M_{\odot}})$ } &  \thead{$Y_{SZ}$ \\ $(10^{-3} ~ {\rm arcmin^{-2})}$} & $c$ & $w ~ (\times 10^{-1})$ & M \\
    \midrule
    \midrule
    PSZ2G008.31-64.74 & 0.31 & 4.51 & $7.42^{+0.39}_{-0.38}$ & $4.5 \pm 0.81$ & $0.21^{+0.02}_{-0.03}$ & $0.19^{+0.02}_{-0.03}$ & 0.82 \\
    PSZ2G008.94-81.22$^{\dag}$ & 0.31 & 4.87 & $8.99^{+0.36}_{-0.34}$ & $4.1 \pm 0.55$ & $0.19^{+0.09}_{-0.01}$ & $0.53^{+0.04}_{-0.05}$ & 1.06 \\
    PSZ2G056.93-55.08 & 0.45 & 3.70 & $9.49^{+0.42}_{-0.42}$ & $3.13 \pm 0.51$ & $0.17^{+0.01}_{-0.01}$ & $0.25^{+0.02}_{-0.03}$ & 0.67 \\
    PSZ2G106.87-83.23$^{\dag}$ & 0.29 & 4.81 & $7.73^{+0.35}_{-0.36}$ & $2.54 \pm 0.39$ & $0.34^{+0.05}_{-0.05}$ & $0.09^{+0.01}_{-0.03}$ & -0.65 \\
    PSZ2G159.91-73.50$^{\dag}$ & 0.21 & 6.63 & $8.46^{+0.28}_{-0.32}$ & $6.98 \pm 0.94$ & $0.31^{+0.04}_{-0.04}$ & $0.19^{+0.03}_{-0.04}$ & 0.17 \\
    PSZ2G172.98-53.55 & 0.37 & 3.91 & $7.37^{+0.54}_{-0.55}$ & $3.25 \pm 0.85$ & $0.23^{+0.03}_{-0.04}$ & $0.09^{+0.02}_{-0.04}$ & 0.36 \\
    PSZ2G205.93-39.46$^{\dag}$ & 0.44 & 3.98 & $11.54^{+0.5}_{-0.52}$ & $3.65 \pm 0.52$ & $0.38^{+0.04}_{-0.05}$ & $0.06^{+0.01}_{-0.02}$ & -0.47 \\
    PSZ2G208.80-30.67$^{\dag}$ & 0.25 & 5.40 & $7.26^{+0.47}_{-0.49}$ & $2.51 \pm 0.6$ & $0.15^{+0.02}_{-0.03}$ & $0.7^{+0.2}_{-0.2}$ & 1.43 \\
    PSZ2G225.93-19.99 & 0.44 & 3.82 & $9.79^{+0.47}_{-0.49}$ & $3.77 \pm 0.65$ & $0.24^{+0.02}_{-0.03}$ & $0.85^{+0.07}_{-0.1}$ & 1.49 \\
    PSZ2G239.27-26.01 & 0.43 & 3.71 & $8.77^{+0.44}_{-0.46}$ & $3.39 \pm 0.54$ & $0.24^{+0.02}_{-0.03}$ & $0.3^{+0.02}_{-0.03}$ & 0.34 \\
    PSZ2G243.15-73.84 & 0.41 & 3.75 & $8.09^{+0.48}_{-0.5}$ & $2.53 \pm 0.43$ & $0.16^{+0.01}_{-0.02}$ & $0.22^{+0.02}_{-0.04}$ & 0.39 \\
    PSZ2G259.98-63.43$^{\dag}$ & 0.28 & 4.87 & $7.45^{+0.33}_{-0.32}$ & $3.11 \pm 0.44$ & $0.44^{+0.04}_{-0.05}$ & $0.06^{+0.01}_{-0.02}$ & -0.78 \\
    PSZ2G262.27-35.38 & 0.30 & 4.98 & $8.76^{+0.24}_{-0.25}$ & $5.3 \pm 0.47$ & $0.14^{+0.02}_{-0.03}$ & $0.35^{+0.08}_{-0.1}$ & 0.97 \\
    PSZ2G263.68-22.55$^{\dag}$ & 0.16 & 7.89 & $7.96^{+0.23}_{-0.21}$ & $7.85 \pm 1.0$ & $0.41^{+0.06}_{-0.06}$ & $0.1^{+0.01}_{-0.02}$ & -0.55 \\
    PSZ2G266.04-21.25$^{\dag}$ & 0.30 & 5.58 & $12.47^{+0.27}_{-0.28}$ & $6.53 \pm 0.53$ & $0.27^{+0.02}_{-0.03}$ & $0.13^{+0.02}_{-0.03}$ & 0.02 \\
    PSZ2G277.76-51.74 & 0.44 & 3.65 & $8.65^{+0.33}_{-0.34}$ & $3.28 \pm 0.44$ & $0.14^{+0.02}_{-0.02}$ & $0.37^{+0.09}_{-0.11}$ & 1.05 \\
    PSZ2G278.58+39.16 & 0.31 & 4.73 & $8.29^{+0.42}_{-0.42}$ & $3.73 \pm 0.6$ & $0.32^{+0.05}_{-0.06}$ & $0.37^{+0.09}_{-0.11}$ & 0.70 \\
    PSZ2G286.98+32.90 & 0.39 & 4.65 & $13.74^{+0.37}_{-0.39}$ & $8.33 \pm 0.77$ & $0.22^{+0.04}_{-0.04}$ & $0.17^{+0.01}_{-0.04}$ & 0.51 \\
    PSZ2G313.33+61.13 & 0.18 & 7.42 & $8.77^{+0.34}_{-0.34}$ & $6.13 \pm 0.81$ & $0.55^{+0.04}_{-0.04}$ & $0.042^{+0.001}_{-0.004}$ & -1.35 \\
    PSZ2G313.88-17.11$^{\dag}$ & 0.15 & 8.37 & $7.86^{+0.26}_{-0.27}$ & $8.61 \pm 1.02$ & $0.5^{+0.07}_{-0.07}$ & $0.034^{+0.01}_{-0.002}$ & -1.09 \\
    PSZ2G346.61+35.06 & 0.22 & 6.20 & $8.41^{+0.39}_{-0.43}$ & $7.25 \pm 1.13$ & $0.14^{+0.04}_{-0.05}$ & $0.6^{+0.3}_{-0.3}$ & 1.51 \\
    PSZ2G349.46-59.95$^{\dag}$ & 0.35 & 4.77 & $11.36^{+0.34}_{-0.34}$ & $4.79 \pm 0.48$ & $0.44^{+0.03}_{-0.03}$ & $0.051^{+0.01}_{-0.014}$ & -0.71 \\
    \bottomrule
    \end{tabular}  
    \tablefoot{The ones marked with $\dagger$ are those belonging to the MGCLS sample.}
    \tablebib{
    (1)~cluster name (the ones with the $\dag$ are part of the MGCLS); (2)~redshift; (3)~$R_{500}$; (4)~$M_{500}$; (5)~SZ signal; (6)~concentration; (7)~centroid shift; (8)~level of dynamical disturbance as estimated by the M morphological parameters from C22.
    }
\end{table*}
\begin{table*}[ht!]
    \centering
    \caption{Best-fit values of the elliptical exponential model for each radio halo considered.}
    \begin{tabular}{lccccccc}
    \toprule
    Name & $P_{RH} ~ ( 10^{24} {\rm W ~ Hz^{-1}} ) $ & $I_0 ~ ( {\rm \mu Jy ~ arcsec^{-2}} ) $ & $r_1$ (kpc) & $r_2$ (kpc) & S/N & $\chi^2$\\
    \midrule
    PSZ2G008.31-64.74 & $7.06 \pm 0.48$ & $0.23 \pm 0.01$ & $1003 \pm 59$ & $376 \pm 25$ & 30.4 & 0.96 \\
    PSZ2G008.94-81.22 & $16.35 \pm 1.06$ & $2.80 \pm 0.09$ & $293 \pm 12$ & $246 \pm 10$ & 41.8 & 0.54 \\
    PSZ2G056.93-55.08 & $10.24 \pm 0.65$ & $0.83 \pm 0.02$ & $364 \pm 23$ & $270 \pm 21$ & 51.0 & 1.53 \\
    PSZ2G159.91-73.50 & $1.96 \pm 0.35$ & $0.55 \pm 0.30$ & $306 \pm 115$ & $205 \pm 58$ & 6.0 & 2.09 \\
    PSZ2G172.98-53.55 & $2.63 \pm 0.21$ & $0.46 \pm 0.04$ & $278 \pm 46$ & $204 \pm 44$ & 19.0 & 0.59 \\
    PSZ2G205.93-39.46 & $11.02 \pm 0.95$ & $1.67 \pm 0.15$ & $310 \pm 29$ & $172 \pm 19$ & 16.1 & 0.53 \\
    PSZ2G208.80-30.67 & $1.97 \pm 0.68$ & $0.38 \pm 1.45$ & $349 \pm 249$ & $226 \pm 145$ & 3.0 & 2.13 \\
    PSZ2G225.93-19.99 & $12.98 \pm 0.78$ & $2.30 \pm 0.02$ & $264 \pm 10$ & $177 \pm 10$ & 148.2 & 3.07 \\
    PSZ2G239.27-26.01 & $7.83 \pm 0.48$ & $1.33 \pm 0.02$ & $273 \pm 11$ & $181 \pm 10$ & 108.0 & 1.32 \\
    PSZ2G243.15-73.84 & $3.87 \pm 0.25$ & $0.57 \pm 0.02$ & $315 \pm 20$ & $193 \pm 11$ & 41.1 & 3.81 \\
    PSZ2G259.98-63.43 & $9.49 \pm 0.65$ & $6.12 \pm 0.30$ & $158 \pm 9$ & $131 \pm 9$ & 30.9 & 0.59 \\
    PSZ2G262.27-35.38 & $5.28 \pm 0.35$ & $0.27 \pm 0.01$ & $731 \pm 37$ & $340 \pm 22$ & 37.1 & 1.01 \\
    PSZ2G263.68-22.55 & $9.66 \pm 0.72$ & $1.30 \pm 0.07$ & $598 \pm 43$ & $251 \pm 14$ & 22.5 & 0.70 \\
    PSZ2G266.04-21.25 & $25.46 \pm 1.60$ & $5.72 \pm 0.15$ & $304 \pm 11$ & $187 \pm 8$ & 52.0 & 0.84 \\
    PSZ2G277.76-51.74 & $6.25 \pm 0.44$ & $0.28 \pm 0.02$ & $440 \pm 31$ & $409 \pm 33$ & 27.2 & 0.94 \\
    PSZ2G278.58+39.16 & $13.55 \pm 0.82$ & $4.74 \pm 0.04$ & $200 \pm 10$ & $177 \pm 10$ & 183.0 & 2.80 \\
    PSZ2G286.98+32.90 & $28.35 \pm 1.71$ & $2.36 \pm 0.02$ & $389 \pm 7$ & $293 \pm 7$ & 177.8 & 1.57 \\
    PSZ2G313.33+61.13 & $2.21 \pm 0.13$ & $1.40 \pm 0.02$ & $204 \pm 10$ & $146 \pm 10$ & 102.3 & 0.83 \\
    PSZ2G313.88-17.11 & $2.95 \pm 0.28$ & $1.54 \pm 0.12$ & $265 \pm 31$ & $153 \pm 16$ & 13.8 & 0.28 \\
    PSZ2G346.61+35.06 & $3.37 \pm 0.21$ & $0.34 \pm 0.01$ & $419 \pm 12$ & $393 \pm 13$ & 61.0 & 0.90 \\
    PSZ2G349.46-59.95 & $3.81 \pm 0.27$ & $1.44 \pm 0.08$ & $190 \pm 12$ & $151 \pm 11$ & 27.0 & 0.48 \\
    \bottomrule
    \end{tabular}
    \tablebib{
     (1)~cluster name; (2) total radio halo power at 1.28 GHz; (3)~radio halo central surface brightness; (4-5) e-folding radii; (6) signal-to-noise ratio of the emission; (7) reduced chi-square of the fit.
    }
    \label{tab:RH_prop}
\end{table*}
\begin{table*}[ht!]
    \centering
    \caption{Radio relics properties derived as described in Sect.~\ref{sec:radio_relics}. }
    \begin{tabular}{lccc}
    \toprule
    Name & $P_{RR} ~ ( 10^{24} {\rm W ~ Hz^{-1}} )$ & LLS (kpc) & $D_{cc-RR}$ (kpc) \\
    \midrule
    PSZ2G008.31-64.74 & $3.36 \pm 0.20$ & $1562 \pm 101$ & $1906 \pm 137$ \\
    PSZ2G008.94-81.22 & $0.74 \pm 0.04$ & $1051 \pm 68$ & $1079 \pm 77$ \\
    PSZ2G008.94-81.22 & $5.41 \pm 0.32$ & $1441 \pm 68$ & $1768 \pm 127$ \\
    PSZ2G056.93-55.08 & $1.51 \pm 0.09$ & $666 \pm 103$ & $2065 \pm 148$ \\
    PSZ2G205.93-39.46* & $1.68 \pm 0.10$ & $1172 \pm 86$ & $3047 \pm 219$ \\
    PSZ2G205.93-39.46* & $0.72 \pm 0.05$ & $382 \pm 86$ & $2767 \pm 199$ \\
    PSZ2G208.80-30.67 & $3.13 \pm 0.19$ & $1264 \pm 58$ & $1233 \pm 89$ \\
    PSZ2G243.15-73.84 & $15.06 \pm 0.90$ & $1097 \pm 55$ & $1229 \pm 88$ \\
    PSZ2G243.15-73.84* & $0.88 \pm 0.05$ & $767 \pm 55$ & $2236 \pm 161$ \\
    PSZ2G243.15-73.84 & $3.33 \pm 0.20$ & $562 \pm 55$ & $844 \pm 61$ \\
    PSZ2G259.98-63.43* & $0.34 \pm 0.02$ & $283 \pm 64$ & $1608 \pm 115$ \\
    PSZ2G259.98-63.43 & $0.68 \pm 0.04$ & $600 \pm 64$ & $2228 \pm 160$ \\
    PSZ2G262.27-35.38 & $1.67 \pm 0.10$ & $1079 \pm 53$ & $1934 \pm 139$ \\
    PSZ2G262.27-35.38 & $12.35 \pm 0.74$ & $1550 \pm 53$ & $2073 \pm 149$ \\
    PSZ2G266.04-21.25 & $23.97 \pm 1.44$ & $1006 \pm 66$ & $1562 \pm 112$ \\
    PSZ2G277.76-51.74* & $0.43 \pm 0.03$ & $925 \pm 143$ & $3970 \pm 285$ \\
    PSZ2G278.58+39.16 & $7.33 \pm 0.44$ & $586 \pm 54$ & $616 \pm 44$ \\
    PSZ2G286.98+32.90 & $14.97 \pm 0.90$ & $1827 \pm 53$ & $3184 \pm 229$ \\
    PSZ2G286.98+32.90 & $29.47 \pm 1.77$ & $2012 \pm 53$ & $529 \pm 38$ \\
    \bottomrule
    \end{tabular}   
    \tablefoot{Objects marked with * are the candidate radio relics.}
    \tablebib{
    (1)~cluster name; (2)~radio relic power at 1.28 GHz; (3)~LLS; (4)$D_{cc-RR}$.}
    \label{tab:RR_prop}
\end{table*}
\end{appendix}

\end{document}